\documentclass{aa} 

\usepackage{natbib}
\usepackage{longtable}
\usepackage{lscape}
\usepackage{graphicx}
\usepackage{txfonts}
\usepackage[dvipsnames]{xcolor}
\usepackage{supertabular}
\usepackage{xfrac}  
\usepackage{multicol}

\bibpunct{(}{)}{;}{a}{}{,} % to follow the A&A style
\providecommand{\abs}[1]{\lvert#1\rvert}

\begin{document} 

   \title{Planetary nebulae in Gaia EDR3: Central Star identification, properties, and binarity} 
   % wide binaries?? : separation>20.000 au (Fran)
   %\title{Astrometric wide binaries among Gaia DR2 Planetary Nebulae}
   
   \titlerunning{Properties of CSPNe in Gaia EDR3}

   \author{I. Gonz\'alez-Santamar\'{\i}a\inst{1,2} \and M. Manteiga\inst{2,3} \and A. Manchado\inst{4,5,6}  \and  A. Ulla\inst{7} \and C. Dafonte\inst{1,2} \and P. L\'opez Varela\inst{3}}
   \authorrunning{I. Gonz\'alez-Santamar\'{\i}a et al.}
   \institute{Universidade da Coru\~na (UDC), Department of Computer Science and Information Technologies, Campus Elvi\~na s/n, 15071 A Coru\~na, Spain \\
  \email{iker.gonzalez@udc.es}
  \and
  CIGUS CITIC, Centre for Information and Communications Technologies Research, Universidade da Coru\~na, Campus de Elvi\~na s/n, 15071 A Coru\~na, Spain                    
  \and
   Universidade da Coru\~na (UDC), Department of Nautical Sciences and Marine Engineering, Paseo de Ronda 51, 15011, A Coru\~na, Spain \\
   \email{manteiga@udc.es}
  \and
  Instituto de Astrofísica de Canarias, 38200 La Laguna, Tenerife, Spain
  \and
  Universidad de La Laguna (ULL), Astrophysics Department, 38206 La Laguna, Tenerife, Spain
  \and 
  CSIC, Spain
 \and
  Universidade de Vigo (UVIGO), Applied Physics Department, Campus Lagoas-Marcosende, s/n, 36310 Vigo, Spain
  \\
           }
   \subtitle{}

   \date{Received 30 July 2021 / Accepted 24 September 2021}

  \abstract
   {The Gaia Early Data Release 3 (EDR3), published in December 2020, features improved photometry 
and astrometry compared to that published in the previous DR2 file and includes a substantially larger 
number of sources, of the order of 2,000 million, making it a paradigm of big data astronomy. Many of the central stars of planetary nebulae (CSPNe) are inherently faint 
and difficult to identify within the field of the nebula itself. Gaia measurements may be 
relevant not only in identifying the ionising source of each nebula, but also in the study their physical 
and evolutionary properties.}
   {We demonstrate how Gaia data mining can effectively help to solve the issue of central star 
misidentification, a problem that has plagued the field since its origin. As we did for DR2, 
our objective is to present a catalogue of CSPNe with astrometric and photometric information in EDR3.
   %, from 
   From that catalogue, we selected a sample of stars with high-quality astrometric 
parameters, on which we carried out a more accurate analysis of CSPNe properties.} 
   { Gaia $G_{BP}-G_{RP}$ colours allow us to select the sources with sufficient temperatures to 
ionise the nebula. In order to estimate the real colour of a source, it is important to take 
into account interstellar extinction and, in the case of compact nebulae, nebular extinction when 
   %it is available. When combined with
   available. 
   %There data combined 
   In addition, distances derived from EDR3 parallaxes (combined with consistent literature values) 
can be used to obtain nebular intrinsic properties from those observed. With 
this information, CSPNe can be plotted in an Hertzsprung-Russell (HR) diagram. From information on 
the spectral classification of the CS (from the literature) and evolutionary models for post-AGB 
stars, their evolutionary state can then be analysed. Furthermore, EDR3 high-quality astrometric 
data enable us to search for objects comoving with CSs in the field of each nebula by detecting 
sources with parallaxes and proper motions similar to those of the CS. 
   }
   %we study the presence of comoving objects to CSs in the field of each nebulae, and we also analyse the possible relationship of the detected red colour stars with the presence of close binaries.} 
   %correlate red colours 
   %we correlate detected red colours for the suspected CS and anomalous astrometry with binarity. }
   %Iker: Incluir esto? - Finally, the properties of stars deficient in H are studied separately by means of evolutionary models suitable for this type of stars. }
   {We present a catalogue  
   %by 
   of 2035 PNe with their corresponding CS identification from among Gaia EDR3 sources. We obtain the distances for those 
   %ones 
   with known parallaxes in EDR3 (1725 PNe). In addition, for a sub-sample (405 PNe) with the most 
accurate distances, we obtain different nebular properties such as their Galactic distribution, radius, 
kinematic age, and morphology. Furthermore, for a set of 74 CSPNe, we present the evolutionary state 
(mass and age) derived from their luminosities and effective temperatures from evolutionary models. 
Finally, we highlight the detection of several wide binary CSPNe through an analysis of the EDR3 
astrometric parameters, and we contribute to shedding some light on the relevance of close binarity in CSPNe.}
   {}

        
\maketitle

\section{Introduction}

This article is an extension of our previous studies on the Galactic planetary nebulae (PNe) 
population using Gaia Data Release 2 (DR2): \citealt{2019A&A...630A.150G} (Paper 1) 
and \citealt{2020A&A...644A.173G} (Paper 2). Here, we make use of the recently 
published Gaia Early Data Release 3 (EDR3) archive, which provides  
a greater quantity of higher quality astrometric and photometric data that allow us to update and enlarge our PNe catalogue 
and analyse its results in greater detail.

The EDR3 archive contains astrometric data (positions, parallaxes and proper motions) and 
photometric data (in three bands) for almost 2000 million sources with improved precision 
compared with DR2. These accurate new data have allowed us to carry out a study of 
the physical and evolutionary properties of PNe with improved statistics and to improve 
the reliability of nebular central star (CS) identifications. Using these
 data, we were also able to detect wide binary systems associated with PNe with greater
 precision. The Gaia mission continues its operations and will release more improved
 data in the near future. Specifically, the full Gaia Data Release 3 (DR3) is planned for 
the first half of 2022 and will contain spectrophotometry in the 
330--1050 nm wavelength
range for a comprehensive astronomical sample.

Recent literature entries studying the Galactic PNe population includes
 \citet{2020A&A...640A..10W}, which provides a catalogue of 620 PNe, analysing several 
parameters of  CSs such as spectral types, effective temperatures, and luminosities. 
Another study by \citet{2020A&A...638A.103C} provides a procedure for selecting central stars of planetary nebulae (CSPNe) from DR2, which was recently updated by EDR3 \citep{2021arXiv210213654C}. 
Several previous studies have taken advantage of the 
extraordinary quality of Gaia DR2 astrometry to search for wide binaries using astrometric
 measurements such as that by \citet{2019AJ....157...78J}, whose methodology was the base for 
the search for wide binaries among DR2 CSPNe that we presented in Paper 2. 
%and Concerning the search for wide binary systems using Gaia astrometry, we have inspired in the work of \citet{2019AJ....157...78J}, in which they have detected more than 3,700 wide binary and multiple star systems. %There is also a recent work by \citet{2021A&A...648A..95C} which is focused in the search of close binary CSPNe based on photometric variability and using Gaia data.

We start in Sect. 2 by explaining  the procedure we used to select a sample of Galactic 
PNe and the method we followed to identify their CSs. As we mentioned before, the literature is 
plagued with CS misidentifications, and in this sense Gaia multi-band measurements are of
great help in finding the source most likely to be the ionising star. From Gaia parallax 
measurements, we obtained distances that endorse the new Bayesian distances catalogue by 
\citet{2021AJ....161..147B}. 
%, and, as we did in Paper 1 we compare such distances with other measurements available in the literature that were historically used to infer distances and properties to PNe. 
In Sect. 3, we select a sub-sample of PNe with the most reliable distance measurements to carry 
out a more detailed study of these sources. 
% in order to study with more detail the properties of PNe derived using such distances, as the physical radii or the kinematic ages, as well as other properties such as their galactic distribution or morphology. 
As in Paper 1, we compare such distances with other measurements available in 
the literature that have historically been  used to attribute distances and properties to PNe. 
%Furthermore, in this section, we compare our distances from Gaia EDR3 with those obtained in previous studies. 
Section 4 is devoted to the study of the evolutionary state of CSPNe 
as derived 
from their luminosities and effective temperatures, and to a comparison of their positions in 
%an 
an Hertzsprung-Russell (HR) diagram with those predicted by evolutionary tracks for post-AGB stars 
(\citealt{2016A&A...588A..25M}). This approach allowed us to estimate masses and evolutionary ages and 
compare their predicted evolutionary state with other properties such as size and  spectral 
classification. 
%We also add a small discussion about their spectral types. 
Finally, in Sect. 5, 
% we analyse the binarity rate
we use the astrometric information available in EDR3 to look for signs of binarity among the selected 
CSPNe. Concretely, we search for wide binaries by using EDR3 astrometric parameters that 
 enable us to detect objects comoving with the CSs. In addition, we carried out a statistical procedure 
in order to detect possible close binaries.

\section{Galactic planetary nebulae catalogue}

The 
%first -- Ana: main, mejor?
main aim of this study is to help consolidate a Galactic CSPNe catalogue by studying the 
information on the properties of point-like sources present in Gaia EDR3. As in 
Paper 1, we started by gathering as many PNe as possible from the literature. Specifically, 
we collected all objects catalogued as True, Likely, or Possible PNe in 
the HASH (Hong Kong/AAO/Strasbourg/$H_{\alpha}$) database (\citealt{parker16}), which already 
contains almost all the objects listed in other PNe catalogues, such as those by \citet{kerber03}, 
\citet{stanghellini10}, and \citet{2020A&A...640A..10W}. 

%In this PNe database, at the same time, are included almost all objects contained in other PNe catalogues as %The first aim of this work is to create a galactic PNe catalogue and to identify their corresponding Central Star (CS) among the sources in the recent survey of Gaia EDR3. For this task, we collected all objects catalogued as \textit{True}, \textit{Likely} or \textit{Possible} PN in the HASH (Hong Kong/AAO/Strasbourg/$H_{\alpha}$) database (\citealt{parker16}). In this PNe database, at the same time, are included almost all objects contained in other PNe catalogues as \citet{kerber03}, \citet{stanghellini10} or \citet{2020A&A...640A..10W}.

In fact, we detected only 21 objects catalogued as True PNe in \citet{2020A&A...640A..10W} 
that are not included in HASH. If we analyse these objects, we see that they are catalogued as 
Symbiotic Stars, HII Regions, or Ionized ISM in HASH. We also 
find that these misclassifications are frequently supported by old entries in the Simbad database.
 Excluding these few objects, we finally gathered a total sample of 3711 PNe.

\subsection{Identification of the central star in planetary nebulae}

Once we collected our general PNe sample, we proceeded to search for their CSs  
in Gaia EDR3. We searched in the field around their nebular centre, set up by their corresponding 
HASH coordinates. The identification of the CS in a PN is not an easy task and has led to 
misidentifications that have remained unnoticed in the literature \citep{2020A&A...638A.103C}. 
Central stars are usually located at the geometric centre of the PN or close to it, but that is not always 
the case. In some instances, it is difficult to establish the geometric centre of the nebula, either 
because of projection effects or because of their interaction with the ISM (\citealt{2012ApJ...748...94V}). 
For extended sources,  there can often be several field stars around the PN centre that hinder 
the CS identification. The nebula itself can also contaminate the field with its brightness; this 
%can happen 
mainly happens in young and compact PNe. In these cases, the detection of the CS is even more complicated. 

 %Furthermore, in some cases it has been observed that the central star is displaced with respect to the observed centre due to the interaction with the interstellar medium. 

Apart from the distance to the nebular centre, another clue that can help to identify  CSs 
is the colour of the star. Gaia multi-band photometry provides colours for most of the stars in the survey,
%, 
and we can use, for instance, the difference between magnitudes measured by the blue photometer 
(BP) and the red photometer (RP),  $G_{BP}-G_{RP}$, or other combinations of the $G$, $G_{BP,}$ and 
$G_{RP}$ bands to define the colour of a source. As we know, the CSPNe must have a minimum temperature 
to be able to ionise the nebula. 
% IKER: Esto de donde saldria ? Yo pondria la referencia de 13,000 K que da Weidmann (2020)
As can be seen in \cite{2006A&A...458..173S}, an A0-type star can give rise to the photoionisation 
of hydrogen, so we can assume a minimum stellar temperature of 13\,000 K for a transition stage 
from a preplanetary nebulae (\citealt{2020A&A...640A..10W}) to about 25\,000 K for a complete 
ionisation of the nebula (\citealt{2000oepn.book.....K}). 
% According to \citet{2020A&A...640A..10W}, this temperature threshold can be set to about 13.000 K to start ionization and over 30.000 K for a complete ionization. 
We can use the relationship between Gaia $G_{BP}-G_{RP}$ and effective temperature obtained in 
\citet{2018A&A...616A...8A} for a reference sample of stars with DR2 photometry to set an
 upper limit on the colour of the ionising source to $G_{BP}-G_{RP}=-0.2$, which corresponds 
to a temperature of around 13\,000 K (see Figure 3 of \citealt{2018A&A...616A...8A}). 

\begin{figure}[h!]
        \includegraphics[width=8.5cm,height=6cm]{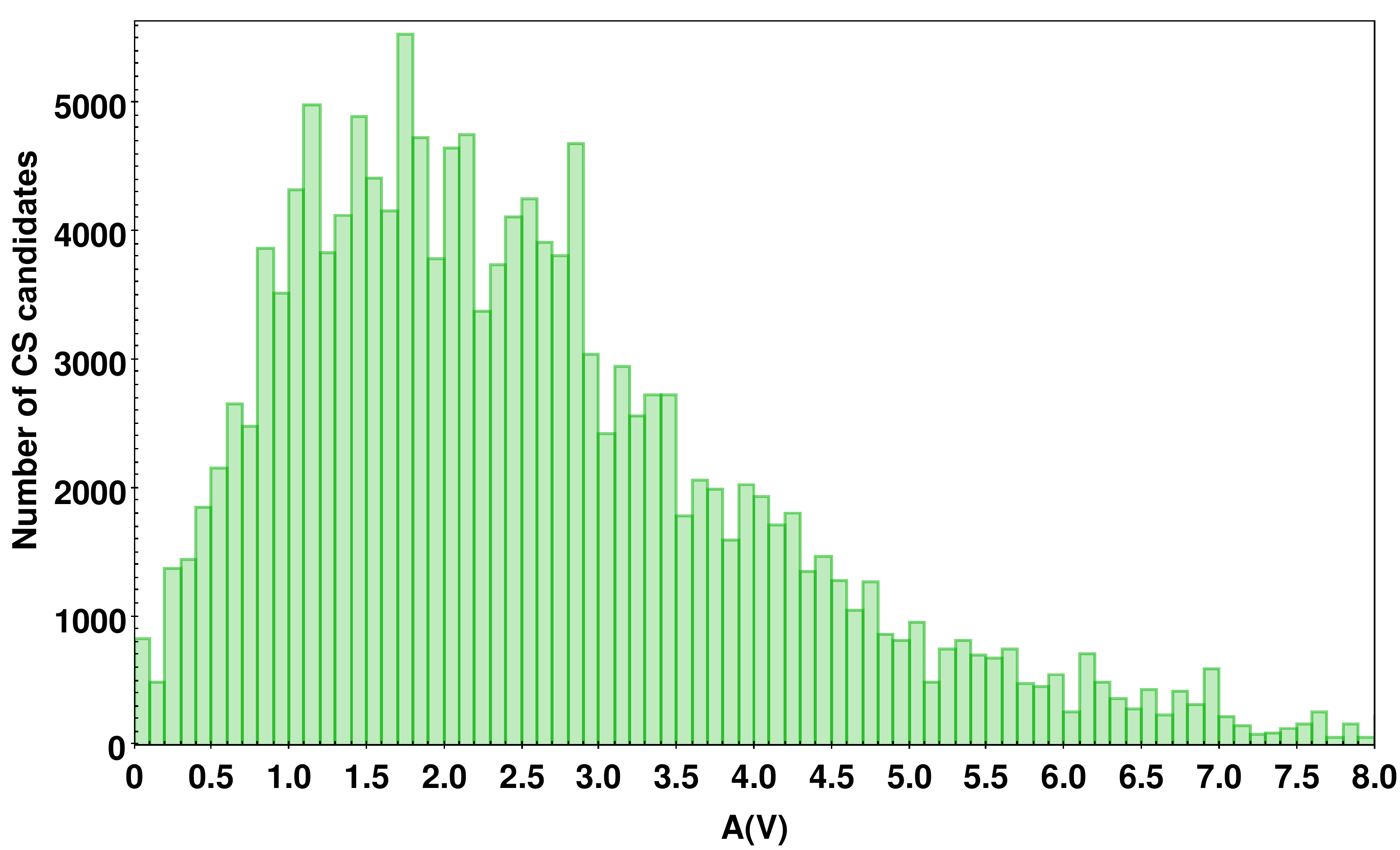}
        \caption{Interstellar extinction distribution for all CS candidates. Obtained from Bayestar and SFD dust maps.}
        \label{fig:dustmaps}
\end{figure}

Interstellar reddening must be accounted for with regard to
the colour of the stars, bearing in mind that we are talking about a population of stars 
mainly from the Galactic disc. In order to quantify 
and correct for interstellar extinction, we made use of the Bayestar (\citealt{2019ApJ...887...93G}) 
and SFD (\citealt{1998ApJ...500..525S}) dustmaps.\footnote{https://dustmaps.readthedocs.io/en/latest/maps.html} 
We obtained the interstellar extinction 
for all our CS candidates
by implementing a Python script to query these two databases. The idea is to correct for extinction before carrying out the search 
procedure for the most probable CS. This extinction distribution (up to a limit of 8 mag) in the visible 
band is shown in Figure \ref{fig:dustmaps}. Using the relations of \citet{2018A&A...614A..19D}, 
we then obtained the extinctions in the Gaia $G, G_{BP}, G_{RP}$ passbands. Finally, we corrected the 
$(G_{BP}-G_{RP})$ colours with the corresponding extinction values. 
%HAY QUE CAMBIAR EL EJE HORIZONTAL DE LA FIGURA 1 Y PONER, NUMBER OF cs CANDIDATES
%OJO, CAMBIAR Y PONER EL SUBÍNDICE O A LOS COLORES DESENRROJECIDOS Y DEJAR SIN SUBINDICE EL COLOR OBSERVADO. TANTO EN EL TEXTO COMO EN LAS GRÁFICAS.

%following a procedure similar to the one described in \citet{2020A&A...638A.103C}, 
We next considered both the distance to the nebular geometric centre (from the HASH catalogue) and the Gaia
de-reddened colour $(G_{BP}-G_{RP})_{\circ}$ to identify the most probable CS in the field of each 
PN. In carrying out this task, we considered all EDR3 sources within a radius of 20 arcsec around each of 
the nebular centres and obtained, 
%in mean  
on average, 40--45 candidate objects per nebula (more than 160\,000 sources in total).

%To find out the most probable central star together with a reliability value, we assigned to the colour ($G_{BP}-G_{RP}$) a weight four times greater than the distance to the geometric centre, achieving a value that we transformed into a confidence index with a subsequent scaling between 0 and 100:

In order to analyse the sources present in each field and 
%decide 
decide upon the one that was most probably the CS, we 
%implement 
implemented an algorithm that 
%allow 
allows us to take into account both the de-reddened colours ($c$) and the distances to the nebular centre 
($d$). We first checked whether there was any star with a 
$(G_{BP}-G_{RP})_{\circ}$
colour lower than $-0.2$ (the colour limit for at least partial hydrogen ionisation) inside a 
region close enough to the nebular centre (defined as a circular region with a radius less 
than the 20\% of the corresponding nebular radius). If we detected more than one 
source within this region, we selected the  one closest to the geometrical centre.

When no star was detected fulfilling these colour and distance thresholds, we built a function 
containing both the value of the angular distance and the colour that allowed us to select 
the source with the
%lower 
lowest value for both parameters (the closest and the bluest star). 
%So, the source whose  parameters minimise such function will be selected as the candidate most likely to be CS of the PN. 
%
%Among the possible factors that can explain the selection of a "red" CSPN, well above the indicated colour limit, one possibility is that it is in fact a binary system in which only the companion was detected by Gaia, that the extinction correction is not adequate, or that it is a false identification, due to the fact that the CS is too weak, too blue, or its colour was not measured by Gaia.
There were numerous cases where a red CSPN was detected, well above the 
indicated colour limit. Among the factors that could explain this type of  selection was 
the possibility  that it corresponded to a binary system where only the red 
companion was detected by Gaia, that the extinction correction was insufficient, or even 
that it was a false identification or the
 true CS was too weak or had a colour not measured by Gaia.

%If we want to consider these two parameters (distance and colour), with such a different nature, in the same function, the possible values for $c$ and $d$ have to be normalised ($N(c)$, $N(d)$) somehow.

The possible values for $c$ and $d$ need to be normalised somehow ($N(c)$, $N(d)$) 
in order properly to consider these two parameters (distance, $d$, and colour, $c$) 
of such different natures within the same function. 
%
%Iker: quitar esto? - We have approximated this function in a simple, but staggered way, to avoid an excessive weight of one of the two conditions when they correspond to extreme values. 
As the CS needed to be located inside its nebula, we only considered those sources 
with a distance to the nebular centre below their corresponding nebular radius ($R$). 
We then normalised their distances to the interval [0,1]. In the case of sources 
located farther than their nebular radius, we assigned them a value of infinity in order to  
discard them with ease. We defined $N(d)$ as follows:

%As we have determined, the angular distances can take values between 0 and 20 arcsec, so we set as the most suitable region the interval which goes from 0 to a distance equal to the mean nebular radius ($R$) and, as a secondary region, 
%the one in the interval from $R$ to 20 arcsecs. Subsequently, we scale the distance, $d$, to the interval [0,1] for the most suitable region while to the interval [1,2] for the outer region: 

%\begin{equation}
     $$N(d) = \left\{
              \begin{array}{ll}
                \dfrac{d}{R}, & d \in [0,R] \Rightarrow [0,1]; \\
                \infty , & d > R. \\ %1 + \dfrac{d-R}{20-R}, & d \in (R,20] \Rightarrow [1,2]. \\
              \end{array}
            \right.$$
%\end{equation}

In a few cases in which the nebular size was unknown, we imposed a minimum radius value of 1
 arcsec. If we did not find any source in this region, we did not assign a CS to that PN.

%a rather neutral intermediate radius value of 10 arcsec.

We also normalized the colour, $N(c)$, in the [0,1] interval. 
From our previous study in Paper 1, we know that most colours move between
the values of $-$3 and 3. For stars with $(G_{BP}-G_{RP})_{\circ}$ in this range, we
%From our previous study in Paper 1, we know that the CSs take colour values not greater than approximately 3. 
%So we decided to set this interval as the most common range of values, and 
scaled their colours to the $[0,\frac{1}{2}]  $interval, which corresponds to the 
most suitable region, whereas for stars with higher colour values (up to 4.82 in the 
whole sample) we scale the colour to the interval $[\frac{1}{2},1]$, which was defined as the 
less suitable region. Finally, for the few stars with very blue colours, 
$(G_{BP}-G_{RP})_{\circ}$ below $-$3, we assigned an $N(c)$ value of 0, since we 
consider these to be sources with a high probability of being the ionising star. We thus obtain 
the following normalised value range for the colour: 

%\begin{equation}
     $$N(c) = \left\{
              \begin{array}{ll}
                \dfrac{3+c}{12}, & c \in [-3,3] \Rightarrow [0,\frac{1}{2}], \\
                0.5 + \dfrac{c-3}{3.64}, & c \in (3,4.82] \Rightarrow [\frac{1}{2},1], \\
                0, & c < -3. \\
              \end{array}
            \right.$$
%\end{equation}

To those objects without measured colour values, we assigned  the mean colour  value 
of all candidates, $c$ equal to 0.3875. Once both the $N(c)$ and $N(d)$ factors were calculated, we summed these to construct the 
function to be minimized in order to determine the  star most likely to be the CS: 
%Instead of just adding up both quantities, we assigned a weight 4 to scaled distances. This value has been empirically adjusted, and it is supported by the fact that the position of the geometric centre of the nebulae is quite uncertain. %In addition, as we consider more significant the colour of the star than its distance to the centre, we think that it is suitable to give more priority to the colour. So we multiply the normalised angular distances by a factor 4, in order to become the selection function more sensitive to the colour. 
%Consequently, we propose the following function:

$$ f (c,d) = N(c) +  N(d).$$

%where $d$ is the angular distance to the nebular centre (in arcsec) and $c$ is the colour $G_{BP}-G_{RP}$. 

%($G_{BP}-G_{RP}$) 4 times we implemented an algorithm that considering also these two factors, gives the most probable Gaia source to be the CS and a reliability value (in percentage) of being it. 

% Algoritmo de identificacion de CSs: angDist & Color
%For this searching procedure, we consider all sources in a radius of 20 arcsecs around each nebular centre, obtaining a mean quantity of 40-45 candidate objects per nebula. Then, we execute the identification algorithm that will select the most probable object to be the CS among all these candidates. This algorithm consists on associating a value to each object depending on its colour and on its angular distance to the nebular centre, that basically is the sum of both factors. So, the lower this value is, the bluer and the closer to the centre would the object be. Then, the algorithm selects the object with the smaller value in each nebular neighbourhood. In addition, it gives a reliability percentage (R) for the object to be the CS, that is a function of both factors and we define it as:

%$$R = 10 \cdot [10 - (\frac{d}{4} + c + 0.8)],$$
% explicar de donde viene esta funcion ??
 
We note that this function takes values in the [0,2] interval. The source that minimises 
this function for each  of the remaining PNe in our sample is identified as the 
most likely CS. 

\begin{figure}[h!]
        \includegraphics[width=9cm,height=6.5cm]{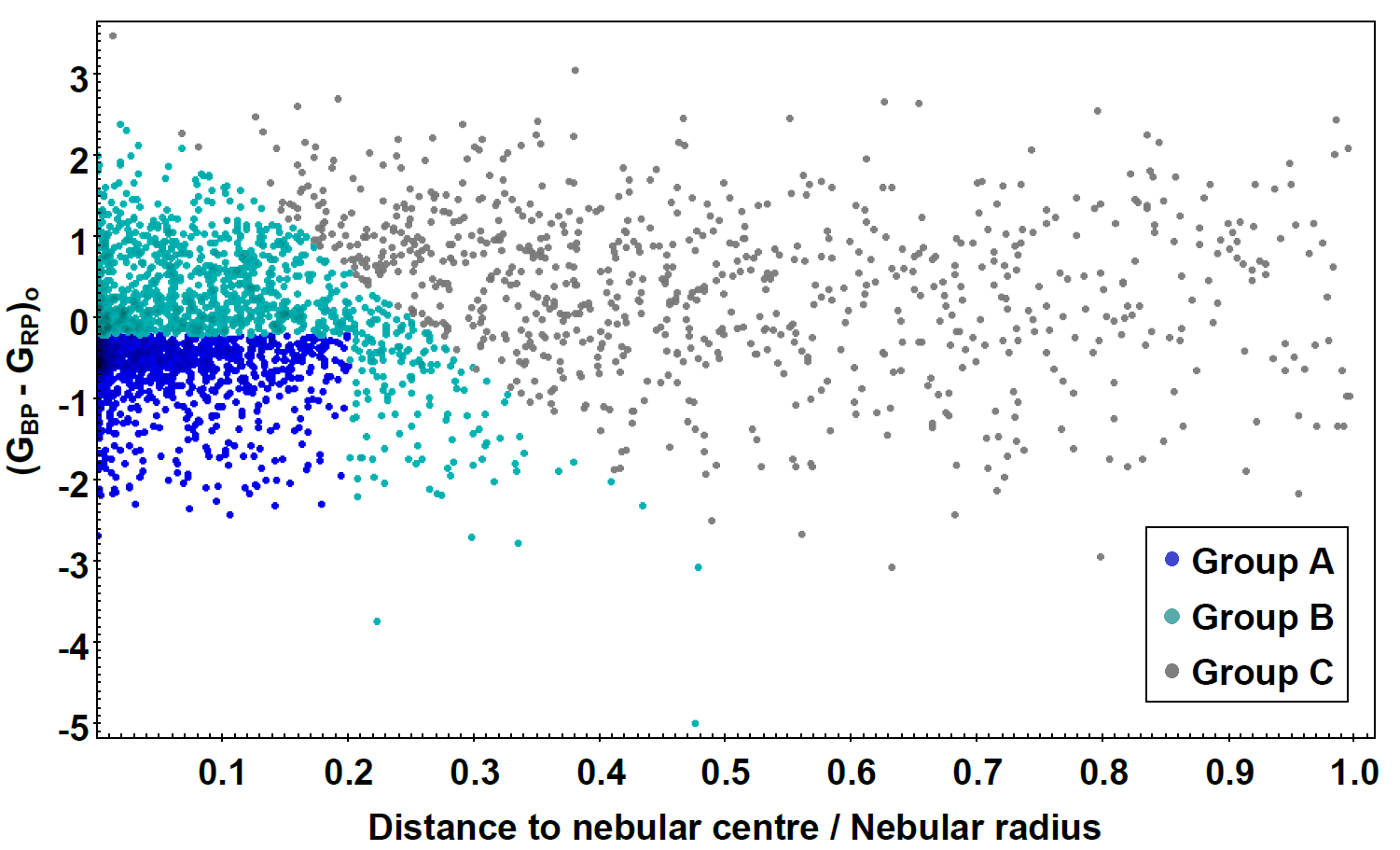}
        \caption{ $(G_{BP}-G_{RP})_{\circ}$ colour versus relative distance to the nebular 
centre for those stars proposed as CSPNe. The objects are divided in three groups (A, B, and C) 
according to the reliability of their identification.}
        \label{fig:reliability}
\end{figure}

%So by this procedure we have been able to identify the most likely CS for each of the 3,692 PNe previously selected. %as Gaia sources, giving a probability percentage of being it. 
After running this algorithm, we obtained CS identifications for a total of 3257 PNe. 
If we compare our identifications with those of \citet{2021arXiv210213654C}, we obtain a coincidence rate of 96\% within the PNe in common in both samples. We note that our algorithm considers colours already corrected for extinction, a fact that the algorithm of \citet{2021arXiv210213654C}  does not take into account. This may be the main explanation for the small mismatch. 

By analysing the few differences in the CS identification between both methods (78 cases), we see that our algorithm gives slightly more priority to colours than the \citet{2021arXiv210213654C} algorithm. Concretely, when both identifications correspond to objects with colours measured by Gaia, in more than 60\% of cases our source is bluer than the one identified by \citet{2021arXiv210213654C}. In addition, when any of the identifications lacks Gaia colour, in around 55\% of cases our source is closer to nebular centre than the other source. 
%Ana: igual estaría bien decir cuántas hay en común; algo como 2021) for the NN PNe in common, we...

%This function can be also used as a measurement of reliability, although we prefer to assign a quality flag to the CS identification because we are not calculating any probability. 
In order to provide the community with an index indicating how reliable these identifications 
might be, we decided to divide the identifications into three groups (A, B, and C), in decreasing order of 
reliability. Stars included in group A all have  colours below $-$0.2 and are located within 
20\% of the nebular radius: these sources are very likely to be CSs. Group B contains stars with values of 
$f(c,d) \le 0.5$, and group C contains all the other stars. In 
group C, we also included those stars selected as CSs without known colours, as they are uncertain identifications.

%according to their $c$ and $d$ values,
%being the group A the one with more reliable identifications and the group C the one with less reliable identifications. Firstly, we defined the group A as the set which contains the CSs selected in the first filtering (stars with colour below -0.2 and closer than 20\% of the nebular radius to the nebular centre). Then, we use the value obtained in function $f(c,d)$ to define the group B and C. Note that this function will give values between 0 and 10 (as N(c) and N(d) are defined between 0 and 2), so we decided to include CSs with values up to 2 in group B and the rest of CSs in the group C.
Figure \ref{fig:reliability} shows, divided in these three groups, the stars proposed as CSs 
in a colour versus distance diagram, where in this case distances are calculated relative to the nebular mean radii.
%we have plotted in a \textit{colour vs distance to the centre} diagram, the identified CSs divided in these three groups.
% calculated a reliability value to each selected star, according to the value obtained in the function.
Of the total of 3257 identified CSPNe, 31.6\% are included in group A, 30.9\% 
in group B, and 37.5 \% in group C. As the CS identifications in group C are not sufficiently 
reliable, we decided not to include them in our CSPNe catalogue. Consequently, our catalogue 
provides information for a total of 2035 CSPNe, corresponding to groups A and B. 

In Table \ref{table:CSPN} we list information on the Gaia EDR3 sources identified as the CS of each of these PNe. 
%We provide different parameters as the reliability flag (Clas.), the coordinates (RA, Dec), 
Apart from their coordinates (RA, Dec) and reliability group (quality label), several other parameters are provided, such as 
the angular distance to 
the 
nebular centre ($D_{\rm ang}$), the Gaia magnitudes and colours ($G$, $(G_{BP}-G_{RP})$, 
$(G_{BP}-G_{RP})_{\circ}$), and the interstellar extinction ($A_V$). The full table is available 
%in 
online at the CDS.

\subsection{Distances to planetary nebulae}

Gaia EDR3 provides parallaxes for approximately 
%the 
81\% of its sources. In our catalogue of 2035 CSPNe, 
%we have 
there are 1725 objects with EDR3 parallaxes. From these parallaxes, we can derive their 
distances, as we did in our Gaia DR2 study (Paper 1), using the new catalogue of 
Bayesian statistical distances of \citet{2021AJ....161..147B}. In comparison with 
%this 
our previous work, we have been able to estimate distances to about 10\% more PNe. With 
DR2, we obtained distances for 1571 PNe, whereas we now have this parameter for 1725 nebulae. 
However, the main improvements are that the new distance determinations are more reliable than 
previous ones and that they come from more accurately selected CSs. 

\begin{figure}[h!]
        \includegraphics[width=9.5cm,height=6cm]{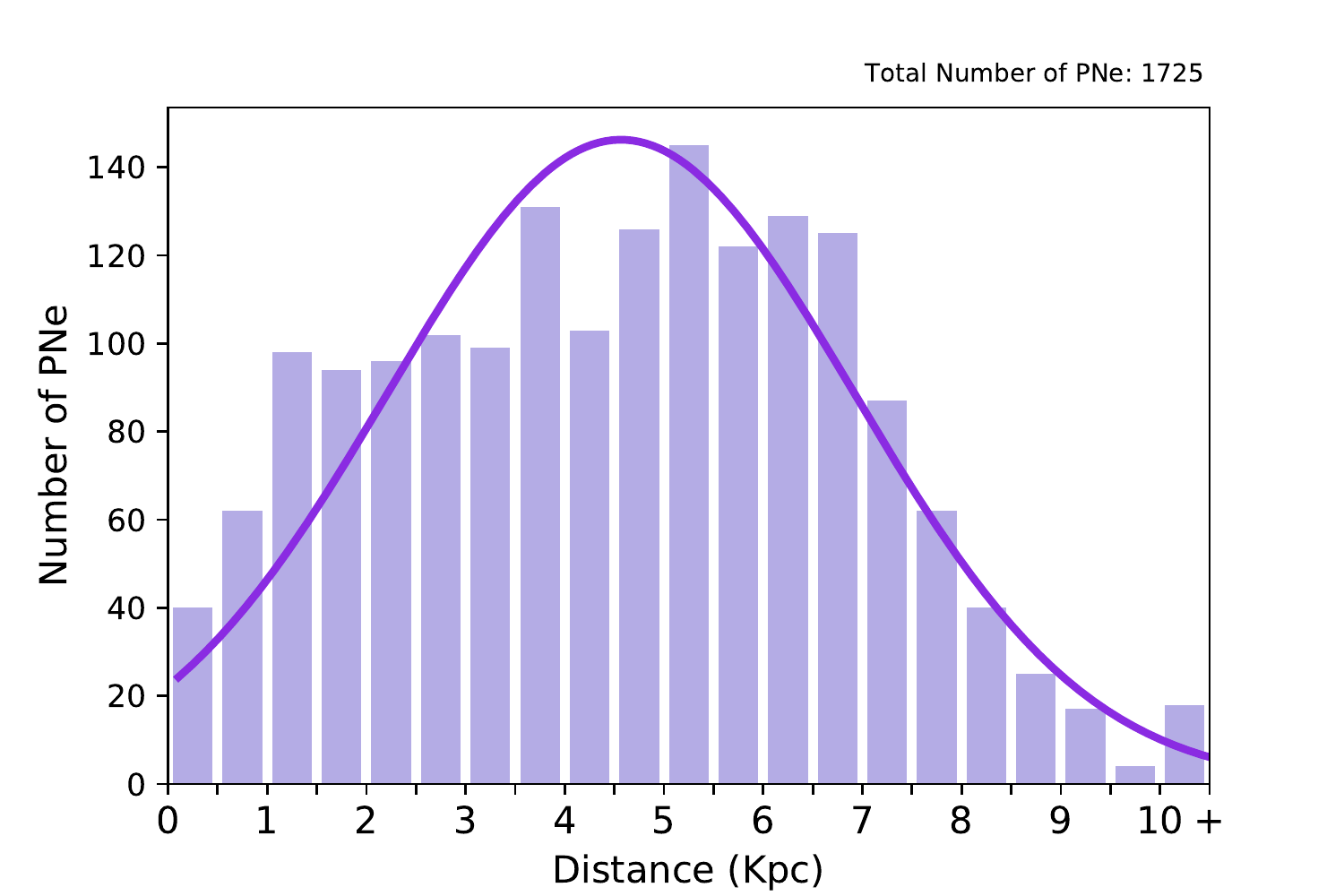}
        \caption{Distance distribution of the general sample of 1725 PNe.}
        \label{fig:distance_general}
\end{figure}

In Figure \ref{fig:distance_general}, we show the obtained heliocentric distance distribution. Its 
peak is located around 5 kpc and   
%From this histogram, we can conclude that most of our PNe are located at distances between 3 and 7 kpc from the Sun. 
%We can also appreciate that around 
%the 
about 6\% of our proposed CSs are located at a distance of less than 1 kpc, while only about 1\% are beyond 10 kpc. 

If we fit a Gaussian function to this distribution, we obtain a mean value of 4.57 kpc, 
with a standard deviation of 2.35 kpc. This means that, in comparison with our results in Paper 1 (mean distance of 3.55 kpc and standard deviation of 1.94 kpc), we detect more distant 
and more equally distributed PNe.  

% Incluir calculos sobre Completitud, Densidad, Poblacion total, Ratio de Nacimiento ??

\section{Catalogue of CSPNe with reliable distances: GAPN-EDR3}

%As we have seen in the previous section, 
As seen in the previous section,
we were able to extract the parallaxes and, consequently, distances for a wide sample of 
CSPNe. Nevertheless, these parallaxes may have quite large uncertainties that  
translate into large distance uncertainties. In order to carry out a detailed study of the 
PNe, we  selected only those objects with the most reliable distances.
% (less uncertainties).% To make this selection, we will get rid of those objects with the highest parallax and distance uncertainties. 

At this point, it is important to note that parallaxes ($w$) in Gaia EDR3 (as was 
also the case for DR2 parallaxes) have a small bias or zero point, $z_{\circ}$, 
which must be corrected for. According to \citet{2021A&A...649A...4L}, this {zero point} 
takes a mean value of $-17\ \mu $as. However, in general there is a certain dispersion
 around this value, depending on the star's magnitude, colour, and celestial position. Lindegren 
provides a recipe \footnote{https://gitlab.com/icc-ub/public/gaiadr3\_zeropoint} for 
estimating the {zero point}, which we have used to correct our CSPNe parallaxes. 
Parallaxes corrected from the {zero point} result from the following:

$$w_{\circ} = w - z_{\circ}.$$

In addition to the uncertainty value of the parallax given in Gaia EDR3 
(internal uncertainty), a systematic uncertainty can also be calculated and applied following the 
prescriptions given in \citet{2021A&A...649A...5F}. %should be considered. So we also applied 
%this correction to our parallax uncertainties as it is explained in \citet{2020arXiv201206242F}. % Revisar esta referencia !    

After applying such corrections, and following the same criteria as in Paper 1, we 
decided to select only those PNe with relative uncertainties below 30\% in parallax and  distance.  
%the 
 We also considered the unit weight error (UWE) and renormalised UWE (RUWE) astrometric quality parameters from Gaia,
which are supposed to take values around 1 for sources 
where the single-star model provides a good fit to the astrometric observations.
According to \citet{lindegren18}, for good quality measurements, they should take 
values fulfilling the condition UWE < 1.96 or RUWE < 1.4.

Following the application of these constraints, we obtained a selection of 405 
PNe. These objects will have  accurate enough distances to enable us to derive 
other properties of the PNe.

%reliability has increase in this sample. 

As in Paper 1, we name this sample the Golden Astrometry Planetary 
Nebulae in EDR3 (GAPN-EDR3). The new sample contains almost twice as many PNe as the previous one,
 which contained 211 objects. If we compare both samples, we observe that there are 
64 PNe from the DR2 sample that are not included in the new one. This might happen 
for a number of reasons. %Concretely, there are 56 PNe that are not in GAPN-EDR3. 
%If we analyse why this is happening, we see that there are different reasons. 
One reason is that our selection of an object as a PN is now based on  HASH 
database cataloguing, whereas for DR2 GAPN we used Simbad database classification. 
In particular, there are 17 objects listed as PNe in the Simbad database that are not included 
in HASH database as True, Likely, or Possible PNe, and these have now been excluded from our sample.

Another reason is that, because of the new astrometric measurements published in Gaia 
EDR3, there is a set of 13 sources that  no longer fulfil the filtering constraints 
(low parallax and distance relative errors, low UWE, RUWE, etc.) required to be included in GAPN-EDR3. 
This might happen for a few cases, but in general we managed to include many more 
objects than with DR2, as the data are more accurate now.  
%This can be partially due to the better astrometric quality of EDR3, but it is important to remark that we are more than doubling the number of objects in GAPN. 

%The last 
Yet another reason is that in the case of some nebulae, the CS identification is 
not the same in both samples. Consequently, 26 of the new sources identified as 
 CSs are not included in GAPN-EDR3 for not having parallax or for not fulfilling 
the filtering constraints. This mismatch in identification might happen 
because for DR2 GAPN we selected the source closest  to the nebular geometric 
centre, whereas now we are using % a selection criteria based in an algorithm that takes into account 
both the distance to the centre and the colour of the source to select the most likely CS.
Moreover, as we are now discarding the less reliable identifications (those from 
group C), we decided not to include a further eight PNe that were present in the DR2 sample.
% (as explained in section 2). %So we hope that in the new GAPN we have corrected some bad identifications done in the old sample. 
%From now on, we will call this new sample as GAPN-EDR3. 

According to their CS identification reliability discussed in the previous section, almost 
%the half of objects 
two thirds of the objects in this sample belong to group A (64 \%) and  
%a 
36\% are from group B. So, we may conclude that the 405 PNe in this 
GAPN-EDR3 sample have more reliable identifications than in the general sample (2035 PNe).

\subsection{Galactic distribution and distances}

\begin{figure}[h!]
        \includegraphics[width=9cm,height=6cm]{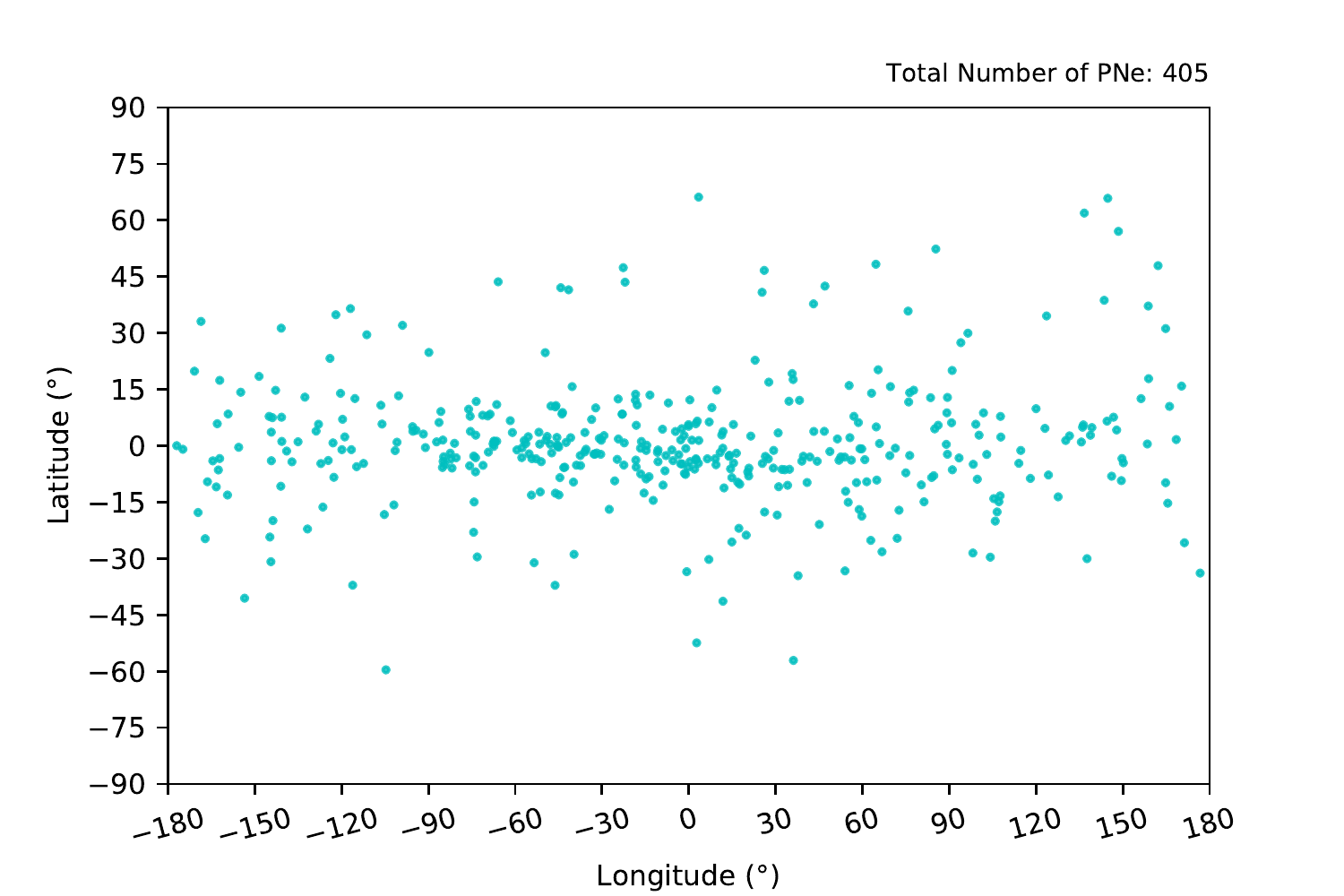}
        \caption{Galactic distribution of objects in the GAPN-EDR3 sample.}
        \label{fig:lat_lon}
\end{figure}

% Distribucion galactica
Once we had selected this new GAPN-EDR3, we were able to analyse some of its general properties. 
Regarding  the Galactic distribution of the PNe, we found that most of them %are located in the area of the
belonged to the Galactic disc population (Figure \ref{fig:lat_lon}), with 77\% of them  
falling within the\ $\pm 15^{\circ}$latitude range. We also observed that, in general, they tended to accumulate towards 
the Galactic centre,  around half of them being located within the longitude range $\pm 60^{\circ}$. 
These results are in agreement with what is expected from the stellar density distribution of the Galaxy.

\begin{figure}[h!]
        \includegraphics[width=9.5cm,height=6cm]{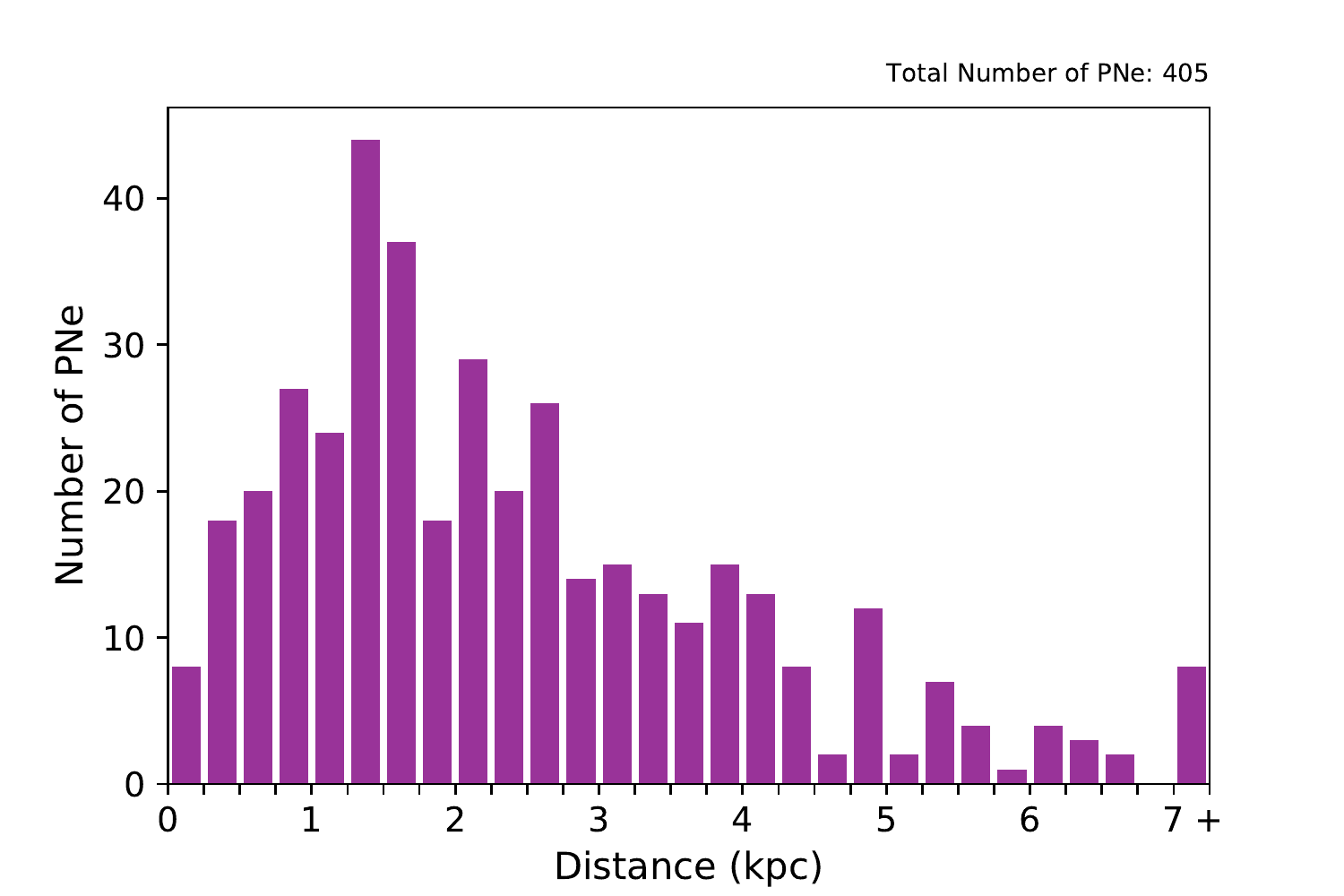}
        \caption{Distance distribution of the GAPN-EDR3 sample.}
        \label{fig:distance_GAPN}
\end{figure}

% Distancias
Figure \ref{fig:distance_GAPN} shows the distance distribution for the 
GAPN-EDR3 sample. In comparison with the general sample distribution, we see that the distances 
tend to be closer, there being only a few PNe farther than 7 kpc; 
around 
%the 
50\% of them are located closer than 2 kpc; and, 
%We also find that the number of objects starts to decrease progressively starting from a distance of approximately 1.5 kpc. 
from a distance of approximately 1.5 kpc onward, we found that the number of 
objects started to decrease progressively.
The mean distance of this distribution is $2.45 \pm 0.96$ 
%pc. 
kpc. In Table \ref{table:GAPN}, we provide the parallaxes and distances (with their 
uncertainties) of  all objects in the GAPN-EDR3 sample, together with the radial velocities 
(Vel$_{ rad}$) for some of them. The full table is available at CDS.

% Comparacion distancias con otros autores

It is interesting to compare the new Gaia EDR3 distances 
to PNe with those obtained by other authors using different methods. %, for the common objects between both samples. 
These comparisons are displayed in 
%Figure \ref{fig:distances_comparison} in different panels.
different graphs.
In each one, the blue line indicates the 1:1 relation between both distance 
derivations, while the red line represents the linear regression between them. 
%Moreover, the 
Individual data points are drawn with error bars for each distance derivation.

\begin{figure}[h!]
        \includegraphics[width=9.5cm,height=6cm]{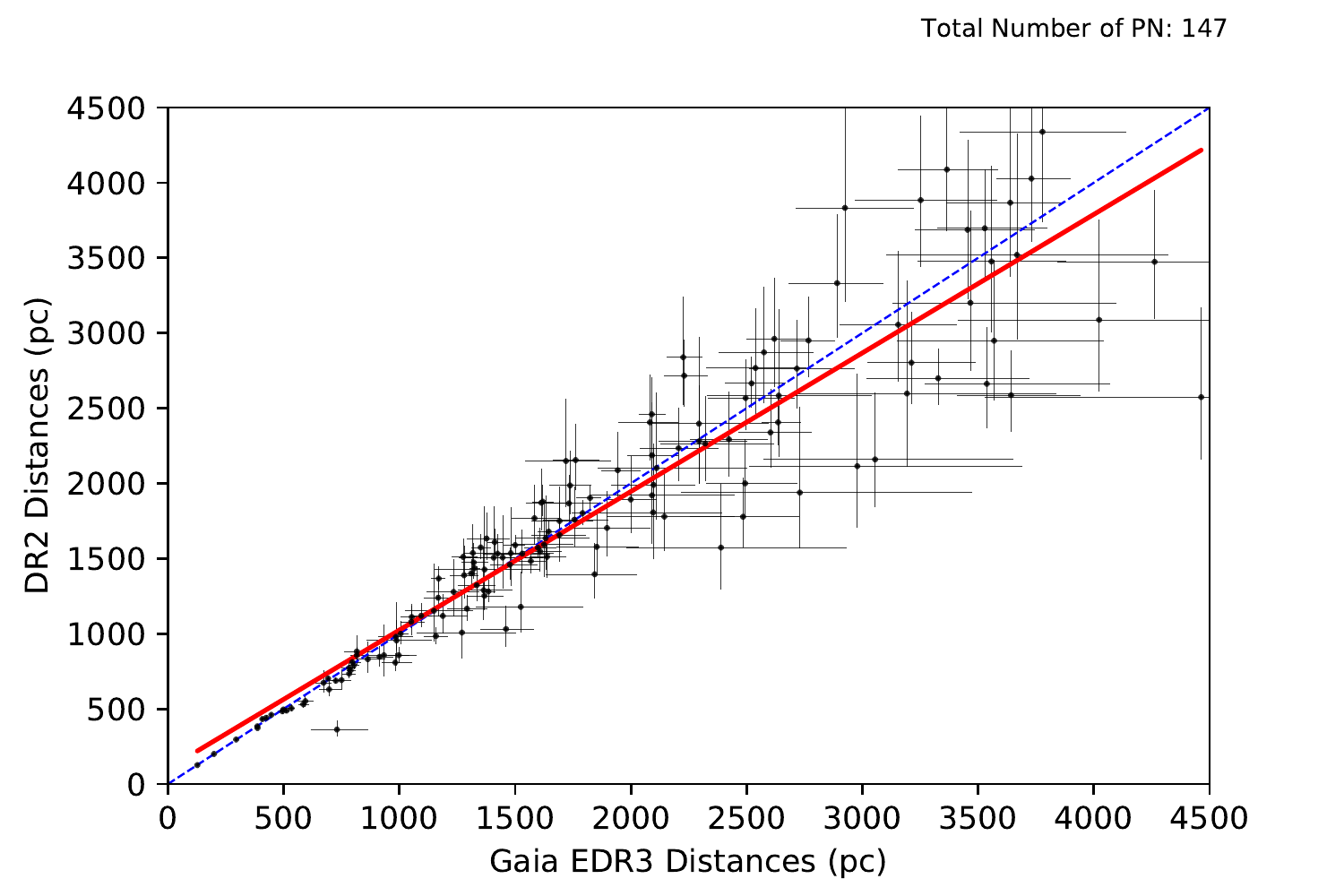}
        \caption{Distance comparison between Gaia DR2 and Gaia EDR3. The blue line 
indicates the 1:1 relation, while the red line represents the linear regression between 
them. These lines are also shown in the figures that follow.}
        \label{fig:DR2_vs_EDR3}
\end{figure}

% Gaia DR2
We first compared EDR3 distances with those obtained in Paper 1 from Gaia DR2. 
In Figure \ref{fig:DR2_vs_EDR3}, we see that despite there being a significant dispersion 
%in 
for some objects, the two distance distributions are mostly equal,
% in mean, %as the solid line lies approximate over the dotted line. This similarity is what we expected, because both distances are derived from the same statistical method (\citealt{bailerJones18}). 
the only difference being that 
the EDR3 distances are more accurate than those of DR2, especially for distances above 2 
kpc. We note, however, that for some stars that have been assigned significantly greater 
distances in EDR3 than in DR2, the errors in EDR3 may be greater than those of DR2. 
%This is why we see a dispersion in some objects distances and a small mismatch for the largest distances. 

\begin{figure}[h!]
        \includegraphics[width=9.5cm,height=6cm]{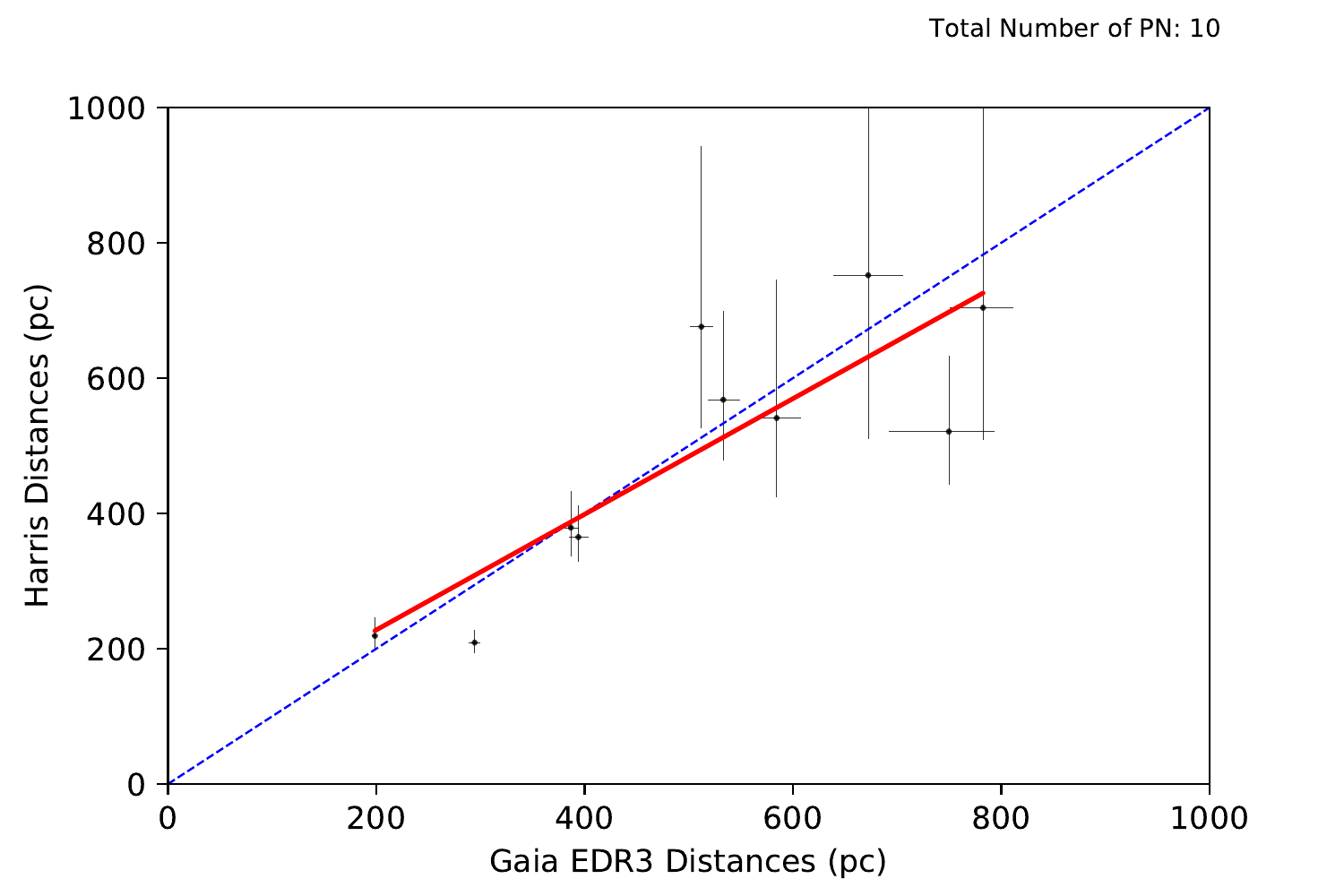}
        \caption{Distance comparison between \citet{harris07} and Gaia EDR3.}
        \label{fig:Harris_vs_EDR3}
\end{figure}

% Harris
If we compare EDR3 distances with others also obtained with astrometric methods, 
such as those  provided by \citet{harris07}, we  see that there is also  reasonable 
agreement between both distance determinations, as  in the case for DR2 distances. Figure
 \ref{fig:Harris_vs_EDR3} displays this comparison, which only depicts
distances below 800 pc.
%We can say that for distances below 500 pc the oincidence is greater than for larger distances.

\begin{figure}[h!]
        \includegraphics[width=9.5cm,height=6cm]{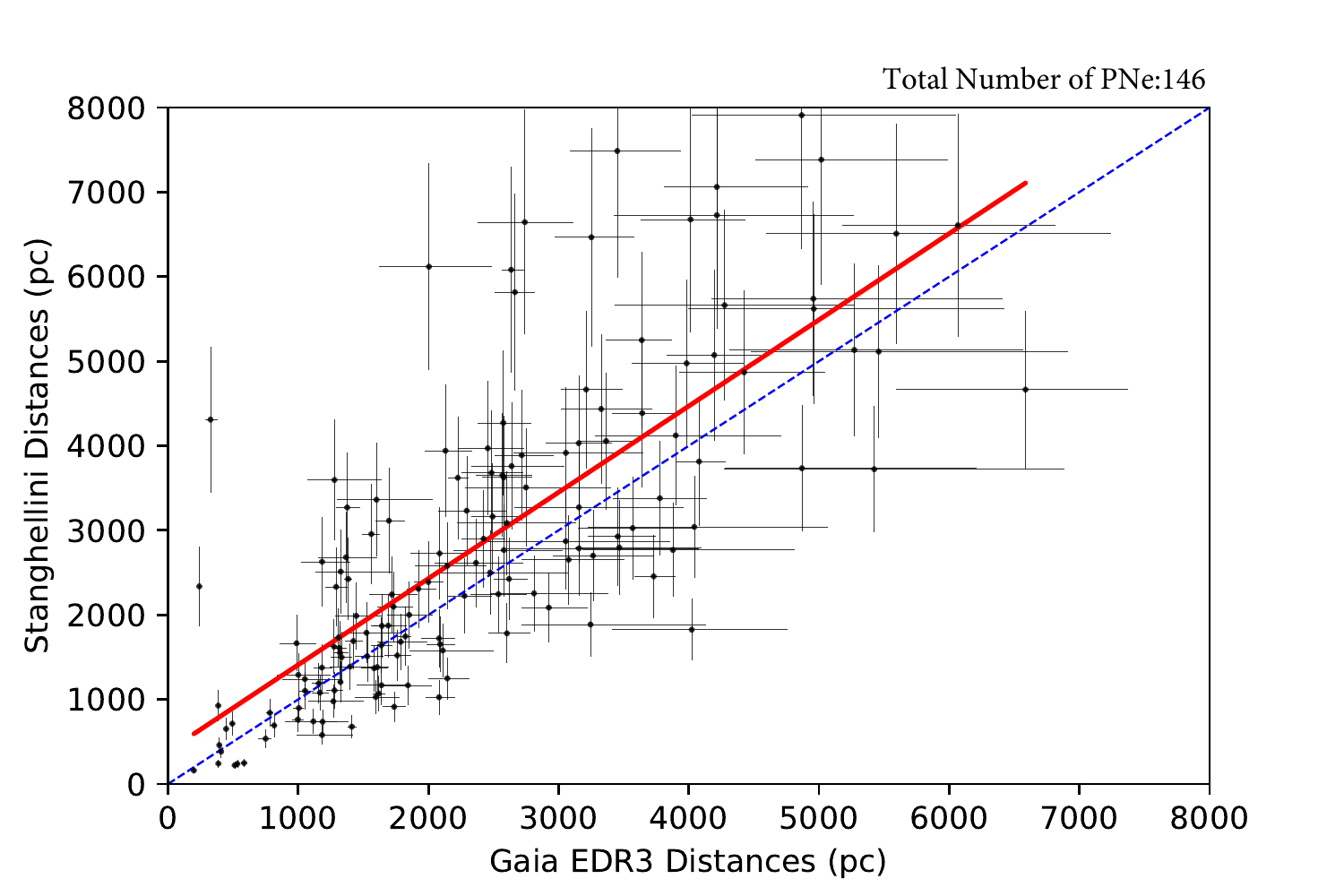}
        \caption{Distance comparison between \citet{stanghellini10} and Gaia EDR3.}
        \label{fig:Stanghellini_vs_EDR3}
\end{figure}

% Stanghellini
Other methods of estimating distances are based %in 
on statistical procedures, such as that used by \citet{stanghellini10}. 
If we compare these distances with those  reported here (Figure 
\ref{fig:Stanghellini_vs_EDR3}), we see that in general they are overestimated, which  had 
already been found with the comparison with distances in DR2 (Paper 1). 
This time, we find a positive bias of around 400--500 pc, a lower value than found
 before. 
%This Statistical distances (Stanghellini & Haywood 2010) do not agree with Gaia 
%distances, showing overestimated values in many cases. A linear fit to these distances 
%leads to a bias of 1 kpc. However, panel c in Fig. 7 shows that such bias is aected by 
%the presence of a marginal group of objects displaying wide discrepancies with DR2. A 
%possible explanation is that those objects are bipolar or butterfly-like PNe, and such 
%a statistical method cannot be applied to those classes of nebulae. 

%\begin{figure*}
%        %\centreing
%        \includegraphics[width=0.52\textwidth]{images/DR2_vs_EDR3.pdf}
%        \includegraphics[width=0.52\textwidth]{images/Harris_vs_EDR3.pdf}
%        \includegraphics[width=0.52\textwidth]{images/Stanghellini_vs_EDR3.pdf}
%        \includegraphics[width=0.52\textwidth]{images/Napiwotzki_vs_EDR3.pdf}
%        \includegraphics[width=0.52\textwidth]{images/Frew_vs_EDR3.pdf}
%        \includegraphics[width=0.52\textwidth]{images/Schonberner_vs_EDR3.pdf}
%        \caption{Comparison between EDR3 and other distance derivations. The dotted line indicates the 1:1 relation between both derivations, while the solid line represents the linear %regression between two derivations.}
%        \label{fig:distances_comparison}
%\end{figure*}

\begin{figure}[h!]
        \includegraphics[width=9.5cm,height=6cm]{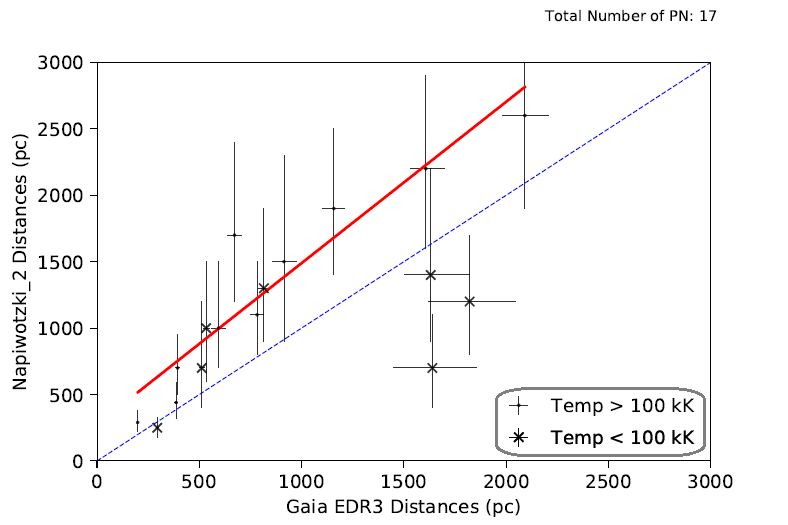}
        \caption{Distance comparison between \citet{napiwotzki01} and Gaia EDR3. Sources are 
divided in two groups according to their effective temperatures.}
        \label{fig:Napiwotzki_vs_EDR3}
\end{figure}

% Napiwotzki
Non-LTE model stellar atmosphere fitting was also used to derive distances to 
CSPNe. 
%A typical method used to estimate distances is the non-LTE model stellar atmosphere fitting. 
This method consists of obtaining the stellar effective temperature and surface gravity 
from  spectra and 
%by a $T_{eff}$ vs $log(g)$ diagram together 
uses evolutionary tracks for post-AGB stars to estimate masses and distances. The method was 
proposed for the first time by \citet{1988A&A...190..113M} to estimate a few PNe distances 
and  was later used by \citet{napiwotzki01} to estimate distances for a large 
sample of PNe. If we compare these distances with those of EDR3 (Figure \ref{fig:Napiwotzki_vs_EDR3}),
 we see that, as in our DR2-based study, Napiwotzki's distances are overestimated in 
comparison with ours. As we did in our previous study, we divided the CSs into high and low 
temperature sets (see legend). We observed that those with lower temperatures conform more closely
in terms of their EDR3 distances with those of Napiwotzki. In contrast, high temperature CSs 
showed a bias between two distance determinations that go from 250 pc (for the closest ones) to 
more than 500 pc (for the farthest ones). It is possible that this effect occurs because non-LTE models do not 
take into consideration line-blanketing for metals. 

%However, for EDR3 distances, we did not find a constant bias between both derivations, while we appreciate a larger overestimation for the closest PNe than for the other ones. 

%ESTO HAY QUE DISCUTIRLO COMO EN EL PRIMER PAPER Y MIRAR SI HAY NOVEDADES EN LA LITERATURA.

\begin{figure}[h!]
        \includegraphics[width=9.5cm,height=6cm]{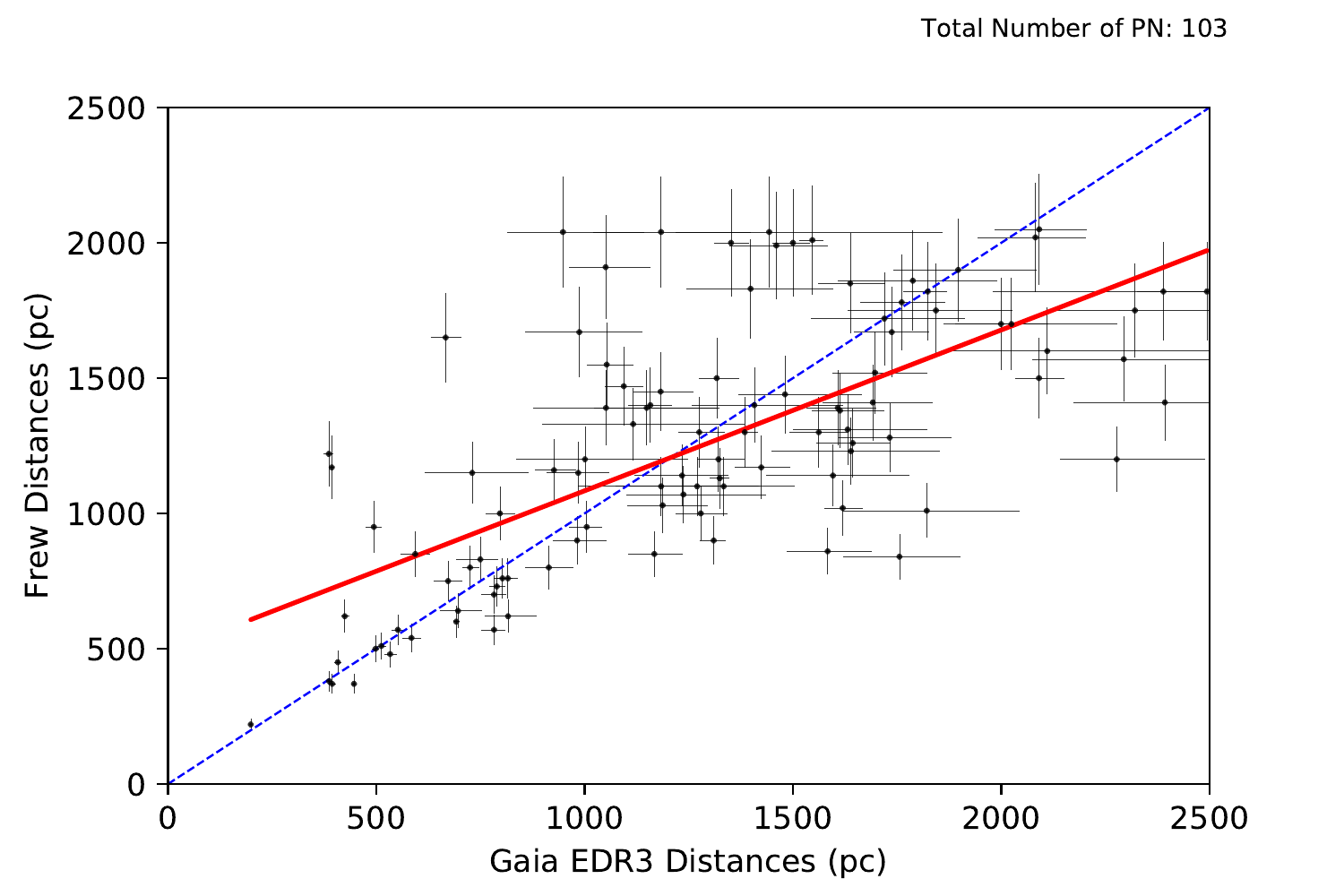}
        \caption{Distance comparison between \citet{frew08} and Gaia EDR3.}
        \label{fig:Frew_vs_EDR3}
\end{figure}

% Frew 
We now compare EDR3 distances with those derived by \citet{frew08}, who used a 
distance scale to CSPNe based on a statistically derived relation of  $H_{\alpha}$ 
surface brightness evolution
with nebular radius. 
%Now, we compare EDR3 distances with those derived by \citet{frew08}. Here, they used a method to estimate distances based on the relation between $H_{\alpha}$ brightness in the stellar surface and the nebular radius. 
As can be seen in Figure \ref{fig:Frew_vs_EDR3}, there is no clear relation between these 
distance derivations. We can  only say that for the closest PNe (below 1,200 pc) Frew's 
distances tend to be overestimated in comparison with ours, whereas they tend to be underestimated
for greater distances 
(over 1200 pc). 

\begin{figure}[h!]
        \includegraphics[width=9.5cm,height=6cm]{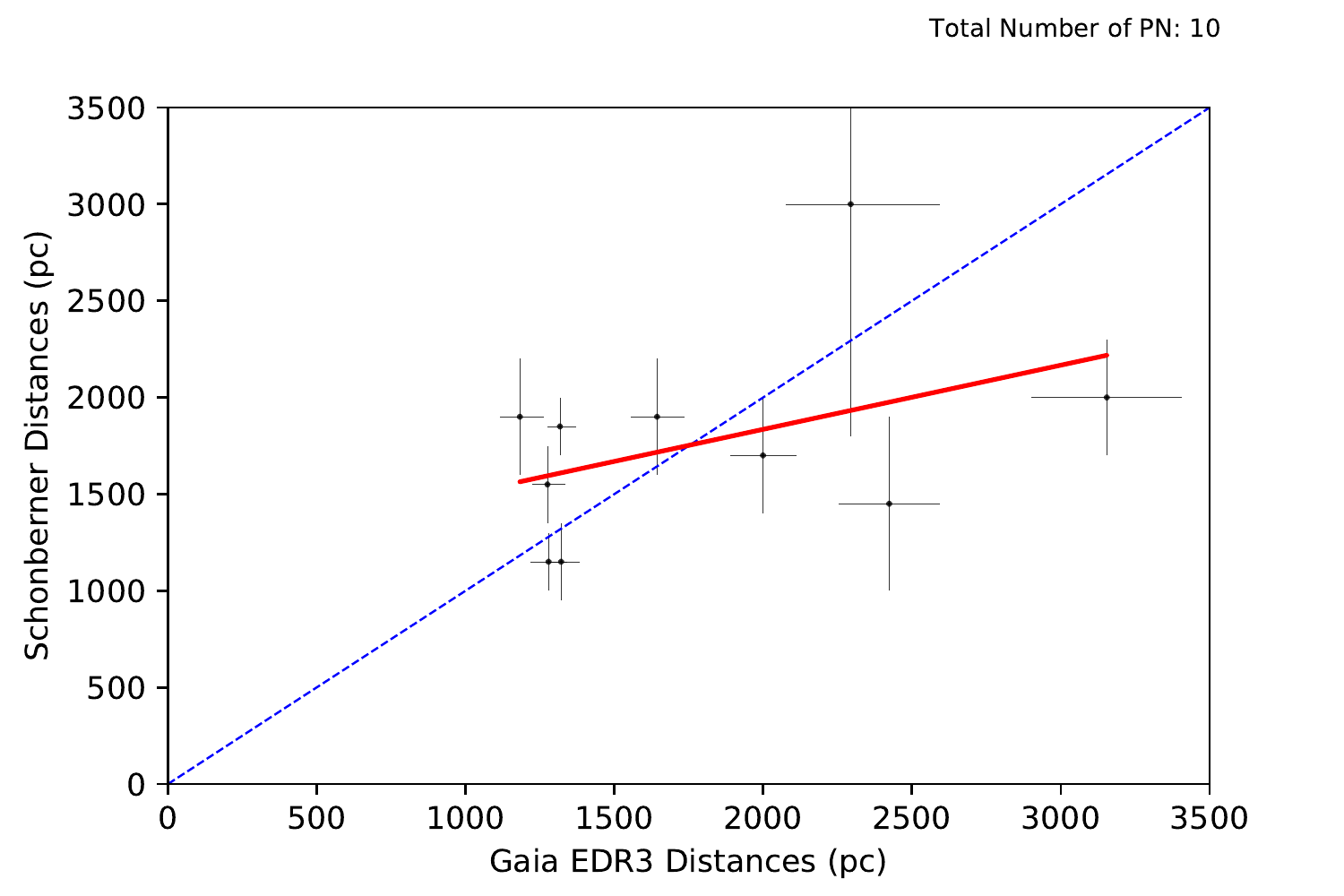}
        \caption{Distance comparison between \citet{schonberner18} and Gaia EDR3.}
        \label{fig:Schonberner_vs_EDR3}
\end{figure}

% Schonberner
Finally, \citet{schonberner18} calculated the distances to 15 round-shaped PNe by measuring the 
expansion velocity of the nebular rim and shell edges, and  correcting the velocities of the 
respective shock fronts with 1D radiation hydrodynamic simulations of nebular evolution. 
%we compare EDR3 distances with those estimated in the recent work of \citet{schonberner18}, where they use a fitting procedure to hydrodynamic models. This method consists on measuring nebular expansion velocities and fitting them to a 1D radiation-hydrodynamics simulations of nebular evolution. 
In Figure \ref{fig:Schonberner_vs_EDR3}, we see a comparison between these distance 
derivations and those from EDR3. The result is very similar to the one reported in 
Paper 1, and we  find no bias or clear relation between these distance derivations.
 The nearest nebulae (below 1800 pc) have overestimated distances in the work of Schonberner 
when compared with EDR3, whereas for distances greater than 1800 pc they seem to be underestimated. 
Nevertheless, the number of objects is fairly small and the errors quite large, so no 
firm conclusions may be drawn.

\subsection{Nebular sizes and morphology}

% Radios nebulares
Knowledge of the distances to PNe, together with the measured angular sizes of the 
nebulae, allows us to estimate their physical radii. We obtained the angular sizes from the HASH 
database, where minor and major nebular diameters are provided for almost all (99\%) of the PNe in our 
sample. Using mean angular radii as a proxy for true angular radii, we then obtained the physical 
radii. 
%were then obtained.%($\phi$
%by the use of distances, 
%converted to physical radii. % $R = \frac{2 \pi}{360} \cdot \frac{\phi}{3600} \cdot D$. Where $D$ is the distance to the nebula.

\begin{figure}[h!]
        \includegraphics[width=9.5cm,height=6cm]{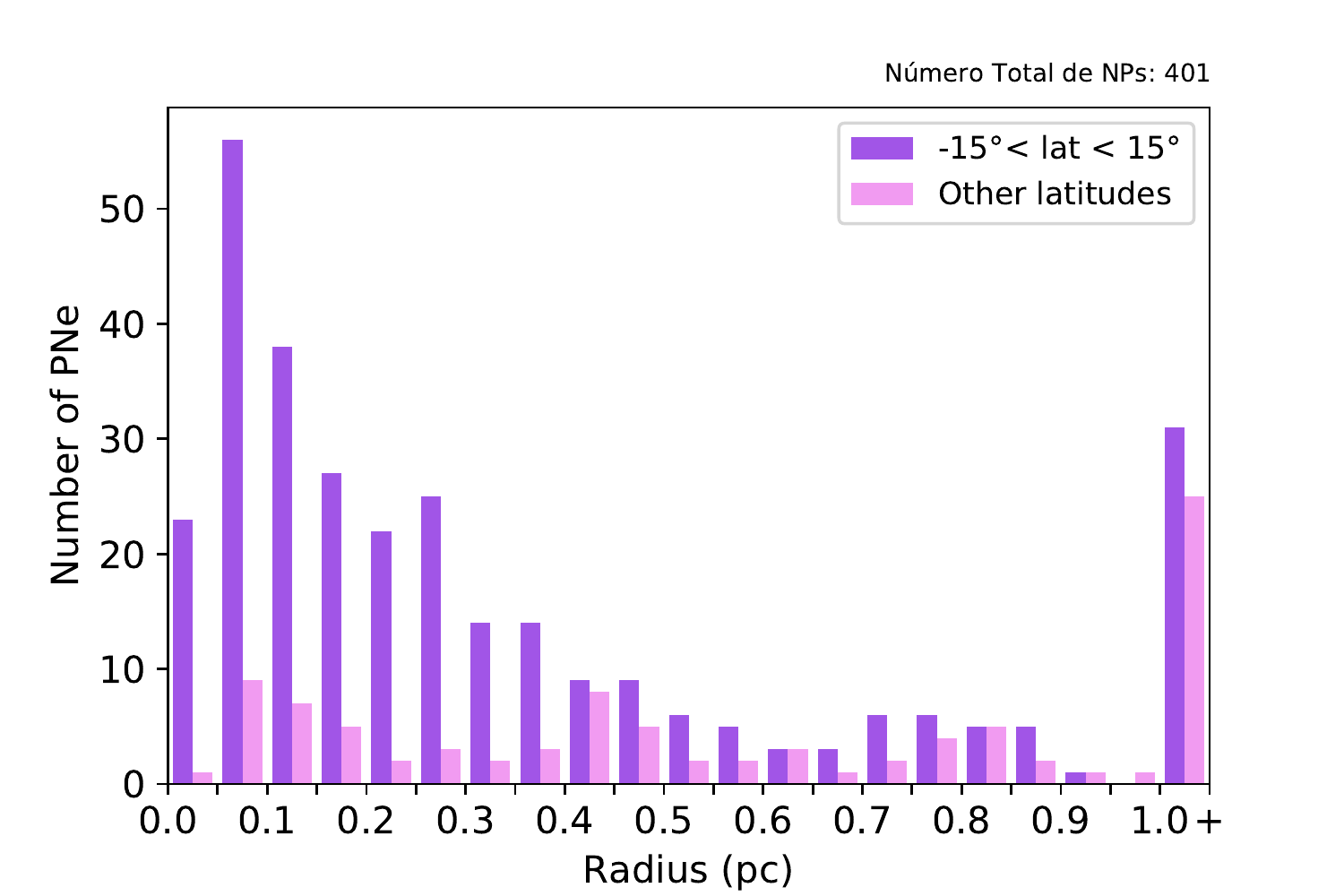}
        \caption{Radius distribution of the GAPN-EDR3 sample.}
        \label{fig:radius}
\end{figure}

%In Figure \ref{fig:radius} we have plotted 
Figure \ref{fig:radius} shows the nebular radius distribution for our objects in the 
GAPN-EDR3 sample. 
%As can be seen, most of the PNe (around
%the 
Fifty-four per cent have a radius shorter than 0.3 pc (1 light-year approximately). However, there are 
PNe with  much greater radii, which illustrates the diversity of evolutionary 
states in the sample. In fact, around 14 \% of the PNe have a radius greater than 1 
pc. These must be evolved nebulae, with their CSs already being at the white dwarf stage. 
%of white dwarf (WD). 
The mean radius of the entire sample is 0.59 pc. 
%So we can say that, in mean, the PNe have a diameter of approximately 1 pc. 
For more information, both angular and physical radii are listed in Table \ref{table:GAPN}.

For a more detailed analysis, we divided the PNe into two sets: those located close 
to the Galactic plane (latitudes in the interval $\pm 15^{o}$) and those located far from the Galactic 
plane (other latitudes). As shown in 
%this 
Figure \ref{fig:radius}, the PNe of the first group tend to be smaller than those of the 
second group, which are more evenly distributed over a wide range of sizes. 
%Iker: As star population in the galactic disk tend to be more massive, their corresponding nebulae should evolve faster and then should be larger than the other ones. However, this is in disagreement with the obtained results.
Thus, we may conclude that PNe in the Galactic plane tend to be younger than the others since they are still smaller in size.
% Iker: la vida nebular es suficientemente larga como para poder sacar esta conclusion ??
%\textcolor{blue}{This result is in agreement with the fact that the stars located in the galactic disk tend to be younger than those located 
%away from it, and consequently smaller yet}. 
%CAMBIARLO PARA 15 GRADOS EN VEZ DE 10 
%(that seems to have equally distributed radii, approximately).  

% Morfologia

As is well known from the pioneering work of \citet{1972A&A....18...70G}, 
PNe display a wide variety of morphologies. The most common ones are round, 
elliptical, and bipolar, but there are also others with star-like (or point-like), 
asymmetrical or irregular shapes. Logically, the morphological classification of PNe is 
subject to the uncertainty because of projection effects, and also because of the resolution and 
sensitivity of the images. If we analyse the morphologies in the GAPN-EDR3 sample (endorsing 
the HASH morphological cataloguing), we find that 40.2\% of the PNe are elliptical, 
22\% bipolar, 18.8\% round, 2.7\% star-like, 2\% irregular, and 1.2\% asymmetrical. 
We note that we have a significant percentage of PNe (13.1\%) of a non-classified 
morphology. In order to check if these percentages might be biased owing to a brightness 
effect (bipolars tend to be brighter on average), we calculated the morphological 
distribution for nebulae at distances of less than 2 kpc, 
%(328 PNe within the GAPN-EDR3 sample), 
% Iker: 328 son todas las que tienen tipo morfologico eliptico, bipolar o redondo
% De estas aproximadamente la mitad estan a menos de 2 kpc
which can be set as our approximate limit of completeness. As a result, we obtained 
a similar morphological distribution (37.2\% elliptical, 24\% bipolar, and 16.8\% round). 
% yo aqui pondrian los valores que se obtienen para cada morfologia si se limita la muestra a 2 Kpc. Tambien incluiria lo de los efectos de proyeccion, (el 7.3 de la elipticas se podrin ver como redondas de acuerdo a los efectos de proyeccion).

% so we can discard any morphological preference. The morphology type of each PNe is shown in Table \ref{table:GAPN}. 

%In Figure \ref{fig:morph}, we can observe this morphologycal distribution. CAMBIAR ESTA FIGURA POR LA DE LAS DISTANCIAS DE ESCALA DE LOS DIFERENTES TIPOS (VER DESPUÉS)

%With the aim of searching for a relation or pattern in this morphologycal distribution,
%we have analysed the PNe morphologies of the sample in function of different properties, like the galactic distribution, the nebular size, the evolutive age, the CS temperature or the stellar mass... In general, we have not found any clear relation between this properties and the nebular morphology. 
%However, we can trough some conclusions about the galactic distribution of elliptical, bipolar and round morphologies.

It is interesting to investigate whether any relation exists between PNe 
morphological types and their location in the Galaxy.
%Figure \ref{fig:morph_lat} help us to analyse if there is any relationship between PNe morphologies and their location in the Galaxy.% in Figure  \ref{fig:morph_lat} we have plotted elliptical, bipolar and round PNe in a galactic height vs. longitude diagram. %At naked eye, we can not see any clear pattern, but we will analyse this with more accuracy. 
Previous studies (see, for instance, \citealt{1983IAUS..103..233P}, 
\citealt{1988ApJ...324..501Z}, \citealt{corradiSchwarz95}, \citealt{2004ASPC..313....3M}) 
have shown that bipolar nebulae tend to be high-excitation, type I nebulae, and that 
they are located closer to the Galactic plane than elliptical ones. This led to the 
conclusion that bipolars  have probably  evolved from a more massive disc population, although this conclusion was based 
%yo quitaria lo de young, porque confunde con young population
%in a rather poor sample of objects.
on a poorly populated sample of objects.
%It seems that the elliptical ones, in general, are located closer to the galactic plane. But we can not assume as a clear pattern. 

%REHACEMOS ESTE PÁRRAFO CUANDO TENGAS LOS RESULTADOS DE <Z>. 
%YO CREO QUE SI SE VE ALGUN PATRÓN EN LA FIGURA 13. PRUEBA A CALCULAR EL SCALE HEITH DE LAS TRES POBLACIONES (DISTANCIA A LA CUAL EL NÚMERO DE OBJETOS DECRECE UNA CANTIDAD e). PUEDES HACERLO CON LA LATITUD, PERO LO MÁS LÓGICO SERÍA PASAR LAS COORDENADAS A RECTANGULARES (COMO HICIMOS PARA LOS CUMULOS) Y CALCULARLO PARA LA COORDENADA Z, QUE TE DA LA DISTANCIA PERPENDICULAR AL PLANO. dEBERÍA VERSE QUE LAS BIPOLARES ESTÁN MÁS CONCENTRADAS EN EL DISCO QUE LAS REDONDAS.

\begin{figure}[h!]
        \includegraphics[width=9.5cm,height=6cm]{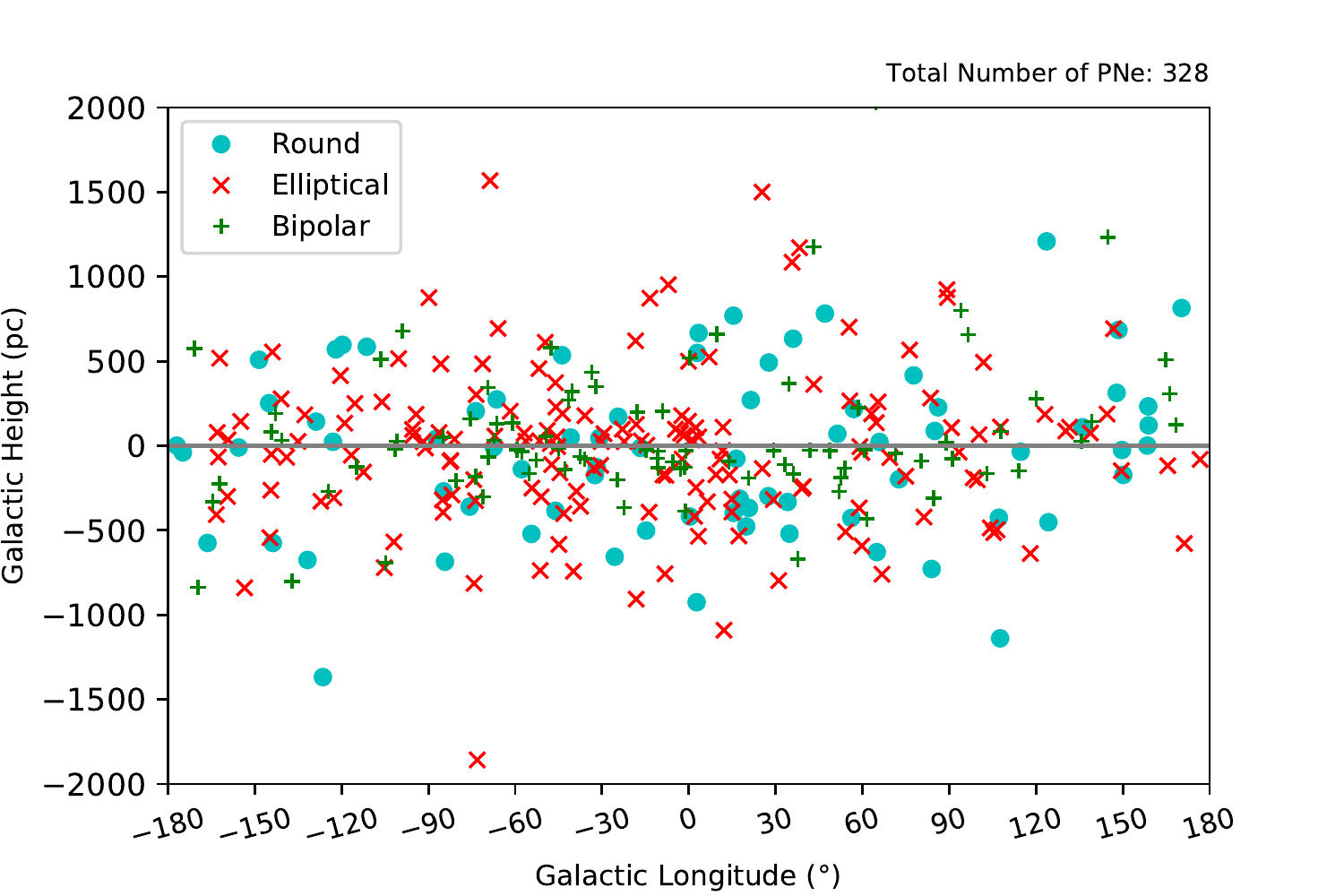}
        \caption{Galactic distribution of round, elliptical, and bipolar PNe.}
        \label{fig:morph_lat}
\end{figure}

%It is important to remark that the determination of the nebular morphology is not an easy task, because it is subject to many uncertainties, as visibility restrictions or projection effects.

\begin{figure}[h!]
        \includegraphics[width=9.5cm,height=6cm]{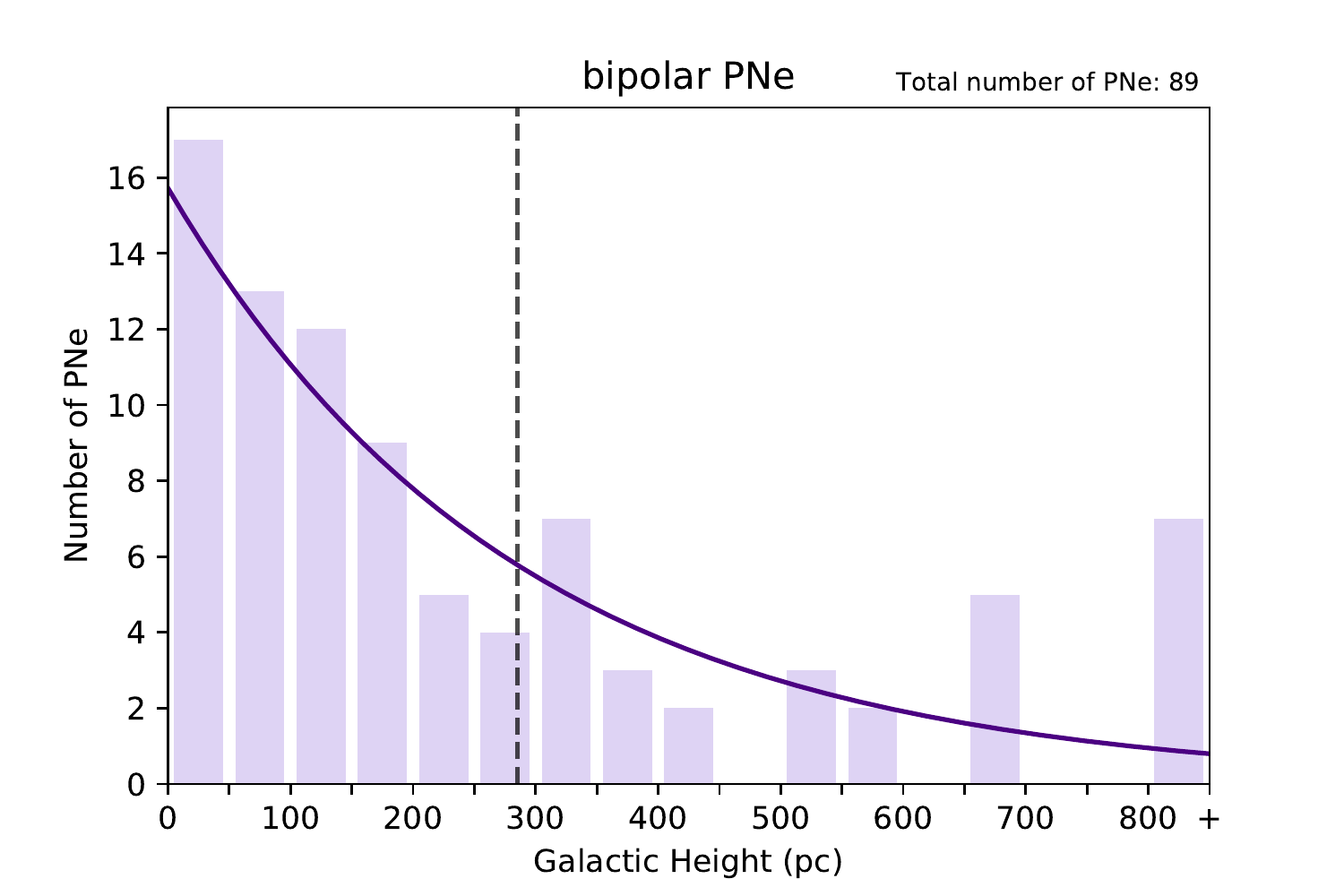}
        \includegraphics[width=9.5cm,height=6cm]{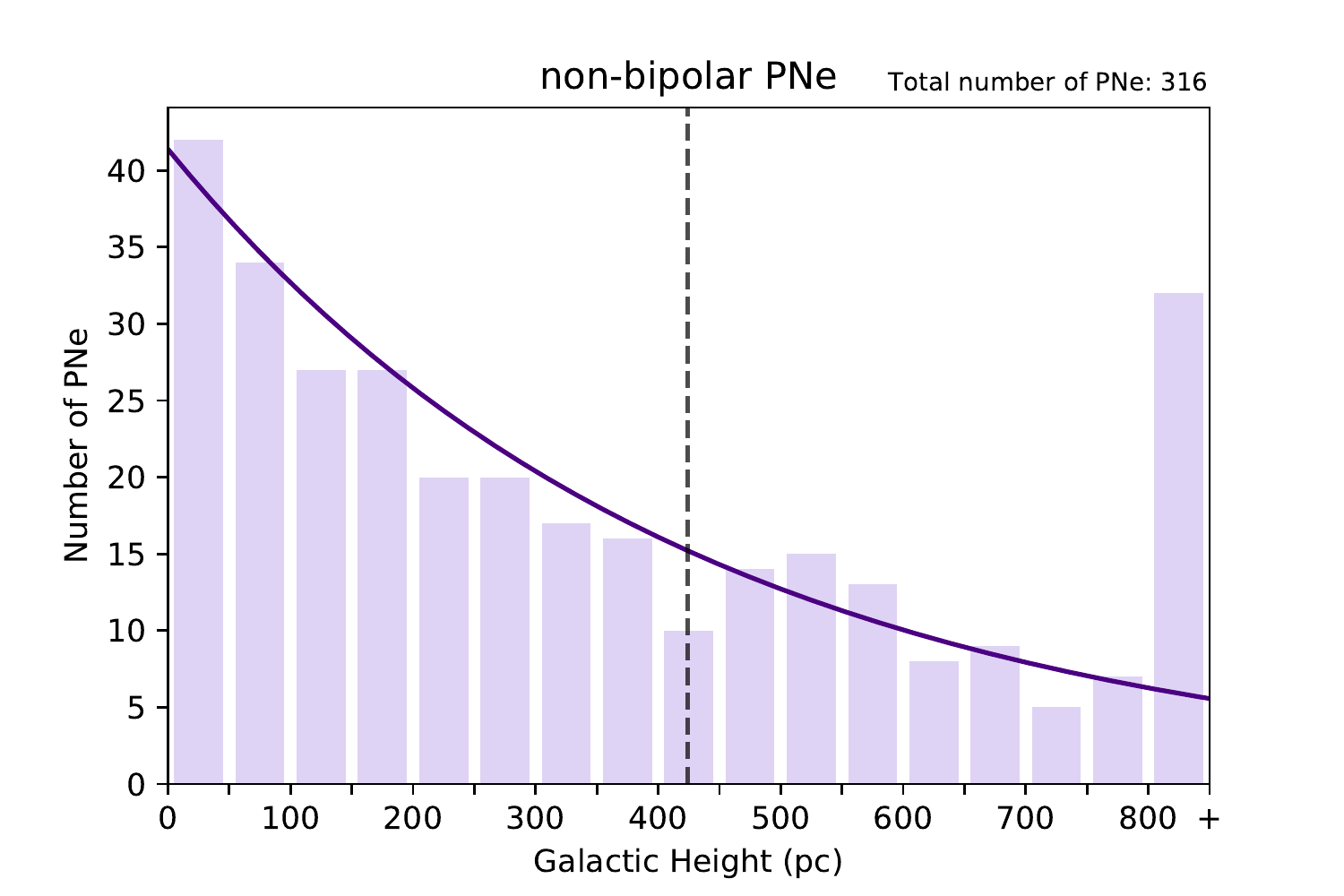}
        \caption{Galactic height distributions for bipolar (up) and non-bipolar (down) PNe. 
The dark curve represents the PNe population as a function of Galactic height per 50 pc bin, and 
the dotted vertical line indicates the scale height.}
        \label{fig:scale_height}
\end{figure}

Our data and accurate distances allow us to study this relationship in detail. 
%By using our GAPN-EDR3 data, we have carried out an analysis to probe this hypothesis. For this task, 
From latitudes and distances, we calculated the Galactic height ($z$) of 
the PNe in our sample. We then plotted (Figure \ref{fig:morph_lat}) the 
PNe from the main morphological categories (elliptical, bipolar, and round) as a 
function of their Galactic longitude and height.
%At a glance, there is not a clear difference between the bipolar PNe distribution and the other ones. 
%However, if we analyse this in more detail, we can draw some conclusions.
%A glance at this figure allows us to intuit that there really is a lack of bipolar nebulae at high galactic heights.
A glance at this figure seems to indicate that there really is a lack of bipolar nebulae at high Galactic heights.
To check this, we divided the PNe into two sub-samples according to their morphology 
(bipolar and non-bipolar). For each set, we displayed the distribution of the nebula 
population as a function of Galactic height (see Figure \ref{fig:scale_height}). 
Then, by fitting a logarithmic curve to these distributions, scale height 
values ($H_{z}$) were calculated for both populations: 
% we fitted the logarithm of these distributions by a lineal regression, and thus, we obtained a relation of the PNe population in function of the galactic height (in bins of 50 pc height). 

%PONER ENTRE PARÉNTESIS LOS VALORES DE LA SIGMA DEL AJUSTE 
%So the aim is to estimate at which height the population has decreased in a factor $e$ from the plane ($z=0$), i. e., to  obtain their scale height (Hz). After doing this calculation for the three morphological sets, we obtained the following values for their scale heights:

%$$H_{z} (elliptical) = 365 \pm \small 7 \thinspace pc,$$
$$H_{z}^{\rm bipolar} \thinspace \thinspace \thinspace = 285 \pm \small 8 \thinspace {\rm pc}, $$
$$H_{z}^{\rm non-bipolar} \thinspace \thinspace \thinspace = 424 \pm \small 7 \thinspace {\rm pc}. $$
%$$H_{z} (round) \thinspace \thinspace \thinspace \thinspace \thinspace \thinspace = 735 \pm \small 3 \thinspace pc. \\ $$

As can be seen, we obtained considerably different scale height values (with small uncertainties) for each morphological group. 
%So we can conclude that bipolar nebulae have clearly lower scale height value than non-bipolar population.
% In conclusion, bipolar PNe population will decrease faster with the height than the elliptical or round one, while these last populations will show a similar behaviour. Then, we have obtained that the bipolar PNe are effectively 
Bipolars are more concentrated in the Galactic disc region, so our data lend support to the conclusion that, in general, they come from a 
different population. 
%more massive progenitor stars.  
%Aqui tambien quito lo de younger
%CUANDO TENGAMOS LOS ERRORES PODREMOS DECIR MEJOR SI ESTAS DIFERENCIAS SON O NO SIGNIFICATIVAS.
Figure \ref{fig:scale_height} shows Galactic height distributions for each morphological 
group, together with their derived population curve. The scale height value for each set is shown by a vertical dotted line. 
% in each graphic.
The particular Galactic height values are listed in Table \ref{table:GAPN}.

We also studied the relationship between morphological type and nebular size.
 If we calculate the mean radius ($< R>$) for bipolar and non-bipolar PNe, we obtain
 the following values:

%$$<R> (elliptical) = 0.40 \pm \small 0.15 \thinspace pc,$$
$$< R>_{\rm bipolar} \thinspace \thinspace \thinspace = 0.31 \pm \small 0.12 \thinspace {\rm pc}, $$
$$< R>_{\rm non-bipolar} \thinspace \thinspace \thinspace = 0.67 \pm \small 0.21 \thinspace {\rm pc}. $$
%$$<R> (round) \thinspace \thinspace \thinspace \thinspace \thinspace \thinspace = 0.47 \pm \small 0.21 \thinspace pc. \\$$

As a result, we obtained that bipolar PNe tend to be considerably  smaller than the others,
 with a mean radius of less than a half that of the non-bipolar mean radius, although the 
uncertainty values are considerably higher in this case. 
%We can see that bipolar PNe 
%are 
%seem to be the 
%smaller 
%smallest ones, although %, while the elliptical and round ones tend to be a bit larger. %LO MISMO AQUI, NOS FALTAN LOS ERRORES. 
%Nevertheless, in this case, 
%the high uncertainties prevent 
%us 
%from obtaining conclusive results. % values are quite large, so we can not conclude to be a clear relation between the morphological type and the nebular size.
In the following section, we study the evolutionary properties (such as temperature, mass, and age) of CSPNe and analyse possible relationships 
between nebular morphology and these other properties.
%In the following section, where we will study the evolutionary properties of the CSPNe, we will analyse the relationships between the morphology and these properties, such as temperature, mass and age. 

% Velocidades Radiales (OPCIONAL)

\subsection{Expansion velocities and kinematic ages}

% Velocidades de Expansion

Another interesting property to study is the kinematic age of PNe. This parameter 
measures the time elapsed since the envelope was ejected from the CS crust, assuming 
that a constant representative expansion velocity for the nebula applies for its entire
life span. In Paper 1, we discussed  the problems with 
measuring expansion velocities and interpreting them in terms of the hydrodynamical 
evolution of real objects. For our new sample, we explored the literature for expansion velocity measurements and took
most of the values from the \citet{frew08} compilation of expansion velocities for a large set 
of PNe. 
% by considering other measurements in the literature. 
%Additionally, we have completed our expansion velocities sample with values from
Kinematic ages from \citet{1989A&AS...78..301W} were also included. From these expansion velocity values 
and physical nebular radii, we were then able to derive kinematic ages 
%as the ratio between radii and expansion velocities, 
under certain assumptions and constraints that 
%but certain assumptions and constraints 
needed to be considered.

We first of all worked on the  supposition that the expansion velocity is constant throughout the nebular 
phase and that this velocity was the same in all directions of the nebula. Hence, we only considered the case of approximately round-shaped nebulae. We proceeded to select a sample 
of PNe with minor semi-axis size 
values with 
at least 
%the 
75\% of the size of the major semi-axis: $R_{\rm min} \geq 0.75 \cdot R_{\rm max}$. In compliance with 
\citet{Jacob13}, we also rejected those PNe considered as H-deficient or 
containing
close binaries because their evolution follows a different path. We thus selected a 
sample of 65 PNe with known expansion velocities that fulfilled these conditions. 

Hydrodynamical modelling of the evolution of PNe by  \citet{Jacob13} indicates that 
expansion velocities will probably not be equal for all their layers, and that it is 
advisable to apply a correction factor to the measured velocities in order to obtain 
more realistic mean expansion velocity values. Following the prescriptions in \citet{Jacob13}, 
we decided to use an overall value of 1.5 as the correction factor, just as we did in Paper 1.

\begin{figure}[h!]
        \includegraphics[width=9.5cm,height=6cm]{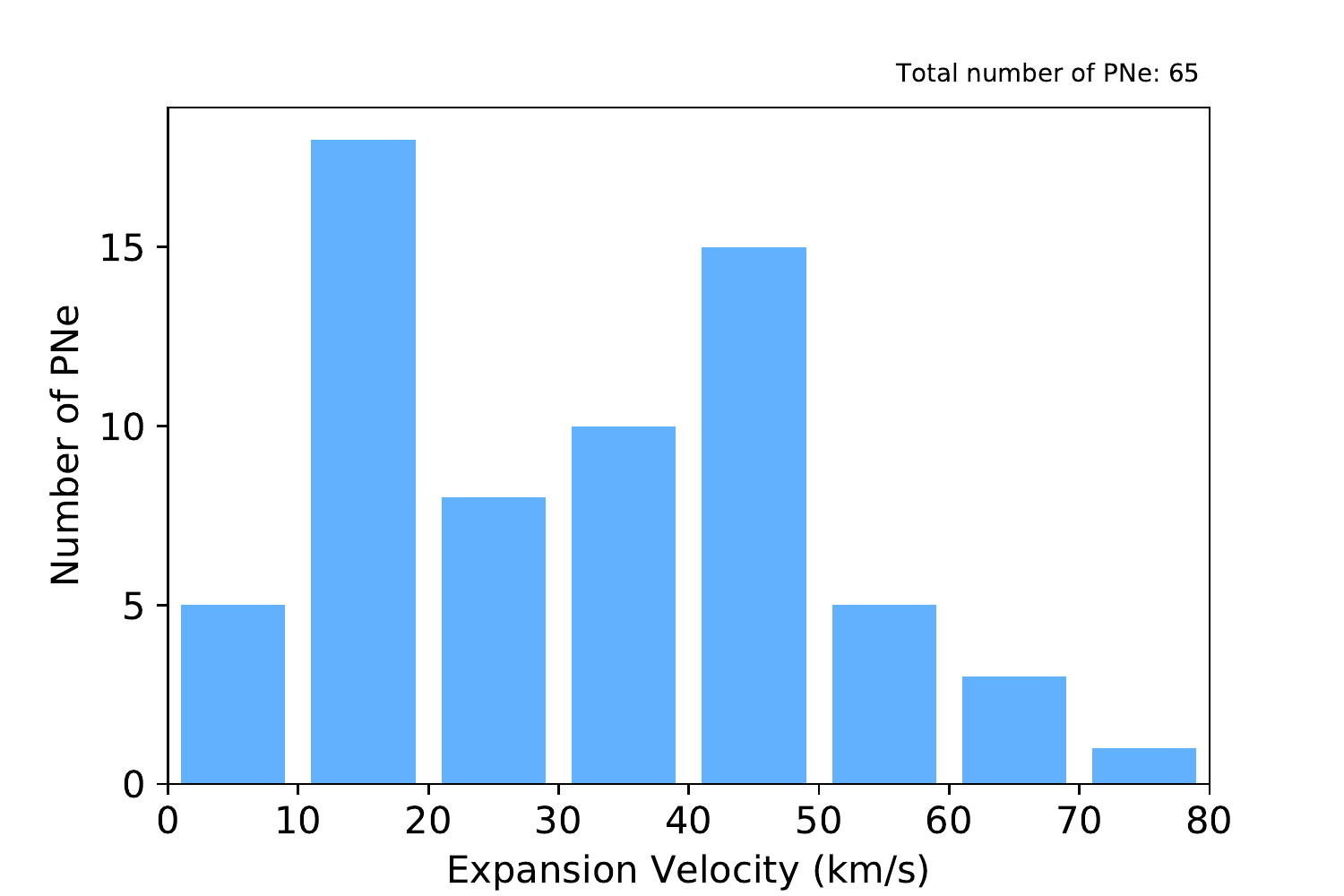}
        \caption{Expansion velocities distribution of 65 selected objects from the GAPN-EDR3 sample.}
        \label{fig:vel_exp}
\end{figure}

%After applying this correction, we get the expansion velocities for these PNe, whose 
The obtained expansion velocity distribution is shown in Figure \ref{fig:vel_exp}. 
Most (78\%) of the selected PNe have expansion velocities between 10 and 50 km/h, while 
the fastest nebula expands with a velocity of almost 80 km/h. The mean expansion velocity 
of this sample is $32 \pm 13$ km/h,
%$31.9 \pm 12.8$ km/h, 
slightly lower than the one obtained in Paper 1 ($38 \pm 16$ km/h), but within the 
uncertainty interval. Particular expansion velocity values for each of the 65 (uncorrected) PNe are listed in Table \ref{table:vel_exp}.

%where we had a poorer statistics.% In that previous work, we did a similar study but over a smaller sample (45 PNe), so we can say that the current results are more consistent than those ones. The extension of this new sample (67 PNe) is mainly due to the fact that GAPN-EDR3 is larger than the old GAPN, but also because in 
%In this new study we have complement the expansion velocities (obtained from \citealt{frew08}) with values obtained from \citet{1989A&AS...78..301W}.

% Edades Cinematicas

From the expansion velocities, we were able to calculate the corresponding kinematic ages using the following simple relation:

$$T_{\rm kin} = \dfrac{R}{V_{\rm exp}}.$$

\begin{figure}[h!]
        \includegraphics[width=9.5cm,height=6cm]{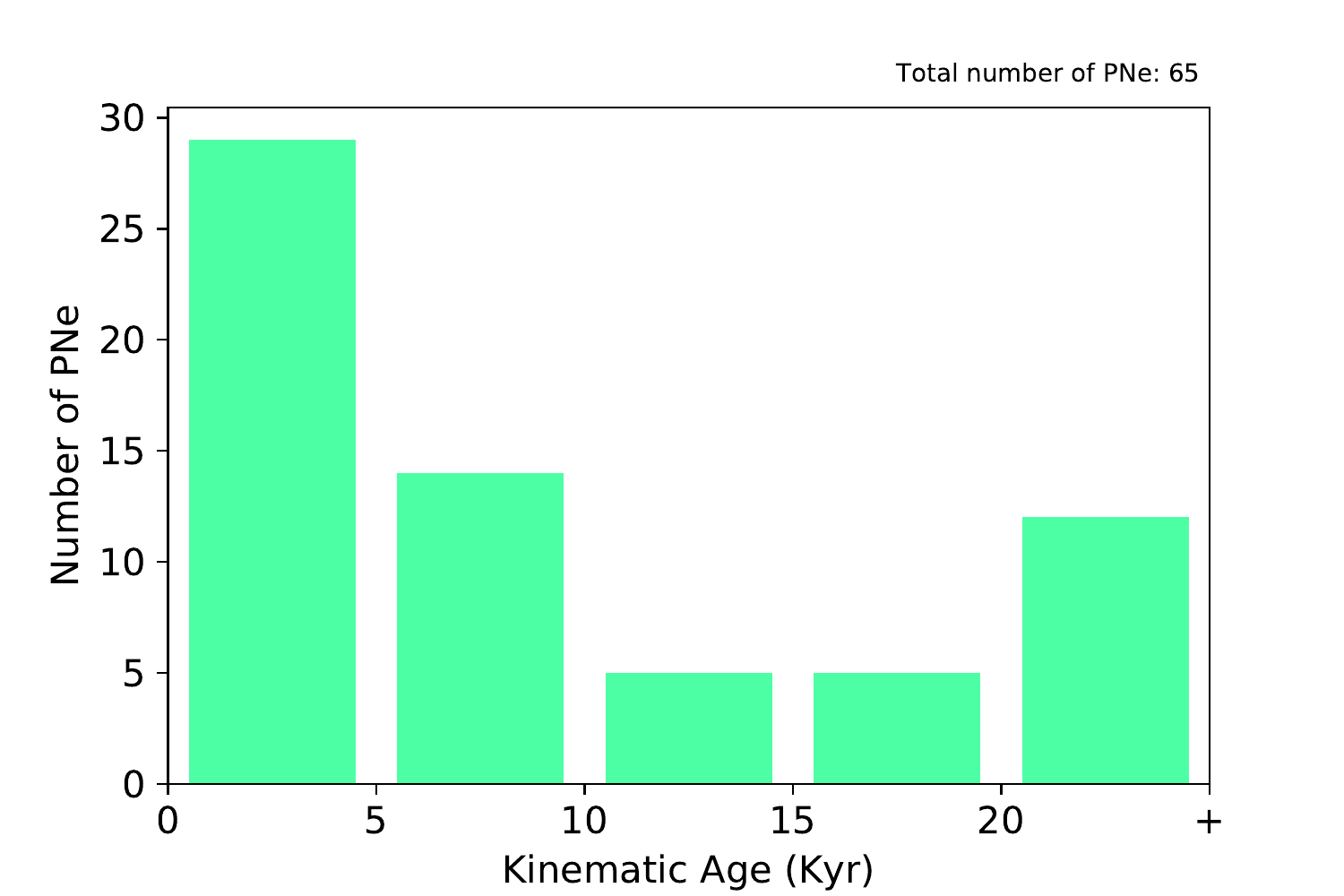}
        \caption{Kinematic age distribution of 65 objects  selected from the GAPN-EDR3 sample.}
        \label{fig:kin_age}
\end{figure}

The resultant kinematic age distribution is shown in Figure \ref{fig:kin_age}. 
As can be seen there, the majority (66\%) of the PNe 
%(the 
 are quite young, with a kinematic age below 10 kyr. However, there is a considerable fraction (18\%) of 
%PNe with older ages, 
older PNe, with ages
above 20 kyr. The reason why our results could be biased 
%to 
towards younger ages 
%we get more young PNe than old ones 
is probably that young PNe are easier to detect than %the 
old ones, as they are generally brighter. 
%Our sample mean %In addition, if we calculate the mean 
%kinematic
The mean kinematic
age of our sample is $17.8 \pm 3.0$ kyr. 
%This value may seem a bit high, this is due to the presence of a few very old PNe.
Table \ref{table:vel_exp} lists the kinematic age values of the PNe in this sample.

We were also able to estimate the maximum age of a PN or, equivalently, the maximum time that a PN is 
%visibly 
visible before becoming completely diluted in the interstellar medium. This concept is known as 
visibility time ($T_{vis}$). For this calculation, we need to consider a maximum mean nebular radius 
for the PNe \citep[reaching, according to ][0.9 pc, which is in good agreement with what we state 
in Sect. 3.2]{Jacob13}. By dividing this value by the obtained mean expansion velocity, we then calculated 
the following visibility time:

$$T_{\rm vis} = \dfrac{R_{\rm max}}{< V_{\rm exp}> } = 27.8 \pm 11.7 \thinspace \thinspace {\rm kyr}.$$

In conclusion, a PN is expected to have a mean lifetime of approximately 28 kyr. This value is higher than that obtained in Paper 1 ($23.4 \pm 6.8$ kyr), but both values are in agreement to within their uncertainties.

%\section{Compilation of complementary information from the literature}

\section{Evolutionary state of CSPNe}

Knowledge of accurate distances for the CSs in the GAPN-EDR3 sample can be 
combined with brightness and effective temperature values from the literature to 
%be able to 
locate them on an HR diagram. Making use of evolutionary tracks for post-AGB stars,
 their evolutionary state can then be estimated (i.e. their masses and ages).

\subsection{Temperatures and luminosities}

Literature about CSPNe effective temperatures mostly contains measurements obtained by the Zanstra method (\citealt{zanstra28}), which is based on measuring HI or HeII nebular fluxes. 
So, similarly to what we did in  Paper 1, we used literature effective temperature values obtained by this method.
%we searched the literature for GAPN-EDR3 effective temperatures 
%After a research on several sources of the literature, we decided to collect 
%for our GAPN-EDR3 objects, effective temperature values from some different works that mainly use the
%prioritizing Zanstra temperatures. 
%Thus, 
In this way, we were able to collect consistent temperature values for a total of 151 stars in our 
sample from \citet{frew08}, \citet{frew16}, \citet{1989A&A...222..237G}, and \citet{2013A&A...553A.126G}. 
%So, we were able to get effective temperatures for 210 CSPNe of our GAPN-EDR3 sample.

To calculate stellar luminosities, we  evaluated the bolometric magnitudes 
from brightness measurements in Gaia or other photometric bands. CSPNe are hot stars, 
with effective temperatures well beyond the limit of 8000 K that Gaia DPAC set for 
bolometric corrections and luminosities derived from DR2 observations (Andrae et 
al.\ 2018). As in Paper 1, we decided instead to use the \citet{vacca96} 
relationship for bolometric corrections in the visible band, so we 
%need 
needed to compile $V$ magnitudes for our sample of CSPNe. 
% Concerning the stellar brightness, the aim is to calculate bolometric agnitudes and luminosities of the CSPNe. To calculate this parameter i
%It will be also necessary to assess the interstellar extinction in the visible band. %So, for the objects of the
GAPN-EDR3 $V$ magnitudes were generally obtained from the studies of Frew (\citealt{frew08} 
and \citealt{frew16}), complemented with data from the APASS 
database\footnote{https://www.aavso.org/download-apass-data}, \citet{1991A&AS...89...77T}, and \citet{2020A&A...640A..10W}. 
We were thus able to obtain $V$ magnitudes from the literature for a total of 326 stars in GAPN-EDR3 (80\% of them).

\begin{figure}[h!]
        \centering
        \includegraphics[width=8.5cm,height=6cm]{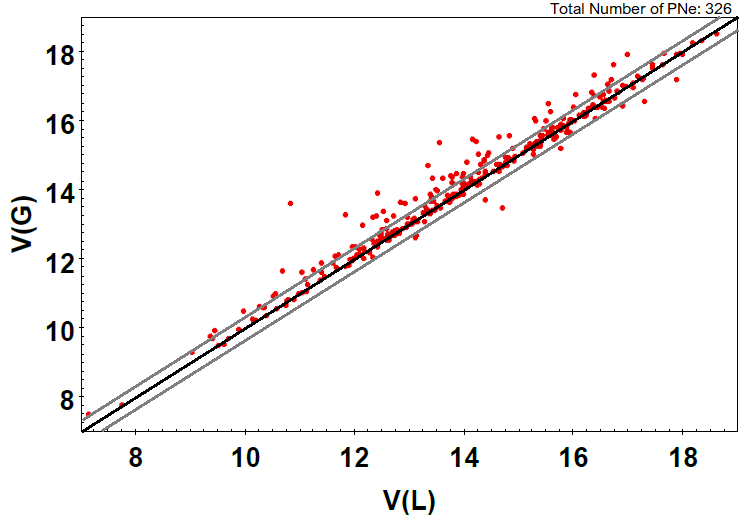}
        \caption{Comparison between $V$ magnitudes from the literature $(V_{L})$ and $V$ magnitudes 
calculated from Gaia passbands $(V_{G})$ for 326 objects of the GAPN-EDR3 sample. The black
 central line indicates the 1:1 relation, while grey lines indicate the threshold for the mild outliers.}
        \label{fig:VL_VG}
\end{figure}

Given the possibility that some of our CS identification did not match those 
in the literature, we decided to compare the de-reddened $G$ and $V$ magnitudes 
for each star. Gaia EDR3 online documentation\footnote{https://gea.esac.esa.int/archive/documentation/GEDR3} 
provides a relationship used to estimate the Johnson $V$ magnitude from the Gaia $G$, $G_{BP}$ and $G_{RP}$ magnitudes 
(henceforth, $V_{G}$). Such a relationship was calculated by Gaia DPAC using data with a lowest colour limit value
%colour low limit 
of $(G_{BP}- G_{RP})=-0.5$, but
%, as we will show next, 
we now show that it also works 
well
with blue stars beyond that limit. In fact, if we calculate the difference between literature $V$
values ($V_{L}$) and $G$-derived $V$ values 
($V_{G}$), 
% with   verify if these V magnitudes corresponds really to the detected CSs, we make use of their Gaia magnitudes ($G, G_{BP}, G_{RP}$) to estimate their corresponding theorical  magnitude by the relation $G-V = f(G_{BP},G_{RP})$ given in the Gaia EDR3 documentation
%Then, we compare the calculated V magnitudes ($V_{G}$) with those ones collected from the literature ($V_{L}$), obtaining quite similar values, see Figure \ref{fig:VL_VG}. Note that the $V = f(G,G_{BP},G_{RP})$ relation is only verified for objects with $G_{BP}-G_{RP}>-0.5$. However, as can be appreciated in this Figure, the relation also seems valid for bluer sources. In fact, if we calculate the difference between both magnitudes in each subsample, $\abs{V_{G}-V_{L}}$, 
we obtain a mean value and standard deviation of $\abs{V_{G}-V_{L}}=0.227 \pm 0.054$ mag 
for the sample of stars with bluer colours ($G_{BP}- G_{RP}\le{-0.5}$), 
and $\abs{V_{G}-V_{L}} = 0.278 \pm 0.098$ mag for the remaining objects in the sample. 
The comparison between the $V_{G}$ and $V_{L}$ magnitudes is shown in Figure \ref{fig:VL_VG}.
In order to filter out those CSPNe with suspect $V$ magnitudes from
 the literature, we use the term `mild outlier' as a threshold for
 such uncertain data.
%Now, our aim is to get rid of those CSPNe with unreliable V magnitude from the literature, and we will use the definition of mild outlier as the threshold for such uncertain data. %difference between both magnitudes greater than a certain threshold. 
To estimate this threshold value, we obtained the differences $\abs{V_{G}-V_{L}}$ and 
calculated their distribution interquartile range; namely, the difference between 
the third and first quartiles: ($Q3-Q1$). 
%Mild outliers are data values which lie between 1.5 times and 3.0 times the interquartile range below the first quartile or above the third quartile. %: To set this threshold we do a statistical analysis of the resulting differences $\abs{V_{G}-V_{L}}$, and we calculate the limit for the mild outliers. 
%Then, the 
As  is well known, the mild outlier limit is set as 1.5 times the interquartile range. 
Following this procedure, we obtained a threshold value of 0.382 mag (see Figure \ref{fig:VL_VG}). 
We then decided to discard those CSPNe with $\abs{V_{G}-V_{L}}$ greater than this value,
 and we finally obtained 253 CSPNe with reliable $V_{L}$ magnitude values. These $V$ magnitudes 
(and also the $G$ magnitudes) are listed in Table \ref{tab:evo}.

%we compare them with the corresponding Gaia G magnitudes (that should have similar values), obtaining differences below 1 magnitude. %Then, for V and G values differing more than 1 magnitude for the same object, we discard their V magnitude. 

\begin{figure*}[h!]
        \centering
        \includegraphics[width=19cm,height=12cm]{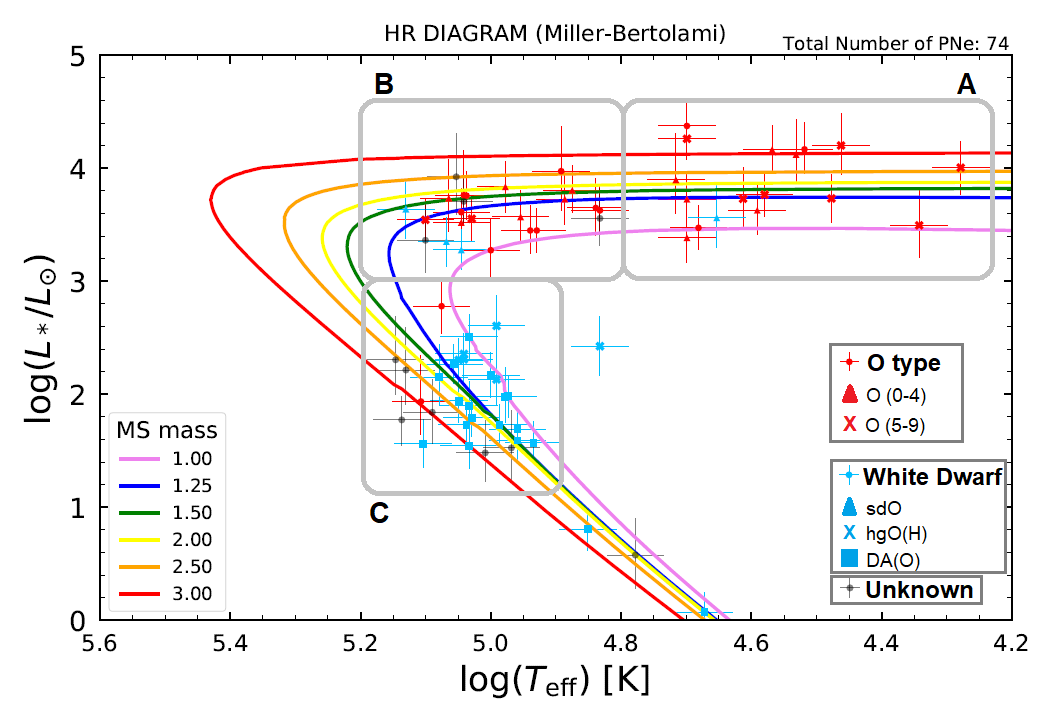}
        \caption{Location in the HR Diagram of the 74 CSPNe with known luminosities and 
temperatures in the GAPN-EDR3 sample, together with evolutionary tracks by \citet{2016A&A...588A..25M}. 
Information about their spectral classification is also provided.}
        \label{fig:HRD}
\end{figure*}

We also  had to consider the interstellar extinction for each star. As explained 
in Sect. 2, we obtained these values from Bayestar and SFD dust maps. 
The Galactic position of each source was employed to estimate the interstellar extinction. 
%These interstellar extinctions were estimated from the galactic position of the sources.
However, we know that the nebula itself might also contribute to the extinction of 
the CS, especially in young and compact nebulae. So, to be more 
%precise, 
thorough, we looked for specific 
%extinction in the literature 
literature extinction information 
for the selected stars. For consistency, we decided to obtain these values from the 
studies of Frew (\citealt{frew08}  and \citealt{frew16}), complemented by 
% In addition, we complemented 
extinction values from \citet{1992A&AS...95..337T} and \citet{1992A&AS...94..399C}. 
For the remaining CSPNe (51\% of them), we used the general extinction obtained from the dust maps. 
All these extinction values are shown in Table \ref{tab:evo}.

With this procedure, we were able to calculate the absolute visible magnitude ($M_{V}$) for the 253 CSPNe
(with reliable $V_{L}$ magnitudes)
cited previously using the following relation:

$$M_{V} = V +5 -5\cdot\log{(D)} -A_{V},$$

\noindent where $V$ is the visible magnitude, $D$ the distance, and $A_{V}$ the interstellar extinction.

%Then, in order to estimate their luminosity, it is necessary to obtain the absolute magnitude corresponding to the stellar radiation in all wavelengths, namely, 
%Then, 
The absolute bolometric magnitude ($M_{\rm Bol}$) could then be calculated from the absolute 
visible magnitude ($M_{V}$) and the effective temperature ($T_{\rm eff}$) via the relation of \citet{vacca96}:

$$M_{\rm Bol} = M_{V} + 27.66 - 6.84\cdot\log{(T_{\rm eff})}.$$

Finally, we obtained the stellar luminosity as:

$$\log{(\dfrac{L}{L_{\odot}}) = \dfrac{M_{{\rm Bol}_{\odot}} - M_{\rm  Bol}}{2.5}},$$

\noindent where $M_{{\rm Bol}_{\odot}}=4.75$ is the Sun's absolute bolometric magnitude.

As we compiled effective temperatures for a limited set of objects in GAPN-EDR3 
sample, we were able to estimate the luminosities for a total of 121 CSPNe. 
%If we want to 
In order to
analyse the evolutionary state of such stars using standard H-rich evolutionary tracks, 
we then had to exclude those objects catalogued as H-deficient stars, as well as known 
close binary stars. 
%, because this type of objects are not valid for these evolutionary models. 
For the identification of these types of objects, we followed the classification of \citet{2019ibfe.book.....B}, \citet{2020A&A...640A..10W}, and  \citet{Jacob13}.
%, obtaining a subset of 102 CSPNe. 
%without evidence of binarity CSPNe. %In addition, if we compare the effective temperatures from different authors, we see that some of Weidmann's ones are not in agreement with the other author's ones, so we decided to discard the stars with temperatures from Weidmann for this analysis in order to have more consistency in the results. 
We finally ended up with a sample 74 CSPNe, whose evolutionary state could be studied 
from their temperature and luminosity values. All these values are presented in Table \ref{tab:evo}.
%
%Within this set of CSPNe, we have an 83\% of them belonging to group A of identification reliability, while only a 17\% are from group B and none of them from group C. So in this sample the reliability is very high.
This sample of CSPNe has a very high reliability rate, as 85\% of them belong to 
  identification reliability group A, and only 15\% belong to group B.

\subsection{Mass and evolutionary age}

%In the same way that we did with the 
As in our previous DR2 study, the next step was to plot our selection of 74 
CSPNe with reliable temperature and luminosity determinations
%determinations of temperature and luminosity
on an HR diagram and compare their location with the predictions of evolutionary 
models for the post-AGB phase. As in Paper 1, we decided to use the post-AGB 
evolutionary tracks of \citet{2016A&A...588A..25M} because of the updated opacities, 
nuclear reaction rates, and  consistent treatment of stellar winds for the C- and 
O-rich regimes in that paper. Miller Bertolami computed the evolutionary sequences from the ZAMS to the post-AGB 
phase for stars in the initial  0.8--4 $M_\odot$ mass range, which corresponds to 
post-AGB masses from 0.5 to 0.85 $M_\odot$, which is the most relevant range for CSPNe. In 
the case of solar metallicity (which we assumed), the upper mass limit for main sequence stars 
is 3 $M_\odot$.

%Then, we proceed to plot these CSPNe in a Luminosity vs Temperature HR Diagram, together with the evolutionary tracks for post-AGB stars by \citet{2016A&A...588A..25M}, as can be seen in Figure \ref{fig:HRD}. These tracks represent the path that a CSPNe should follow for different masses of their progenitor star, that go from 1 $M_{\odot}$ to 3 $M_{\odot}$. These paths start at the right upper region of the diagram, that indicates the beginning of the post-AGB phase, after the AGB phase. 
In these models, a temperature limit of 7000 K ($\log{(T_{\rm eff})} = 3.85$) is set as 
the reference for the beginning of the fast part of the post-AGB evolutionary process, when 
winds play only a secondary role in setting
the timescales. The CSPNe evolutionary age is calculated by adding a transition time 
that begins earlier, when the envelope mass reaches  1\% of the stellar mass 
and the AGB stage ends (see \citet{2016A&A...588A..25M} for details). 
%the beginning of the post-AGB evolutionary stage (or late AGB phase) , as it corresponds to the beginning of the fast part in the post-AGB phase, concretely when the CS reaches a value of $\log{(T_{eff})} = 3.85$. Thus, the CSPNe evolutionary age will be calculated from this point, adding a transition time value that begins a bit earlier, when the post AGB phase starts. This moment is set in this model when the envelope mass reaches a value greater than the 1\% of the total progenitor star mass. 
Following the evolutionary paths, the stars warm up to an approximately constant 
luminosity. After reaching the maximum temperature, they  start to cool down and 
lose luminosity until they eventually
 %will 
reach a stable phase as white dwarf stars. 
%So each point in the tracks indicates an evolutionary time since the beginning of the post-AGB phase. 
As we have seen (depending on the mass), the whole PN phase time span can last 
tens of thousands of years.
%As we have seen before, the whole time of the PN phase can last several tens of thousands of years, with a value that depends on the mass.
%The more massive a star is, the faster it will evolve. So the tracks associated to most massive progenitor stars will increase the evolutionary age faster than the other ones. 

In Figure \ref{fig:HRD}, the location of our CSPNe is set in the HR diagram, 
together with the evolutionary tracks. As can be appreciated, all the points in this diagram 
are plotted together with their corresponding error bars. We assumed relative errors of 10\% in 
temperature. Taking into account the uncertainty in luminosity, we  calculated 
them by error propagation from the uncertainties in distance (low and high), $V$ magnitude, 
and temperature. Then, by interpolation between the tracks, we were able to estimate 
the progenitor mass and evolutionary age for all the CSPNe.

% masas

\begin{figure}[h!]
        \includegraphics[width=9.5cm,height=6cm]{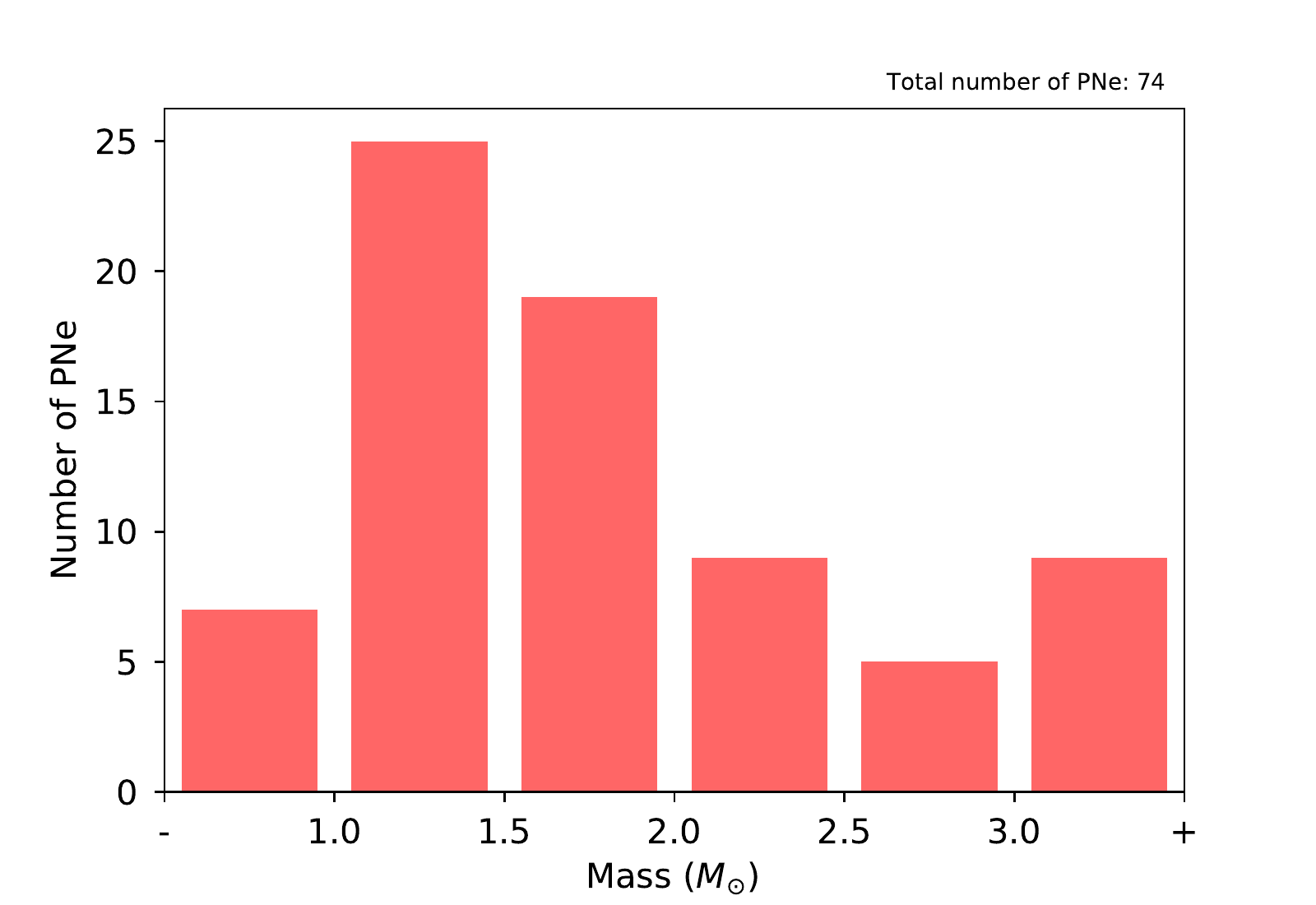}
        \caption{Estimated progenitor masses ($M_{\odot}$) for the 74 CSPNe in the HR diagram in Figure \ref{fig:HRD}.}
        \label{fig:mass}
\end{figure}

%Now, for each object in the HR diagram, by interpolation, we calculate its corresponding progenitor mass and its evolutionary age. 
Figure \ref{fig:mass} shows %these 
the mass distribution for the 74 CSPNe selected. Around 
%the 
70\% of the CSPNe in our sample come from progenitor stars with masses below 2 $M_{\odot}$, 
and we obtain a mean value for the progenitor mass of $1.8 \pm 0.5 M_{\odot}$. 
%OJO HE REDONDEADO ESTOS VALORES A UNA CIFRA SIGNIFICATIVA, HAY QUE COMPROBAR TAMBIEN LA TABLA. On the other hand, we obtain a few CSPNe coming from massive progenitors, 
Only a few stars (12\% of the sample) lie slightly beyond the 3 $M_{\odot}$ track. 
%While we obtain a mean value for the progenitors mass of $1.76 \pm 0.46 M_{\odot}$. 
Particular mass values are listed in Table \ref{tab:evo}.

% edad evolutiva

\begin{figure}[h!]
        \includegraphics[width=9.5cm,height=6cm]{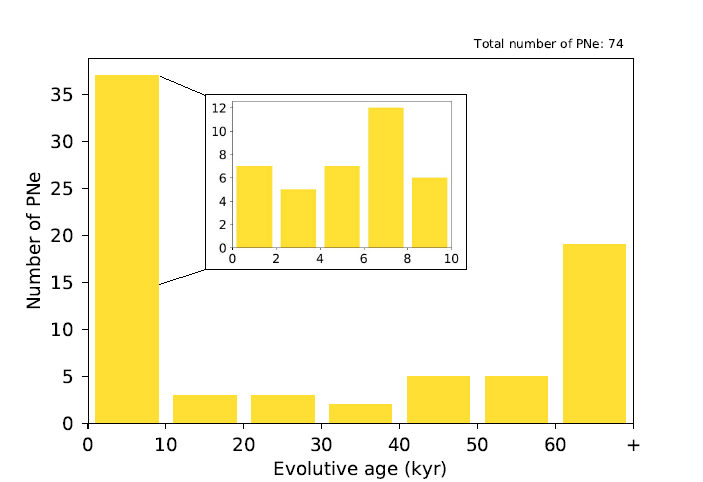}
        \caption{Estimated evolutionary ages (kyr) for the 74 CSPNe in the HR diagram in Figure \ref{fig:HRD}.}
        \label{fig:age_evo}
\end{figure}

Concerning 
%to 
the evolutionary age (see Figure \ref{fig:age_evo}), 
%we obtain a large variety of values, whose distribution can be observed in Figure \ref{fig:age_evo}. 
although we found a large variety of values, most of the CSPNe are quite young, as around %the 
50\% 
have ages below 10 kyr. Within this group, the age distribution is quite constant, as can be seen in the inset of the figure.
% Esto se cumple con las edades de Vassiliadis !
%In addition, we can say that the population of CSPNe decreases with the increase of the evolutionary age. As we saw with the kinematic ages, this could happen because the older the PNe are the more difficult are to be detected, due to the brightness loss with the time. 
On the other hand, we detected several objects with ages far above 60 kyr. 
These ages correspond to CSPNe located 
towards the end of the evolutionary tracks
%at the final region of the evolutionary tracks 
(see Figure \ref{fig:HRD}). These stars have 
lost almost 
all of their nebulae and have already reached the
white dwarf
phase. 
%of white dwarf. 
%While 
%Meanwhile, 
The remaining CSPNe present intermediate evolutionary ages between 10 and 60 kyr. We note that, 
following the prescriptions of \citet{2016A&A...588A..25M}, we added a certain quantity corresponding to the transition time to the ages obtained 
from the tracks. As we already mentioned, 
this time extends from the beginning of the post-AGB phase to the instant 
%which 
at which
the CS reaches the  temperature of 7000 K set by this model. This time interval lasts 
between approximately 1 and 36 kyr, depending on the progenitor masses. All the 
evolutionary ages are listed in Table \ref{tab:evo}.

In general, the location of the CSPNe in the HR diagram adequately fits  the region covered 
by the models, with only one clear outsider: 
%One of them is the CS of the Abell 46 nebula. This is a O9 star with a temperature of 66\,000 K and  luminosity  $\log{(\sfrac{L}{L_{\odot}})}=3.08$. This luminosity is rather low for this type of star, which is why it falls below the tracks of the corresponding region. 
%The other outlier is 
the CS of the HaWe 13 nebula. This is an hgO(H) white 
dwarf with a temperature of 68\,000 K and luminosity  $\log{(\sfrac{L}{L_{\odot}})}=2.43$. 
Being a white dwarf star, it should  have yet a higher temperature for that luminosity or
should otherwise already have a lower luminosity for that temperature.

% Tipos espectrales

\subsection{Spectral types}

We  retrieved the spectral classifications for those stars located in the HR diagram from 
\citet{2020A&A...640A..10W}. 
In Figure \ref{fig:HRD}, we plot those 
objects that are classified as O-type stars with a red symbol, 
%blue symbols 
 white dwarf stars with a blue symbol, and those of unknown classification with a grey symbol.
\citet{2020A&A...640A..10W}  reviewed the evolutionary paths followed by stars until the PN stage 
from an observational point of view. Although there is no  generally accepted theoretical model, it is agreed that 
%in general, single stars seem genrally to follow an  evolutionary path that divides post-AGB evolution into 
two groups of H-rich and H-poor stars, single stars seeming to follow two evolutionary post-AGB channels, according 
to their H-rich and H-poor composition
 (\citealt{2019MNRAS.489.1054L}). \citet{2020A&A...640A..10W} discuss the spectral types 
associated with different masses and the uncertainties concerning interpretation of  spectral 
sequences. For their representation on an HR diagram, we discarded objects with spectral types 
 associated with H-deficient stars, as well as all objects susceptible to binarity. 
Figure \ref{fig:HRD} shows spectral types divided into early O-type (0--4) stars, late O-type (5--9) stars,
 O-type hot subdwarfs (sdO), intermediate evolved objects (hgO(H)), and H-rich white dwarfs (DA(O)).

A fairly coherent sequence of spectral types in
our HR diagram became notably apparent: 
%A notable result is that the arrangement of spectral types in our HR diagram shows a fairly coherent sequence: 
O-type stars were located in the constant luminosity region of the tracks, 
%and, 
with 
 late O-type stars 
%are 
generally being
located in the right part of this region (cooler temperatures) and early O-type 
stars tending to be located at a later stage (hotter temperatures): a scenario that supports a good 
correspondence between an object's position in the diagram and its spectral classification. 
%Stars 
Stars classified as white dwarfs were also located in the region where they were 
expected to be, near the elbow that marks 
%the 
maximum temperatures and in the region where the central star begins to cool down and 
drastically decrease in luminosity. %Subdwarfs 
O-type hot subdwarfs tended to be located in the most luminous zone. We then found hgO(H) 
intermediate objects and in the most evolved region there were H-rich white dwarfs. 
%Of course, there are a few objects that are
Only a few objects appeared not to conform  to these rules.
The particular spectral types of GAPN-EDR3 objects are listed in Table \ref{table:GAPN}, 
particularly for those objects located in the HR Diagram, in Table \ref{tab:evo}.

%Regarding to white dwarfs, we see that they are located in the more advanced region of the PN phase, when the CSPNe have started loosing temperature and luminosity. Within this region, it can be appreciate that subdwarfs tend to be located in the more luminous zone, then we find the hgO(H) intermediate objects and in the most evolved region there are located the H-rich white dwarfs. 

% Analisis por regiones del DHR

\subsection{HR diagram analysis per region}

For a more detailed study, we  separated the HR diagram into three regions (see Figure \ref{fig:HRD}) 
according to the evolutionary status of the objects. As in Paper 1, we defined the regions as follows: 
region A corresponds to an early PN phase, with stars increasing their temperatures (till 
$\log{(T_{\rm eff})} = 4.8$) at a fairly
constant luminosity ($\log{(\frac{L}{L_{\odot}})} > 3$); region B corresponds to the same flat 
luminosity part, but for higher temperatures (until the maximum temperature); and region C 
corresponds to the late-PN phase with lower luminosities than the previous regions and decreasing temperatures.   
%We define the region A as the area with $\log{(\frac{L}{L_{\odot}})} > 3$ and $\log{(T_{eff})} < 4.8$, that corresponds to the early PN phase at high luminosities and relatively low temperatures. Then, the region B is the area  with $\log{(\frac{L}{L_{\odot}})} > 3$ and $\log{(T_{eff})} > 4.8$, that corresponds to the intermediate PN phase with also high luminosities but higher temperatures than the previous region. Finally, we have the region C, that is the area  with $\log{(\frac{L}{L_{\odot}})} < 3$ and $\log{(T_{eff})} > 4.9$, and corresponds to the late PN phase with lower luminosities than the previous region and decreasing temperatures. 
Almost all objects located in region A are O-type stars. On the other hand, almost all those
 located in region C are white dwarf stars. While in region B, we find several objects of
 both types,  the majority of them being O-type stars.

\begin{table} [h!]
\caption{Mean values (with uncertainties) of different parameters in three regions of the HR diagram.}  
\label{tab:HRD}      
\centering  
%\large

\begin{tabular}{l| l| l| l}          
\hline                      
\textbf{Parameter} & \textbf{Region A} & \textbf{Region B} & \textbf{Region C} \\    
\hline \hline
    \textit{Number of PNe}  & \textit{17} & \textit{24} & \textit{29} \\ 
    \hline 
    $< R >$ (pc) & 0.08 \scriptsize{(0.02)} & 0.23  \scriptsize{(0.13)} & 0.75 \scriptsize{(0.32)}\\
    \hline 
    $< T_{\rm eff}$> (kK) & 39 \scriptsize{(9)} & 100  \scriptsize{(17)} & 109 \scriptsize{(11)} \\
    \hline 
    $< M >$ ($M_{\odot}$) & 1.98 \scriptsize{(0.50)} & 1.57  \scriptsize{(0.42)} & 1.87 \scriptsize{(0.62)} \\
    \hline 
    %<$M^{Vas}$> ($M_{\odot}$) & 2.17 \scriptsize{(0.61)} & 1.52  \scriptsize{(0.34)} & 2.11 \scriptsize{(0.57)} \\
    %\hline 
    $< T_{\rm evo}> $ (kyr) & 16.0 \scriptsize{(3.8)} & 22.0 \scriptsize{(1.5)} & 53.5 \scriptsize{(30.3)}\\
    \hline 
    %<$Age_{evo}^{Vas}$> (kyr) & 4.7 \scriptsize{(2.3)} & 12.2 \scriptsize{(7.8)} & 59.8 \scriptsize{(22.1)}\\
    %\hline     
    $< T_{\rm kin} > $ (kyr) & 5.5 \scriptsize{(2.5)} & 6.4 \scriptsize{(1.9)}   & 34.7 \scriptsize{(8.9)} \\
    %\hline 
    
\end{tabular}

\tablefoot{%Parameter $<V_{exp}^{mod}>: $ mean expansion velocity from evolutionary age 
%and nebular size;   $<V_{exp}^{obs}>: $ mean expansion velocity from emission lines observations.  
Region A: $ \log\left( \frac{L}{L_{\odot}}\right)  >3.0$ \& $\log\left( {T_{\rm eff}}\right)  < 4.8$; 
Region B: $\log\left( \frac{L}{L_{\odot}}\right)  >3.0$ \& $\log\left ({T_{\rm eff}}\right)  > 4.8$;
Region C: $\log\left( \frac{L}{L_{\odot}}\right) < 3.0$ \& $\log\left( {T_{\rm eff}}\right) > 4.9$.
}

\end{table}

Table \ref{tab:HRD} shows the mean values and standard deviations of the evolutionary 
parameters in each region (nebular radius, CSPNe temperature and age, and progenitor star mass).
%, and we compare the mean values of this parameters %between 
%among different regions. 
%Table \ref{tab:HRD} shows mean values, and standard deviations, of these evolutionary parameters in each region.
%In Table \ref{tab:HRD} we show the mean values of these evolutionary parameters in each region, together with the standard deviation. 
Regarding the nebular radius mean value ($< R> $), the 
clear and expected increase in size from region A to C is significant, going from less than 0.1 pc to almost 
1 pc. So, the properties of stars and the derived size of their nebulae are in agreement with what 
is expected for these rapidly evolving objects. 
%So we verify that as the nebulae evolve, they tend to increase their physical size. We can also appreciate how the effective temperature mean value (<$T_{eff}$>) is much higher in regions B and C than in the early A region, where the stars have not already reached so high temperatures. 
On the other hand, we can observe that the progenitor star mass mean value ($< M> $) 
%keeps 
remains approximately constant 
%through 
throughout the three regions, with a mass around almost 2 $M_{\odot}$. This is what we 
expected, as this parameter does not depend on the evolutionary phase reached by the PN. 
%According to the evolutionary age, there is a clear increment on its mean value ($<T_{evo}>$) from region A to B, and especially from region B to C. Nevertheless, these 
Evolutionary age mean values ($< T_{\rm evo}> $) tend to increase considerably from region
 A to C, although there is great dispersion, as their standard deviations are considerably 
high, especially in region C. That is probably due to the presence of very old objects that 
have almost entirely lost their nebulae and have already become a white dwarf. Finally, we 
have also included the corresponding mean values for the kinematic ages 
($< T_{\rm kin}> $) obtained in the previous section in this table. As expected, these also show a similar tendency to
 increase as evolutionary ages. We note that there are four CSPNe located outside the 
three regions; these are generally stars that are too evolved to be considered for this evolutionary analysis.    

%, in order to compare these ages with the evolutionary ones (that should be similar) in each region. 
%We find a high similarity between both ages in the region A. 
%Mainly, we see that kinematic ages are underestimated comparing with the evolutionary ages. However, they show a similar tendency of increment from one region to another. kinematic ages mean values tend to present lower uncertainty values than evolutionary ages ones.

%We can observe that the kinematic ages mean values are underestimated comparing with evolutionary ones, however, we can conclude that they tend to increase in a similar way from one region to another. 

As we said previously, in Paper 1 we performed a similar analysis using Gaia DR2
parameters, and now we have been able to locate more CSPNe on the HR diagram  with
 better quality  parameters, so these results should be more consistent. 
Furthermore, some discrepant objects located away from the evolutionary tracks in
the previous study are no longer present.
%Furthermore, we have got rid of some controversial objects that appeared out of the evolutionary tracks in the previous study.
These objects were discarded for several reasons: misclassification as PNe according to 
the HASH database, a new source in Gaia EDR3 identified as  CSPNe, or their being
catalogued as a close binary or H-deficient star in \citet{2020A&A...640A..10W}. 
%Another important difference between both studies is that in Paper 1 we made use of \citet{millerbertolami17} models and now we are using \citet{2016A&A...588A..25M} models (as they include a wider range of masses) to estimate the progenitor masses and evolutionary ages. Nevertheless, the results are quite similar with both procedures. 
%In the main, 
In summary, in Paper 1 we had slightly higher values 
for nebular radii and lower values for the evolutionary age, but 
the mean values generally follow a similar trend and are compatible to within their uncertainties.

% Morfologia

\begin{table} [h!]
\caption{Mean values (with uncertainties) of different parameters from the main three morphological types.} 
\label{table:morph}      
\centering  
%\large

\begin{tabular}{l| l| l| l}          
\hline                      
\textbf{Parameter} & \textbf{Elliptical} & \textbf{Bipolar} & \textbf{Round} \\    
\hline \hline
    \textit{Number of PNe}  & \textit{27} & \textit{20} & \textit{24} \\ 
    \hline 
    $< |z|> $ (pc) & 450 \scriptsize{(223)} & 413  \scriptsize{(140)} & 522 \scriptsize{(222)} \\
    \hline 
    $< R> $ (pc) & 0.42 \scriptsize{(0.12)} & 0.21 \scriptsize{(0.07)} & 0.51 \scriptsize{(0.21)}\\
    \hline 
    $< T_{\rm eff}> $ (kK) & 90 \scriptsize{(18)} & 85  \scriptsize{(31)} & 90 \scriptsize{(13)} \\
    \hline 
    $< M> $ ($M_{\odot}$)  & 1.81 \scriptsize{(0.34)} & 1.83 \scriptsize{(0.47)} & 1.80 \scriptsize{(0.55)}\\
    \hline 
    $< T_{\rm evo}> $ (kyr) & 53.4 \scriptsize{(3.3)} & 20.5 \scriptsize{(2.9)} & 66.3 \scriptsize{(48.1)}\\
    \hline
    $< T_{\rm kin}> $ (kyr) & 14.0 \scriptsize{(2.8)} & 8.4 \scriptsize{(3.1)}   & 31.9 \scriptsize{(5.1)} \\
    %\hline     

\end{tabular}

%\tablefoot{%Parameter $<V_{exp}^{mod}>: $ mean expansion velocity from evolutionary 
%age and nebular size;   $<V_{exp}^{obs}>: $ mean expansion velocity from emission lines observations.  
%}

\end{table}

Returning to the morphological study, we can now analyse the evolutionary state of each of 
the main morphological groups (elliptical, bipolar, and round) by obtaining the mean value of 
their different evolutionary parameters,
%. As 
as shown in Table \ref{table:morph}.
We note that 96\% of the CSPNe in this sample are catalogued within these main morphological types. 
According to the results, 
we are able to confirm that bipolar PNe are notably smaller than the other 
%ones 
types 
(as we postulated in Sect. 3.2), with a mean radius of only 0.2 pc. Consequently, 
our sample of bipolar PNe tend to be younger than  elliptical or round ones. 
Regarding the CSPNe temperature and progenitor mass, the three morphological types 
show similar mean values, and
no definitive conclusions may be drawn concerning a possible relationship between these 
properties and  nebula morphological type. 
It should be taken into account that a round nebulae count may be influenced by
 projection effects ($7\%$ of ellipticals can be seen as round,  \citealt{2004ASPC..313....3M}).

%we cannot conclude anything about the relation of the morphological type with these parameters.
%This result is in agreement with what we predicted in section 3.2, that as bipolar PNe tend to be closer to the galactic disk than the others, they should come from more massive progenitor stars which have not had enough time to leave this galactic primitive region. 

\section{Binary CSPNe}

It is well known that the large variety of PNe shapes has been related to binary companion interaction (Jones and Boffin 2017). Gaia astrometry may provide some clues to assess the importance of binarity and stellar multiplicity in the formation and evolution of CSPNe. Some authors have proposed that even distant binary companions may provide an alternative mechanism for the formation of highly aspherical morphologies by influencing the direction of collimated winds from the parent star (\citealt{GarciaSegura97}, see discussion in \citealt{1998AJ....116.1357S}). Therefore, we found that it may be useful to search for binaries by using Gaia EDR3 precise astrometry.

In Paper 2, we performed 
a search for comoving objects in the fields of PNe with good determinations of parallaxes and proper motions. 
%We found that several cases exist, 
Several cases were found and we were able to estimate 
%masses 
mass values for both the CSs and their binary companions compatible with a 
joint evolutionary scenario. It should be noted here that none of the systems found in such work was close enough to generate the necessary gravitational torque that could result in an impact on the morphology of the nebula, according to the models by \citet{GarciaSegura97}.
%However, all our comoving binary companions are too far from the central star to influence their nebular morphology. 
More precise astrometry in the EDR3 archive allowed us to update 
the aforementioned search of comoving companions in our GAPN-EDR3 sample.

The Gaia object detection pipeline and the astrometric solution for multi-epoch 
observations in the EDR3 catalogue provides users with parameters that can be used to search for evidence of close binary pairs. 
We performed 
a statistical test to look for such evidence of binarity 
%between our samples 
among our sample targets with `red' and `blue' ($G_{BP}-G_{RP} \le{ -0.2}$) CSs. 

%At least  In some cases, the PNe are evolved from a binary CS instead of from a single CS. The study of these cases can provide more information to better understand the formation and evolution of the PNe (\citealt{2019ibfe.book.....B}). In addition, the presence of binary systems could be related with the aspherical morphologies in PNe (\citealt{2001ApJ...558..157S}). In this section we will focus in the searching of binary systems among our GAPN-EDR3 sample objects using astrometric parameters from EDR3.

\subsection{Search for wide binaries in GAPN-EDR3}

\begin{table*}[h!]
    \caption{Data from new binary systems detected in GAPN-EDR3.}
    \label{tab:binaries}
    \begin{tabular}{ l c c c c c c c c} 
    %\begin{tabular}{ l l l l l l l l}
    %\begin{tabular}{ l r r r r r r r}
    
\hline\hline
$Object$ & RA & Dec & Separation & Distance & Parallax & $PM_{RA}$ & $PM_{Dec}$ & $G_{BP}-G_{RP}$\\ 
& (º) & (º) &  (AU) & (pc) & (mas)  & (mas $\cdot$ yr$^{-1}$) & (mas $\cdot$ yr$^{-1}$) & (mag)\\ 
\hline
     
%\object{Abell 24} (CS)& 117.9065 & 3.0059 & - & 750$^{+44}_{-57}$ & 1.36 $\pm$ \small{0.10} & -4.27 $\pm$ \small{0.12}& -0.70    $\pm$ \small{0.09}\\ %& -                            \\
%\object{Abell 24} - B^{(*)}^{(+)}& 117.9067 & 3.0021 & 10,254 & 743$^{+21}_{-24}$ & 1.35 $\pm$ \small{0.04}& -4.36 $\pm$ \small{0.08}& -0.95 $\pm$ \small{0.07} \\ %& 0.06 $\pm$ \small{0.01} \\
%\object{Abell 33} (CS)& 144.7880 & -2.8084 & - & 997$^{+47}_{-46}$ & 1.02 $\pm$ \small{0.06}& -14.85 $\pm$ \small{0.09}& 9.55 $\pm$ \small{0.08} \\ %& -          \\       %60
%\object{Abell 33} - B^{(*)}& 144.7878 & -2.8089 & 1,665 & 926$^{+53}_{-45}$ & 1.09 $\pm$ \small{0.07}& -14.86 $\pm$ \small{0.09}& 9.71 $\pm$ \small{0.08}  \\ %& 0.12 $\pm$ \small{0.03}\\
%\object{Abell 34} (CS)& 146.3973 & -13.1711 & - & 1,187$^{+108}_{-84}$ & 0.84 $\pm$ \small{0.07}& 3.20 $\pm$ \small{0.09}& -9.16 $\pm$ \small{0.09}  \\ %& -           \\
%\object{Abell 34} - B^{(*)}& 146.3955 & -13.1693 & 10,761 & 1,126$^{+32}_{-27}$ & 0.89 $\pm$ \small{0.03}& 3.30 $\pm$ \small{0.07}& -9.09 $\pm$ \small{0.07}  \\ %& 0.04 $\pm$ \small{0.01}\\
%& 0.42 $\pm$ \small{0.11}      \\

%\object{NGC 3699} (CS)& 171.9910 & -59.9579 & - & 1,446$^{+198}_{-164}$ & 0.74 $\pm$ \small{0.10} & -3.22 $\pm$ \small{0.10}& 1.11 $\pm$ \small{0.10} \\ %& -         \\  %-16
%\object{NGC 3699} - B^{(*)}& 171.9922 & -59.9585 & 4,634 & 1,521$^{+129}_{-95}$ & 0.66 $\pm$ \small{0.04}& -3.27 $\pm$ \small{0.08}& 1.10 $\pm$ \small{0.08}   \\ %& 0.04 $\pm$ \small{0.01} \\
\object{NGC 6720} (CS)& 283.3962 & 33.0291 & - & 783$^{+29}_{-32}$ & 1.29 $\pm$ \small{0.05}& 1.73 $\pm$ \small{0.08}& 2.36  $\pm$ \small{0.08} & -0.79\\ %& 1.54$^{+0.01}_{-0.47}$ \\ %& -            \\
\object{NGC 6720} - B& 283.3915 & 33.0324 & 14,448 & 1,093$^{+346}_{-220}$ & 1.05 $\pm$ \small{0.26} & 1.22 $\pm$ \small{0.24}& 2.56 $\pm$ \small{0.24} & -0.23 \\ %& > 0.20 \\ %& 0.05 $\pm$ \small{0.01}       \\
\object{NGC 6781} (CS)& 289.6170 & 6.5387 & - & 494$^{+19}_{-19}$ & 2.03 $\pm$ \small{0.07} & -6.93 $\pm$ \small{0.11}& -4.17 $\pm$ \small{0.09} & -0.51\\ %& 2.21$^{+0.51}_{-0.89}$ \\ %& -           \\       %4
\object{NGC 6781} - B& 289.6189 & 6.54012 & 4,139 & 470$^{+34}_{-29}$ & 2.15 $\pm$ \small{0.19}& -6.64 $\pm$ \small{0.19}& -3.79 $\pm$ \small{0.16} & 2.66\\ %& 0.35$^{+0.05}_{-0.05}$ \\ 
\object{PN G030.8+03.4a} (CS)& 278.8805 & -0.2640 & - & 546$^{+82}_{-53}$ & 1.92 $\pm$ \small{0.27} & 2.01 $\pm$ \small{0.22}& -8.85$\pm$ \small{0.18} & 0.43\\ %& -     \\
\object{PN G030.8+03.4a} - B& 278.8800 & -0.2612 & 5,425 & 533$^{+169}_{-98}$ & 2.18 $\pm$ \small{0.44} & 2.66 $\pm$ \small{0.35}& -8.51 $\pm$ \small{0.29} & 1.66\\ %& 0.11 $\pm$ \small{0.03}      \\
\hline \hline
\textit{Possible Triple System} \\
\hline
\object{Fr 2-42} (CS)& 298.4000 & -10.3255 & - & 129$^{+1}_{-1}$ & 7.76 $\pm$ \small{0.06}& -11.48 $\pm$ \small{0.09}& -16.52 $\pm$ \small{0.08} & -0.23 \\ %& -       \\
\object{Fr 2-42} - B& 298.4001 & -10.3249 & 304 & 130$^{+1}_{-1}$ & 7.68 $\pm$ \small{0.06}& -10.92 $\pm$ \small{0.09}& -15.78 $\pm$ \small{0.08} & -0.20\\ %& 1.95 $\pm$ \small{0.49}           \\
\object{Fr 2-42} - C& 298.3880 & -10.3320 & 6,528 & 128$^{+1}_{-1}$ & 7.78 $\pm$ \small{0.10}& -10.96 $\pm$ \small{0.12}& -16.14 $\pm$ \small{0.09}  & -0.03\\

%& 0.14 $\pm$ \small{0.04}       \\
%\object{NGC 6853} (CS)& 299.9016 & 22.7212 & - & 387$^{+7}_{-6}$ & 2.58 $\pm$ \small{0.04}& 10.58 $\pm$ \small{0.07}& 3.57  $\pm$ \small{0.08} \\ %& -         \\     %-42
%\object{NGC 6853} - B^{(*)}^{(+)}& 299.9005 & 22.7197 & 2,484 & 383$^{+4}_{-5}$ & 2.62 $\pm$ \small{0.05}& 10.26 $\pm$ \small{0.07}& 3.81  $\pm$ \small{0.08}  \\ %& 0.23 $\pm$ \small{0.06}\\
%\object{NGC 6853} - C^{(*)}& 299.8997 & 22.7202 & 2,809 & 538$^{+184}_{-89}$ & 2.05 $\pm$ \small{0.44}& 10.07 $\pm$ \small{0.25}& 3.04 $\pm$ \small{0.37}   \\ %& 0.22 $\pm$ \small{0.05} \\
%\object{Sh 2-123} (CS)& 325.5646 & 44.4679 & - & 688$^{+58}_{-48}$ & 1.50 $\pm$ \small{0.11}& -2.65 $\pm$ \small{0.12}& -6.11  $\pm$ \small{0.13} \\ %& -       \\
%\object{Sh 2-123} - B^{(+)}& 325.5645 & 44.4685 & 1,432 & 811$^{+41}_{-33}$ & 1.23 $\pm$ \small{0.07}& -3.08 $\pm$ \small{0.09}& -6.15 $\pm$ \small{0.09}  \\ %& 0.17 $\pm$ \small{0.04}   \\

\hline
    \end{tabular}
    
\tablefoot{All parameters are obtained or derived from Gaia EDR3. Colours are extinction corrected.}

%(*): Companion star previously identified in Paper 2 with Gaia DR2.

%(+): Companion star detected due to the possible orbital velocity influence.}

\end{table*}    

In Paper 2, we showed that high-precision Gaia  parallaxes and proper motions allow
us
to find wide binary companions as comoving objects. In a similar way, we used the improved Gaia EDR3 astrometry to search for such
types of 
resolved binaries in our GAPN-EDR3 sample. We define sources that 
%present 
have celestial coordinates and astrometric parameters (parallaxes ($w$) and proper 
motions in RA ($PM_{RA}$) and Dec ($PM_{Dec}$) coordinates) compatible with a gravitational 
bond to within the observational errors as comoving objects. We note that as radial velocities are, in general, unavailable 
for the companion stars, we limited our study to movements 
%in 
on the plane of the sky.

%The procedure we pursued is equivalent to the one that has been followed in our previous work. 
The procedure pursued is equivalent to that in our previous work.
We first selected all the Gaia EDR3 sources within a radius of 120 arcsec around each of our CSPNe. We then
%get rid of 
discarded those objects without parallax and obtained 630 objects on average in each PNe neighbourhood. 
%For such sources we corrected their parallaxes (and their uncertainties) as we explained at
The parallaxes (and uncertainties) for these objects were corrected as explained at
the beginning of Sect. 4. In this case, we also needed to correct the proper motions and their uncertainties,
%. This 
and this was done following the relations %explained 
in \citet{lindegren18}. Then, 
%we only selected 
only objects with accurate astrometry were selected, that is, those with relative errors  below 30\%
in parallax,  distance and proper motions. On average, we ended up with around 50 
%candidates for 
binary companion candidates per PN. We note that the central stars in the GAPN-EDR3 sample also had to
pass such filtering in proper motions, which is why we lost a few of them and finished  with 357 CSPNe.

The next step was to implement an algorithm to detect the comoving objects to any CSPNe. 
%We 
As in Paper 2, we adapted the procedure used by \citet{2019AJ....157...78J}.
%, in the same way we did in Paper 2. 
The method consists of selecting objects for which the three astrometric parameters 
($w$, $PM_{RA}$, $PM_{Dec}$) differ less than 2.5 times $\sigma$ in comparison with those of 
their corresponding CSs, where $\sigma$ is the 
%higher 
highest uncertainty value between the candidate object and the CS for each parameter. 
After executing the algorithm, we 
%have been 
were able to detect 85 possible binary (or multiple) systems as CSs in the GAPN-EDR3 sample.

However, many of those binary companions might be too far from the CSPNe to be truly gravitationally 
%linked 
bound, and they might only share the same astrometric parameters by chance. 
%So it is necessary to define a maximum separation distance which we can consider that both stars could be gravitationally linked, and consequently form a wide binary system. Furthermore, it is more realistic to use the physical distance than the angular one. 
In Paper 2, we included a discussion of the expectation for the maximum 
projected physical separation between binaries. Following \citet{2020AJ....159...33Z}, 
we set this maximum value at 20\,000 AU. By using the same constraint
here, we finally isolated eight wide binary or multiple systems. 

In comparison with the binary CSPNe found in Paper 2, we confirm the presence 
of five wide binary systems (Abell 24, Abell 33, Abell 34, NGC 3699, and NGC 6853). For more 
details, we invite the reader to consult Sect. 3 of Paper 2. On the other hand, we could not, for different reasons, include in the present study
three possible wide binary systems found in Paper 2: NGC 246 (because
its CS did not fulfil the astrometric quality constraints to be included in GAPN-EDR3), SB 36 (because
it is not catalogued as a PN in the HASH database), and PHR J1129-6012 (because the identification of the CS 
was not clear). Apart from the five systems mentioned, we detected three possible new  wide binary systems 
%in 
for PN G030.8+03.4a, NGC 6720, and NGC 6781. In Table \ref{tab:binaries}, we show the astrometric 
parameters (coordinates, separation, distance, parallax, and proper motions) and Gaia colours of
 both components belonging to the systems cited before. 
%In addition, their images can be seen in Appendix \ref{fig:images_1}.

% NGC 6720

\begin{figure}[h!]
        \includegraphics[width=9cm,height=7.5cm]{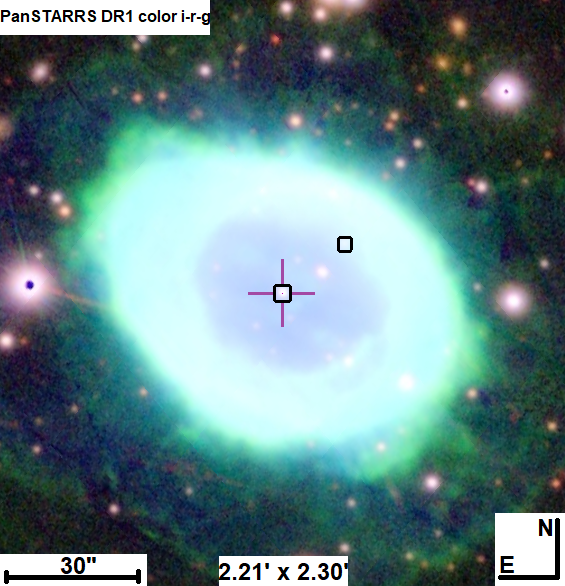}
        \caption{Image of a possible binary system in \object{NGC 6720}, showing the location of 
both the CSPN (with a cross) and the comoving companion. Image from the \textit{Aladin} sky atlas (CDS).}
        \label{fig:ngc6720}
\end{figure}

The first new binary system detected 
%belongs 
corresponds to the planetary nebula \object{NGC 6720}  (Figure \ref{fig:ngc6720}), also known as the 
Ring Nebula. This nebula is located at a distance of almost 800 pc from the Sun and  has an 
elliptical shape, with a mean radius of around 30\,000 AU (0.147 pc). We detected a binary 
companion with a projected separation of more than 14\,000 AU from the CS. According to its absolute 
$G$ magnitude ($M_{G}= 8.16$ mag) and its colour ($G_{BP}-G_{RP}= -0.23$), it must also be a white dwarf. 
%So, from evolutionary models for WDs\footnote{https://www.astro.umontreal.ca/~bergeron/CoolingModels}, we have estimated for it an effective temperature of 20,000 K and a up-limit mass of 0.2 $M_{\odot}$. On the other hand, from Miller-Bertolami models, we obtained a progenitor mass of 1.54 $M_{\odot}$ for the CS, that corresponds to a CSPN mass of 0.58 $M_{\odot}$. The masses values, with the uncertainties, are available in Table \ref{tab:binaries}. 

% NGC 6781

\begin{figure}[h!]
        \includegraphics[width=9cm,height=7.5cm]{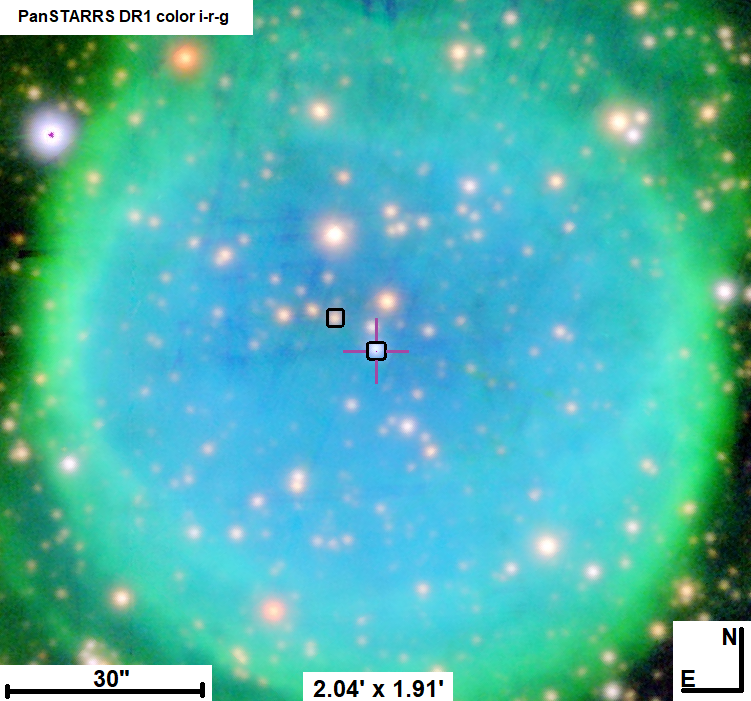}
        \caption{Image of a possible binary system in \object{NGC 6781}, showing the location of 
both the CSPN (with a cross) and the comoving companion. Image from the \textit{Aladin} sky atlas (CDS).}
        \label{fig:ngc6781}
\end{figure}

We detected another binary system in the bipolar
planetary nebula \object{NGC 6781}
%. This is a bipolar planetary nebulae 
(Figure \ref{fig:ngc6781}),
which is 
located 
%at 
almost 500 pc 
from the Sun. It has a similar size 
%than 
to that of the Ring Nebula, with a mean radius of around 31\,000 AU (0.15 pc). 
In this case, we detected a binary companion with a projected separation of 
less than 5000 AU from the CS. Using 
the
Virtual Observatory SED Analyser (VOSA) tool from the Spanish Virtual Observatory 
(SVO) platform\footnote{http://svo2.cab.inta-csic.es/theory/vosa}, we were able to 
obtain some evolutionary parameters of the companion star. From its 
coordinates, distance and extinction, we first retrieved its photometry and built its 
spectral energy distribution (SED). By fitting this SED to NextGen models 
(\citealt{2012RSPTA.370.2765A}), we were then able to calculate its effective temperature 
($T_{\rm eff}= 3400$ K) and luminosity ($\log[L/L_{\odot}]= -1.18$). These values 
correspond to an M-type star. Finally, by plotting the star in an HR diagram and using 
NextGen evolutionary tracks, we estimated a mass of 0.35 $M_{\odot}$ for the 
companion
star. 
%So, we obtain that the CS is quite more massive than the companion star, which has a mass about 0.60 $M_{\odot}$, 
The CS with 0.60 $M_{\odot}$ is considerably more massive than the companion star, as its 
mass in the MS would be about 2.21 $M_{\odot}$ according to the Miller Bertolami models.
%its mass (0.35 $M_{\odot}$) and age (10.5 Myr).
The presence of a binary system in this nebula was proposed a few years ago by 
\citet{2015MNRAS.448.3132D}, who catalogued it as an M3-type star, in good 
agreement with the values found here. 
%Now, with Gaia EDR3 astrometry, we have been able to throw more light about the existence of this binary CSPN.

% IPHAS J183531

\begin{figure}[h!]
        \includegraphics[width=9cm,height=7.5cm]{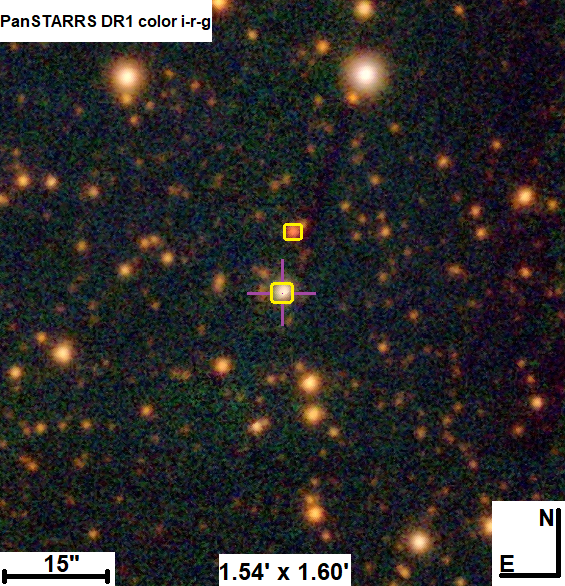}
        \caption{Image of a comoving system in \object{PNG 030.8+03.4a}, showing the 
location of both the CSPN (with a cross) and the companion. Image from the \textit{Aladin} sky atlas (CDS).}
        \label{fig:iphas}
\end{figure}

The third binary system detected by our algorithm is located in the nebula 
\object{PNG 030.8+03.4a} (Figure \ref{fig:iphas}). This %planetary nebula 
PN has a star-like morphology and is also close 
%from 
to the Sun, at a distance of around 550 pc. Here, we detected a binary 
companion with a projected separation of
%a bit more than 
slightly over 5000 AU from the CS. In this case, by fitting its SED through the VOSA tool, 
we estimated its temperature at 3500 K with a luminosity of $\log[L/L_{\odot}]= -0.94$, 
which corresponds to an M-type star. In addition, by using the models evolutionary tracks, 
we obtained a mass of 0.32 $M_{\odot}$ for this companion
star.

% Fr 2-42

\begin{figure}[h!]
        \includegraphics[width=9cm,height=7.5cm]{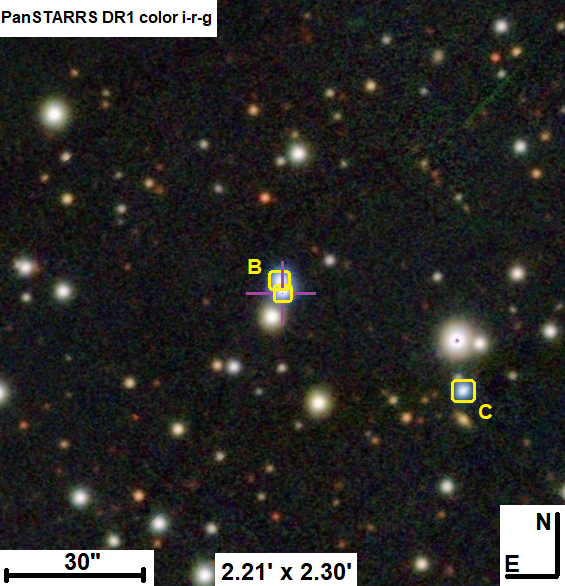}
        \caption{Image of a possible triple system in \object{Fr 2-42}, showing the 
location of the CSPN (with a cross) and the two comoving companions. Image from the \textit{Aladin} sky atlas (CDS).}
        \label{fig:fr_2-42}
\end{figure}

%While we were analysing our systems candidates to wide binaries, 
During the analysis of our wide binary systems candidates,
a group of stars that did not strictly meet our astrometric criteria attracted our 
attention in the nebula Fr 2-42 (Figure \ref{fig:fr_2-42}). %This nebula 
As this PN is located a fairly short distance from the Sun (around 130 pc),  
%so in the case that 
if any companion object orbited the centre of mass of the system, the orbital 
movement might explain, to a certain degree, certain differences in the observed proper motions between the system
components.
%of the system. 
In particular, we  found two possible companions to the proposed CSPN. Their coordinates, 
Gaia colours, and other astrometric properties are listed in Table \ref{tab:binaries}. 
All three stars in the system have colours and brightnesses corresponding to hot white dwarf stars, thus 
%making 
rendering this possible triple system interesting for further study.  
Despite the stability of this type of system in the nebular phase being rather dubious (\citealt{10.1093/mnras/stz2293}), the existence of at least one of them (NGC 246) has been proven.

%Note that due to the consideration of the influence of the orbital velocity, two systems have been added (Fr 2-42 and Sh 2-123), as well as the B companion in the NGC 6853 system. The first one is a very close system located just over 100 pc, whose orbital velocity may have a high influence in the overall proper motion. 

\subsection{Discussion concerning 
%type of 
close binaries}

\begin{table*} [h!]
\caption{Mean values (with uncertainties) of the astrometric and image-fitting quality parameters for blue and red CSPNe samples.} 
\label{table:close_binaries}      
\centering  
%\small

\begin{tabular}{l| l| l| l| l}          
\hline                      
\textbf{Parameter} & \textbf{Blue stars} & \textbf{Red stars} & \textbf{p value} & \textbf{D value} \\    
\hline \hline

    \textit{Astrometric excess noise} & 0.287 \scriptsize{(0.060)} & 0.393 \scriptsize{(0.076)} & 0 & 0.358 \\ 
    \hline 
    \textit{IPD harmonic amplitude} & 0.043 \scriptsize{(0.015)} & 0.043 \scriptsize{(0.013)} & 0 & 0.217 \\ 
    \hline     
    \textit{RA error} & 0.047 \scriptsize{(0.013)} & 0.053 \scriptsize{(0.017)}  & 0.001 & 0.177 \\ 
    \hline 
    \textit{Dec error} & 0.044 \scriptsize{(0.010)} & 0.049 \scriptsize{(0.015)} & 0.001 & 0.189 \\ 
    \hline 
    \textit{RUWE} & 1.013 \scriptsize{(0.051)} & 1.059 \scriptsize{(0.042)} & 0.080 & 0.117 \\ 
    
\end{tabular}   

\tablefoot{P-values and D-values from a Kolmogorov--Smirnov statistical 
test between both samples and over those parameters, are provided.}

\end{table*}

Even though binary stars with separations of less than 0.18--0.60 arcsec are not included 
as separated sources in Gaia EDR3 \citep{2021A&A...649A...4L}, the excellent quality of 
Gaia's source detection algorithm, the crossmatch procedure, and the joint astrometric 
solution for parallaxes and proper motions allow us to address the detection of close 
binarity signs in our set of CSs by analysing the statistics of these measurements.
 %so we have to use another method to detect them. An option could be to analyse some EDR3 parameters which measure the quality of detection. 
We were able to relate cases with statistically noisy measurements  with the presence 
of a close binary companion that might be interfering with the CS detection and 
its astrometry. In particular, we  analysed the following parameters that are 
available in the EDR3 archive: astrometric excess noise (measures the disagreement between 
the observations of a source and the best-fitting standard astrometric model), 
image parameter determination procedure (IPD) harmonic amplitude 
(measures the deviation in the image centroid fitting), RUWE, and coordinate uncertainties in RA and Dec. 

Apart from anomalies in the values of parameters related to image detection
 or astrometry quality, the presence of a close companion star could explain the 
red colour of some of the CSs detected. So it could be hypothesised that red CS quality
 parameters would be statistically different from those measured for blue CSs.
 Consequently, for this study, we decided to recover some objects from reliability
 group C that were discarded mainly because of their deep red colour 
(see Figure \ref{fig:reliability}), and that might be useful for this analysis. 
In particular, we included those group C objects located within 50\% of their nebular radius,
 as this is the maximum permitted distance for objects in groups A and B.
%These 850 extra objects are available at CDS as an extension of Table \ref{table:CSPN}. 
We then applied to this extended sample the same filtering constraints that were
 applied to select the GAPN-EDR3 sample (see Sect. 3), so we finally obtained a set of 
464 stars (405 from GAPN-EDR3 and 59 from group C). These 59 extra objects 
are available at CDS as an extension of Table \ref{table:CSPN}.

We then divided this sample into blue ($G_{BP}- G_{RP}\le{-0.2}$) and red stars
 ($G_{BP}- G_{RP}>{-0.2}$) and calculated the mean values of the above-mentioned 
parameters in each subset. As a result, we found that stars in the red CS sample 
had slightly higher mean values for these parameters, as is shown in Table 
\ref{table:close_binaries}. This means that their image centroid fit gives a 
statistically more smeared solution, that the disagreement between the observations of a 
source and the best-fitting standard astrometric model is higher, and that the 
%renormalised unit weight errors
RUWE values, RA, and Dec coordinates are all 
%more noisy. 
noisier. This might be because red stars, compared to blue ones, tend to 
have parameters more related to the presence of unresolved binary systems.

%To quantify 
In order to  test this hypothesis quantitatively and give a measure in statistical 
terms of the difference between both samples of CSs, we decided to carry out a 
statistical test to measure the significance of the similarity or dissimilarity of 
both samples with reference to the aforementioned parameters. 
%Both samples are rather balanced in number: 269 objects for the blue CSPNe sample and 270 objects for the red CSPNe one. 
%and the new binary CSPNe sample by \citet{2021A&A...648A..95C}. As the red CSPNe should have more probability to be close binaries than the blue ones, we separated our CSPNe in these two subsamples (reds and blues) and we analysed separately with the Chornay sample and also between them trough the statistical method. 
We used the Kolmogorov--Smirnov\footnote{https://en.wikipedia.org/wiki/Kolmogorov-Smirnov\_test} 
statistical test. This is a non-parametric test that quantifies the distance between the 
empirical cumulative distribution functions of two samples. The null distribution of this 
statistic is calculated under the null-hypothesis that the samples are drawn from the same 
distribution. From this analysis, two statistical parameters are obtained that provide
information on the similarity between both samples: the p value and the D value. If the p value 
is below a certain $\alpha$ value (usually 0.01 or 0.05), the two samples are said to be different; otherwise, there could be some similarities. 
%In this case 
We considered $\alpha=0.01$ for a 
%more 
highest 
level of significance, 
%so we will have 
corresponding to a 99\%  confidence in the results. 
On the other hand, if the D value is greater than a certain value ($D_{\alpha}$), we can 
say that the two samples are different; otherwise, 
the samples could be similar. This $D_{\alpha}$,
which 
depends on the sample populations and 
%in 
on $\alpha$ itself,
%. In this case it 
in this case takes a value of 0.153. After running this test, we obtained for each parameter 
the p and D values shown in Table \ref{table:close_binaries}.

In conclusion, our results exclude the possibility that the samples are drawn from the same population and therefore 
point to very low coincidences between the parameters of our blue and red CSPNe. Only 
%in 
for the RUWE parameter did we obtain any evidence of a possible similarity between both samples,
as shown in Table \ref{table:close_binaries}. However, it may be pertinent to recall that 
objects in the analysed samples were previously filtered in terms of the quality of the 
RUWE parameter, which probably influenced the relevance of this parameter in the test. 
For the other parameters we obtain results that reject any similarity. This led us to the 
conclusion that red CSPNe properties are drawn from a very different population from that of the blue CSPNe, which could be interpreted as evidence of a greater incidence of close binarity among red CSPNe. 

We find it useful to discuss the results by \citet{2021A&A...648A..95C}, in which Gaia DR2 was used to identify close binary central star candidates based on the released multi-epoch photometry and the excess photometric uncertainty. A Kolmogorov-Smirnov test between our red sample and a sample from \citet{2021A&A...648A..95C} of close binary candidates (EDR3 data) results in four out of five of the astrometric parameters coming from a similar distribution, while a comparison between our blue sample and that from those authors results in similarity for only two out of five astrometric parameters. These results reinforce our conclusion about the possible binary nature of red CSPNe.

%could have values in these parameters more related with a possible binarity than the blue ones. 

\section{Conclusions}

Using the recently published Gaia EDR3
catalogue, 
we identified, with different degrees of reliability, the CS for a total of 2035 Galactic PNe.
To achieve this, we developed an algorithm that, considering both the colour of the star and its 
distance to the nebular centre, selects the source with the best parameters for the CS to be located 
within a radius of 20 arcsec around each planetary nebula. 
% Moreover, we separate the selected CSs in 3 different reliability groups according 
%its colour and distance to the centre.  
For those PNe with known parallaxes in EDR3 (1725 PNe), we obtained their distances from 
the Bayesian statistical method of \citet{2021AJ....161..147B}. 

A set of 405 CSs, which we call GAPN-EDR3, the distances are accurate enough to
determine their nebular physical sizes,  
%and subsequently, the distance to their corresponding PNe. The knowledge of the distance has allowed us to 
luminosities (by combining with other measures from the literature), and  
evolutionary status (for 74 of them). 
%of a subsample of 405 of them.
%(GAPN-EDR3, composed by 405 PNe), t
%Th, with the most accurate distance values, constitute the so-called GAPN-EDR3 sample. 
The  objects in the GAPN-EDR3 sample are mostly located near the Galactic plane and in the direction of the Galactic 
centre, around half of them being located at distances shorter than 2 kpc. Hence, we estimate 
that this sample is approximately complete to within such a distance. For greater distances, the 
quantity of PNe starts to decrease,  with only a few of them reaching distances greater than 7 kpc. 
If we compare EDR3 distances with those obtained in Paper 1 using DR2 data, we obtain similar values. 
%maybe for large distances DR2 ones are a bit
Only for the greatest distances did we find that DR2 estimations tend to be somewhat underestimated. 
In comparison with other distance estimates, we may conclude that Gaia distances are in 
agreement with  those obtained using other astrometric measurements and from statistical 
methods (with a bias of no more than 500 pc), whereas non-LTE model stellar atmosphere fitting 
provides biased distances for high-temperature CSs. 
% tend to be overestimated for close distances (less than around 1.5 kpc) and underestimated for large distances, in comparison with EDR3 ones.  

Concerning 
%to (Ana: creo que es sin to)
nebular physical properties, we note that our GAPN-EDR3 sample shows a mean nebular radius of around 0.5 pc,
 %but there tend to be 
%more smaller PNe, as 
with more than 60\% of them having radii below 0.3 pc. We also studied the morphological classification of the GAPN-EDR3 sample and found that
%According to the nebular morphology, we can say that 
most PNe show an elliptical shape, followed by bipolar and round 
%PNe. 
ones. We found that bipolar 
%ones 
PNe are more concentrated in the region of the Galactic disc than the 
%other ones, 
rest and that they also tend to be smaller. %Furthermore, we can say that 
The majority of PNe are expanding at velocities between 10 and 50 km/h, with
%While Meanwhile, we obtain 
a mean expansion velocity  of 32 km/h. 
%which is considerably lower than which we the value obtained with DR2, of 38 km/h. With 
%this all
Expansion velocities and physical sizes allowed us to estimate a mean kinematic age of about 
17.5 kyr for the GAPN-EDR3 sample and a visibility time of almost 28 kyr.
%(a bit greater 
%(slightly larger than in DR2).

Using the evolutionary models of \citet{2016A&A...588A..25M} for H-rich CSPNe and from their 
temperatures and luminosities, we were able to estimate the masses and evolutionary ages 
for a significant sub-sample of 74 CSPNe from the GAPN-EDR3 sample. 
%At this point, we had to make sure that brightness values from the literature were in agreement with our CSs brightness. 
We conclude that most of them come from progenitor stars of masses below 2 $M_{\odot}$ and  that 
almost 50\% are estimated to have evolutionary ages below 10 kyr.
 %., despite 
%being having a considerable number of them 
%with  (Ana: igual aquí también iría bien poner el number cuántas son) ages far above 60 kyr. This could happen because young PNe tend to be easier to detect (as they use to be brighter) than old ones. 
By analysing the mean 
%values of the 
radii and kinematic ages in different regions of the HR diagram, we clearly 
observe an increase in these mean parameters from 
%early 
the earliest region to 
%late 
the latest one, as expected. 
%These results are what we %expected, we would expect for, as a PN evolves, it tends to increase its nebular radius and hence its kinematic age. 
In addition, we analysed the spectral types from the literature and obtained a fairly consistent scenario. 
%The late 
Late O-type stars are mainly located in the first region of the HR diagram, while early 
O-type stars tend to appear at a more  intermediate stage. Finally, the most evolved region 
is dominated by the presence of white dwarf stars.

Finally, in 
%the study of 
studying CSPNe binarity 
properties, we were able to detect several wide binary or multiple systems with 
%a 
separation distances below 20\,000 AU. Some of these were also detected in Paper 2
 and have now been confirmed with EDR3. In addition,  the 
Gaia object detection pipeline and astrometric solution for multi-epoch 
observations in the EDR3 catalogue provide some parameters that allowed us to search for evidence of binarity, 
%evidences, 
%evidences, 
in this case related to possibly unresolved pairs. 
 %by analysing some EDR3 parameters related with %the a close binarity scenario, 
%We found that red CSPNe properties are drawn from a very different population than that of the blue CSPNe, which can be interpreted as an evidence of a mayor incidence of close binarity among red CS. %
After analysing these parameters, we may conclude that red CSPNe tend to show more 
affinity with a close binary nature for their CSs than the blue ones. Furthermore, a 
Kolmogorov--Smirnov statistical test over these parameters, between red and blue stars,
 allowed us to prove that both samples are statistically different. This result therefore
reinforces our hypothesis.

%the recently published close binary CSPNe sample by \citet{2021A&A...648A..95C}, we have observed several similarities in some astrometric parameters that lead to confirm a possible binarity, but without any clear or concrete discovery.

\begin{acknowledgements}
        This work has made use of data from the European Space Agency (ESA) Gaia mission and 
processed by the Gaia Data Processing and Analysis Consortium (DPAC). Funding for the DPAC has 
been provided by national institutions, in particular the institutions participating in the 
Gaia Multilateral Agreement. This research has made use of the Simbad database and the
Aladin sky atlas, operated at CDS, Strasbourg, France. 
The authors have also made use of the
VOSA tool, developed under the Spanish Virtual Observatory project supported by the Spanish 
MINECO through grant AYA2017-84089, and partially updated thanks to the EU Horizon 2020 
Research and Innovation Programme, under grant 776403 (EXOPLANETS-A). 
%The INT is operated on the island of La Palma by the Isaac Newton Group of Telescopes in the Spanish Observatorio del Roque de los Muchachos of the Instituto de Astrofísica de Canarias. 
Funding from Spanish Ministry project RTI2018-095076-B-C22, Xunta de Galicia ED431B 2021/36, PDC2021-121059-C22, and AYA-2017-88254-P is acknowledged by the authors. 
We also acknowledge support from CIGUS-CITIC, funded by Xunta de Galicia and the European Union (FEDER Galicia 2014-2020 Program) through grant ED431G 2019/01. 
IGS acknowledges financial support from the Spanish 
National Programme for the Promotion of Talent and its Employability grant BES-2017-083126 
cofunded by the European Social Fund. 
\end{acknowledgements}

\bibliographystyle{aa} 

\bibliography{bibliopn}

% APENDICE

\begin{appendix}

\onecolumn

\section{Tables}

% TABLA Identificacion CSPNe

\begin{table}[h!]   
\caption{Gaia EDR3 sources identified as 2035 central stars of planetary nebulae.}  
\label{table:CSPN}      
%\centering  
%\small
%\footnotesize
\scriptsize

\begin{tabular}  {l l l c c c c c c c c}    
\hline\hline                   
PNG name & Other name & Gaia EDR3 ID & Quality & RA & Dec & $D_{ang}$ & G & $G_{BP}-G_{RP}$ & $A_{V}$ & $(G_{BP}-G_{RP})_{\circ}$\\   
 &  &  & label & (º) & (º) & (as) & (mag) & (mag) & (mag) & (mag)\\ 
\hline    

PN G000.0-01.0 & JaSt69 & 4057224264723323264 & B & 267.5416 & -29.3183 & 0.40 & 19.65 & 3.17 & 3.22 & 1.86                                       \\
PN G000.0-02.5 & K6-36 & 4056266379788466304 & B & 268.9702 & -30.2613 & 0.83 & 19.45 & 1.91 & 2.62 & 0.72                    \\
PN G000.0-02.9 & MPAJ1757-3021 & 4056203849243193728 & B & 269.3093 & -30.3645 & 2.52 & 18.42 & 1.85 & 3.57 & 0.11            \\
PN G000.0-06.8 & H1-62 & 4045771305065496832 & B & 273.3248 & -32.3286 & 0.19 & 14.36 & 0.77 & 0.95 & 0.28                    \\
PN G000.1+02.6 & Al2-J & 4061303281130808448 & B & 263.8977 & -27.4015 & 0.58 & 17.94 & 2.41 & 3.31 & 0.99                    \\
PN G000.1+04.3 & H1-16 & 4109679250049965312 & A & 262.3473 & -26.4346 & 0.22 & 18.75 & 1.29 & 3.63 & -0.57                   \\
PN G000.1+17.2 & PC12 & 4130784921205604736 & B & 250.9741 & -18.9533 & 0.08 & 15.17 & 0.62 & 1.13 & 0.03                     \\
PN G000.1-01.9 & JaSt93 & 4056400558791818624 & B & 268.3492 & -29.8300 & 0.80 & 17.64 & 2.33 & 2.32 & 1.37                   \\
PN G000.1-05.6 & H2-40 & 4049045783774253696 & B & 272.1281 & -31.6098 & 0.58 & 18.46 & 1.44 & 1.38 & 0.81                    \\
PN G000.1-08.0 & SB1 & 4045583322840954496 & A & 274.7019 & -32.7986 & 0.48 & 20.01 & -0.17 & 0.45 & -0.42                    \\
PN G000.2-01.9 & M2-19 & 4056495151130724224 & B & 268.4402 & -29.7297 & 0.72 & 16.88 & 1.34 & 2.53 & 0.07                    \\
PN G000.2-01.9a & JaSt2-14 & 4056495971440817408 & B & 268.4195 & -29.7107 & 2.07 & 16.67 & 1.47 & 2.58 & 0.20                \\
PN G000.3+03.2 & PHRJ1733-2655 & 4061390245613519232 & A & 263.4691 & -26.9242 & 0.46 & 20.36 & 1.23 & 3.15 & -0.39           \\
PN G000.3+04.5 & PHRJ1729-2614 & 4109701927482024320 & B & 262.2806 & -26.2455 & 1.31 & 20.69 & 2.13 & 3.44 & 0.56            \\
PN G000.3+07.3 & PHRJ1718-2441 & 4110995812169990400 & B & 259.6779 & -24.6898 & 2.18 & 20.47 & 1.02 & 1.79 & 0.10            \\
PN G000.3+12.2 & IC4634 & 4126115570219432448 & A & 255.3899 & -21.8259 & 0.28 & 13.85 & -0.15 & 0.71 & -0.55                 \\
PN G000.3-04.2 & MPAJ1803-3043 & 4049954873646647040 & B & 270.8405 & -30.7265 & 0.33 & 15.38 & 2.65 & 1.83 & 1.89            \\
PN G000.3-04.6 & M2-28 & 4049886772877581696 & B & 271.2611 & -30.9713 & 0.77 & 19.63 & 1.06 & 1.88 & 0.11                    \\
PN G000.4+04.4 & K5-1 & 4109691718340733568 & B & 262.4682 & -26.1871 & 0.81 & 19.18 & 1.76 & 3.75 & -0.10                    \\
PN G000.4-02.9 & M3-19 & 4056250956464755712 & A & 269.5808 & -30.0109 & 0.68 & 18.34 & 0.82 & 2.70 & -0.60                   \\
PN G000.5-03.1a & MPAJ1759-3007 & 4056239308511338240 & B & 269.8549 & -30.1217 & 0.23 & 19.08 & 1.81 & 2.35 & 0.74           \\
PN G000.6-01.0 & JaSt77 & 4057342599736948992 & B & 267.7980 & -28.9406 & 0.08 & 16.10 & 1.81 & 1.11 & 1.33                   \\
PN G000.6-01.3 & Bl3-15 & 4056579538679085952 & B & 268.1502 & -29.1110 & 0.16 & 19.24 & 2.03 & 4.21 & -0.03                  \\
PN G000.7+08.0 & MPAJ1717-2356 & 4114088875802922880 & A & 259.2877 & -23.9416 & 0.37 & 18.93 & 0.74 & 1.98 & -0.31           \\
PN G000.7-03.7 & M3-22 & 4050168629923554944 & A & 270.5802 & -30.2405 & 0.51 & 18.35 & 0.51 & 1.61 & -0.35                   \\
PN G000.7-06.1 & SB3 & 4048968994096810880 & A & 273.0606 & -31.3335 & 0.06 & 19.03 & 0.32 & 1.06 & -0.25                     \\
PN G000.8+05.2 & PBOZ5 & 4109999173694500992 & B & 261.9106 & -25.3741 & 0.65 & 19.68 & 1.06 & 1.17 & 0.49                    \\
PN G000.8-07.6 & H2-46 & 4045845625148544640 & B & 274.6561 & -31.9126 & 0.17 & 18.87 & 0.37 & 0.61 & 0.05                    \\
PN G000.9+01.3 & MGE000.9363+01.3962 & 4060883679948752384 & B & 265.6120 & -27.4072 & 0.23 & 19.17 & 3.67 & 5.08 & 1.67      \\
PN G000.9-02.0 & Bl3-13 & 4056540677880158208 & B & 269.0116 & -29.1880 & 0.46 & 18.05 & 1.15 & 2.51 & -0.14                  \\
PN G000.9-03.3 & PHRJ1801-2947 & 4050246420409992704 & B & 270.3061 & -29.7837 & 1.56 & 19.02 & 1.34 & 2.00 & 0.36            \\
PN G000.9-04.2 & PHRJ1804-3016 & 4049992845542438912 & B & 271.2006 & -30.2804 & 0.84 & 19.39 & 1.32 & 1.76 & 0.47            \\
PN G000.9-04.8 & M3-23 & 4049925328633027712 & A & 271.7756 & -30.5714 & 0.10 & 19.22 & 0.65 & 1.77 & -0.29                   \\
PN G001.0+01.4 & JaSt2-4 & 4060890513224820352 & B & 265.6169 & -27.2255 & 0.50 & 15.99 & 1.78 & 1.66 & 1.05                  \\
PN G001.0+01.9 & K1-4 & 4060952605654047232 & A & 265.1142 & -27.0173 & 0.21 & 20.20 & 0.94 & 4.63 & -1.22                    \\
PN G001.0-01.9 & K6-35 & 4056546037855269120 & B & 268.9297 & -29.0677 & 0.99 & 16.97 & 1.67 & 3.06 & 0.18                    \\
PN G001.0-02.0 & MGE001.0098-02.0666 & 4056542116661761408 & B & 269.0264 & -29.1259 & 0.67 & 19.41 & 2.15 & 2.64 & 1.00      \\
PN G001.1+02.2 & MPAJ1739-2648 & 4060976176417193856 & B & 264.9571 & -26.8126 & 0.49 & 19.59 & 2.52 & 4.62 & 0.40            \\
PN G001.1-02.6 & MPA1758-2915 & 4062332454138572416 & B & 269.7151 & -29.2663 & 0.66 & 18.64 & 1.91 & 1.92 & 1.07             \\
PN G001.1-06.4 & SB4 & 4048986173916372608 & B & 273.5589 & -31.1861 & 0.85 & 18.76 & 1.43 & 0.65 & 1.14                      \\
PN G001.2+01.3 & JaSt45 & 4060981021062265728 & B & 265.8470 & -27.1868 & 1.39 & 19.46 & 3.65 & 4.74 & 1.76                   \\
PN G001.2+02.1 & Hen2-262 & 4060978306741619968 & B & 265.0535 & -26.7394 & 0.45 & 18.69 & 1.72 & 3.68 & -0.12                \\
PN G001.2+02.8 & PPAJ1737-2621 & 4061796583872776576 & B & 264.3763 & -26.3621 & 0.46 & 16.94 & 1.87 & 3.47 & 0.19            \\
PN G001.2-01.2a & JaSt95 & 4057385514989465216 & B & 268.3970 & -28.4808 & 0.11 & 18.16 & 2.78 & 1.06 & 2.31                  \\
PN G001.2-01.4 & JaSt2-15 & 4063373176122812416 & B & 268.5985 & -28.5804 & 1.80 & 18.26 & 2.54 & 4.78 & 0.34                 \\
PN G001.2-02.6 & PHRJ1759-2915 & 4062333794139284096 & B & 269.7617 & -29.2506 & 1.30 & 16.71 & 1.90 & 1.94 & 1.06            \\
PN G001.2-05.6 & PHRJ1811-3042 & 4049240298544263936 & B & 272.7613 & -30.7033 & 0.10 & 18.61 & 0.35 & 1.00 & -0.19           \\
PN G001.3-01.0 & JaSt89 & 4063395960932807680 & B & 268.2778 & -28.3031 & 0.91 & 17.53 & 1.65 & 4.46 & -0.59                  \\
PN G001.3-05.6 & SB5 & 4049264242907686016 & A & 272.8142 & -30.6306 & 0.10 & 15.50 & 0.24 & 1.02 & -0.32                     \\
PN G001.4+00.5 & [GKF2010]MN66 & 4060739369060593152 & B & 266.7717 & -27.4261 & 0.47 & 17.86 & 2.43 & 2.46 & 1.41            \\
PN G001.4-00.7 & JaSt82 & 4057456261747292672 & B & 268.0220 & -28.0965 & 1.24 & 20.63 & 0.41 & 25.34 & -2.21                 \\
PN G001.5+00.9 & [GKF2010]MN65 & 4060808874483585536 & B & 266.4199 & -27.1542 & 0.97 & 17.30 & 2.21 & 2.31 & 1.23            \\
PN G001.5+01.5 & JaSt46 & 4061009264777169920 & B & 265.8765 & -26.7924 & 0.57 & 19.97 & 2.17 & 5.47 & -0.51                  \\
PN G001.5+03.1 & PHRJ1737-2559 & 4062011744524241536 & B & 264.3188 & -25.9941 & 0.74 & 18.58 & 2.67 & 4.52 & 0.68            \\
PN G001.5+05.3 & KnJ1729.1-2443 & 4110412899193532032 & B & 262.2857 & -24.7194 & 1.27 & 20.23 & 1.40 & 2.85 & -0.04          \\
PN G001.5-01.0 & JaSt2-12 & 4063398469194112128 & A & 268.3648 & -28.1969 & 0.69 & 19.20 & 1.65 & 4.62 & -0.66                \\
PN G001.5-01.6 & JaSt2-17 & 4062630563426969856 & B & 268.9029 & -28.4244 & 0.61 & 19.91 & 2.20 & 3.29 & 0.74                 \\
PN G001.5-06.7 & SwSt1 & 4049331244394134912 & B & 274.0511 & -30.8689 & 0.24 & 11.84 & 0.85 & 0.39 & 0.66                    \\
PN G001.6+01.5 & K6-10 & 4061013383646090624 & B & 265.8206 & -26.7382 & 0.14 & 19.32 & 2.21 & 4.81 & -0.13                   \\
PN G001.7+01.3 & JaSt52 & 4061018988593063040 & B & 266.1551 & -26.7904 & 0.32 & 18.13 & 2.57 & 4.97 & 0.26                   \\
PN G001.7+05.7 & H1-14 & 4110479900747426944 & A & 262.0073 & -24.4232 & 0.31 & 19.64 & 0.82 & 2.76 & -0.62                   \\
PN G001.7-04.4 & H1-55 & 4050131349653595392 & B & 271.8107 & -29.6902 & 0.24 & 16.60 & 1.13 & 1.23 & 0.53                    \\
PN G001.7-04.6 & H1-56 & 4050126711087206016 & A & 271.9745 & -29.7429 & 0.12 & 15.98 & 0.38 & 1.27 & -0.31                   \\
PN G001.8-00.4 & MB4515 & 4063669086488636928 & B & 267.9773 & -27.6029 & 0.04 & 19.97 & 2.06 & 2.58 & 0.93                   \\
PN G001.8-03.7 & PHRJ1804-2913 & 4050381621840757760 & B & 271.1188 & -29.2326 & 0.17 & 16.18 & 1.36 & 1.56 & 0.62            \\
PN G001.9+02.3 & K5-10 & 4061866089277230848 & A & 265.3522 & -26.0648 & 0.60 & 20.41 & 1.40 & 4.19 & -0.71                   \\
PN G001.9+08.2 & PM1-139 & 4114675327843171456 & B & 259.8593 & -22.8032 & 1.20 & 19.69 & 0.67 & 1.75 & -0.25                 \\

$\cdots$ & $\cdots$ &$\cdots$ &$\cdots$ &$\cdots$ &$\cdots$ &$\cdots$ &$\cdots$ &$\cdots$ &$\cdots$ &$\cdots$  \\
\hline \\

\end{tabular}

\tablefoot{Full table is available at CDS, with an extended version containing 850 objects from group C located within 50\% of the nebular radius.}

\tablebib{Extinction values $A_{V}$ are taken from \citet{1992A&AS...94..399C}, \citet{frew08}, \citet{frew16}, \citet{2019ApJ...887...93G}, \citet{1998ApJ...500..525S}, 
and \citet{1992A&AS...95..337T}. }

\end{table}

\newpage

% TABLA Muestra GAPN

\begin{table}  
\caption{Astrometric parameters of the 405 PNe from the GAPN-EDR3 sample.}  
\label{table:GAPN}    

%\centering  
%\scriptsize
%\small
\footnotesize

\begin{tabular}{l c c c c c c c c c c l}  

\hline\hline                    
PNG Name & Parallax & ER paral. & Distance & Low dist. & High dist. & $<z>$ & $<R>_{ang}$ & $<R>_{phy}$ & $Vel_{rad}$ & Morph. & Spec. type\\   
 & (mas) & (mas) & (pc) & (pc) & (pc) & (pc) & (as) & (pc) & (km/h) &  & \\ 
\hline   

PN G000.3+12.2 & 0.414 & 0.053 & 2456 & 2224 & 2735 & 520 & 6.8 & 0.081 & -34 & B & O(H)3 If                                              \\
PN G000.3-04.2 & 0.207 & 0.047 & 5676 & 4518 & 7107 & -418 & 18 & 0.495 & $\cdots$ & R &  $\cdots$                        \\
PN G001.0+01.4 & 0.749 & 0.067 & 1342 & 1228 & 1468 & 35 & 16.1 & 0.105 & $\cdots$ & E &   $\cdots$                  \\
PN G002.0-06.2 & 0.283 & 0.051 & 3852 & 3251 & 4559 & -418 & 2.6 & 0.049 & -119 & E & O5f(H)                        \\
PN G002.4+05.8 & 0.963 & 0.06 & 1053 & 1005 & 1116 & 107 & 14.8 & 0.075 & -106 & E & [WO 3]                        \\
PN G002.4-03.4 & 0.256 & 0.033 & 4084 & 3624 & 4701 & -247 & 7.8 & 0.153 & $\cdots$ & E &  $\cdots$                                 \\
PN G002.4-03.7 & 0.163 & 0.031 & 6067 & 5181 & 6819 & -396 & 1.8 & 0.051 & -92 & S & O(H)7-8                        \\
PN G002.7-52.4 & 0.862 & 0.056 & 1167 & 1104 & 1236 & -925 & 65.1 & 0.368 & -26 & R & hgO(H)                        \\
PN G002.9+06.5 & 0.238 & 0.048 & 4803 & 3896 & 5714 & 549 & 2.5 & 0.058 & 70 & R & $\cdots$                                 \\
PN G003.3+66.1 & 1.482 & 0.242 & 729 & 603 & 905 & 667 & 23.8 & 0.084 & -17 & R & DAO                               \\
PN G003.3-04.6 & 0.157 & 0.025 & 6586 & 5595 & 7376 & -535 & 5.3 & 0.168 & 157 & E & B0I-III           \\
PN G003.4+01.4 & 0.508 & 0.078 & 2131 & 1870 & 2477 & 52 & 7.5 & 0.077 & $\cdots$ & E &  $\cdots$                                   \\
PN G006.4-03.4 & 0.204 & 0.049 & 5473 & 4484 & 6638 & -331 & 13.5 & 0.358 & $\cdots$ & E & $\cdots$                                 \\
PN G006.9-30.2 & 2.916 & 0.048 & 342 & 337 & 346 & -172 & 7650 & 12.691 & $\cdots$ & $\cdots$  &  $\cdots$                                \\
PN G007.0+06.3 & 0.219 & 0.05 & 4761 & 4147 & 5761 & 525 & 3.2 & 0.074 & -26 & E & $\cdots$                                \\
PN G007.9+10.1 & 0.269 & 0.047 & 3954 & 3432 & 4742 & 697 & 3.6 & 0.069 & -45 & S &  $\cdots$                               \\
PN G009.1-03.4 & 0.28 & 0.062 & 4099 & 3446 & 5172 & -245 & 44.5 & 0.884 & $\cdots$ & I & $\cdots$                                 \\
PN G009.4-05.5 & 0.523 & 0.024 & 1943 & 1873 & 2043 & -171 & 8 & 0.076 & 13 & E & [WC5/6]                         \\
PN G009.6+14.8 & 0.408 & 0.078 & 2580 & 2195 & 3030 & 660 & 8.8 & 0.11 & -40 & B & O(He)                           \\
PN G010.8-01.8 & 0.412 & 0.049 & 2484 & 2255 & 2728 & -79 & 6 & 0.072 & 5 & E & Of                                \\
PN G011.3+02.8 & 0.293 & 0.053 & 3668 & 3103 & 4321 & 182 & 2.5 & 0.044 & -62 & $\cdots$  &  $\cdots$                               \\
PN G011.7-00.6 & 0.392 & 0.043 & 2568 & 2362 & 2798 & -29 & 3.6 & 0.045 & 120 & E & [WC5/6]                         \\
PN G011.7-41.3 & 2.34 & 0.048 & 428 & 420 & 437 & -283 & 600 & 1.245 & $\cdots$ & $\cdots$  & sdO                                \\
PN G011.8+03.7 & 0.606 & 0.036 & 1662 & 1572 & 1750 & 110 & 19 & 0.153 & $\cdots$ & E & $\cdots$                                  \\
PN G012.1-11.2 & 0.183 & 0.036 & 5605 & 4588 & 7160 & -1091 & 9.6 & 0.262 & $\cdots$ & E & O(H)6III-V                       \\
PN G013.8-02.8 & 0.572 & 0.077 & 1913 & 1676 & 2214 & -94 & 47.5 & 0.441 & $\cdots$ & B & $\cdots$                                  \\
PN G014.0-02.5 & 0.277 & 0.07 & 3924 & 3181 & 4828 & -173 & 3 & 0.057 & $\cdots$ & E & $\cdots$                                  \\
PN G014.4-06.1 & 0.185 & 0.054 & 5610 & 4558 & 7251 & -601 & 5.4 & 0.146 & 72 & S & wels                            \\
PN G014.8-08.4 & 0.395 & 0.055 & 2654 & 2378 & 3050 & -392 & 9 & 0.116 & -24 & E & O(H)3-4                        \\
PN G014.8-25.6 & 1.512 & 0.245 & 731 & 612 & 877 & -316 & 32.3 & 0.114 & $\cdots$ & E & $\cdots$                                    \\
PN G015.4+05.6 & 0.11 & 0.031 & 7808 & 6529 & 9766 & 769 & 4.5 & 0.17 & $\cdots$ & R & $\cdots$                                   \\
PN G015.4-04.5 & 0.225 & 0.053 & 4957 & 4173 & 6409 & -397 & 3 & 0.072 & 46 & R & O(H)3 If                        \\
PN G016.4-01.9 & 0.449 & 0.017 & 2226 & 2154 & 2309 & -77 & 5.9 & 0.063 & 29 & R & O(H)7I(fc)                       \\
PN G016.9-09.7 & 0.368 & 0.058 & 2896 & 2465 & 3428 & -488 & 29 & 0.407 & $\cdots$ & $\cdots$  & $\cdots$                                 \\
PN G017.3-21.9 & 0.699 & 0.039 & 1424 & 1360 & 1494 & -533 & 59.5 & 0.411 & 13 & E & O(H) + ?                       \\
PN G017.6-10.2 & 0.584 & 0.032 & 1761 & 1662 & 1865 & -313 & 29.6 & 0.252 & 3 & R & O(H)3-5Vz                       \\
PN G019.8-23.7 & 0.863 & 0.19 & 1184 & 985 & 1420 & -477 & 139.5 & 0.8 & $\cdots$ & R & $\cdots$                                 \\
PN G020.3-06.9 & 0.228 & 0.022 & 4471 & 4118 & 4973 & -543 & 11 & 0.238 & $\cdots$ & $\cdots$  & $\cdots$                                 \\
PN G020.7-05.9 & 0.294 & 0.028 & 3538 & 3268 & 4071 & -368 & 3.5 & 0.06 & 46 & R & O(H)4-8III                      \\
PN G020.7-08.0 & 0.754 & 0.066 & 1364 & 1253 & 1487 & -191 & 105 & 0.694 & $\cdots$ & B & O(H)3-5Vz                       \\
PN G021.4+02.5 & 0.161 & 0.035 & 6018 & 4973 & 6929 & 270 & 30 & 0.875 & $\cdots$ & R & $\cdots$                                  \\
PN G022.9+22.7 & 1.125 & 0.131 & 906 & 800 & 1050 & 351 & 180 & 0.791 & $\cdots$ & A & $\cdots$                                   \\
PN G025.3+40.8 & 0.434 & 0.048 & 2294 & 2074 & 2595 & 1500 & 7.5 & 0.083 & 22 & E & O(H)5f                          \\
PN G025.4-04.7 & 0.649 & 0.068 & 1639 & 1449 & 1853 & -135 & 49.8 & 0.395 & -13 & E & hgO(H)                        \\
PN G026.0+46.6 & 0.338 & 0.04 & 2949 & 2617 & 3537 & 2144 & 60 & 0.858 & $\cdots$ & $\cdots$  &  $\cdots$                                \\
PN G026.1-17.6 & 1.018 & 0.015 & 983 & 970 & 999 & -298 & 247.5 & 1.179 & $\cdots$ & A &  $\cdots$                                  \\
PN G026.5-03.0 & 0.264 & 0.078 & 4464 & 3528 & 5498 & -220 & 2.2 & 0.048 & $\cdots$ &  $\cdots$ &  $\cdots$                                 \\
PN G027.4-03.5 & 0.209 & 0.058 & 4873 & 4077 & 5859 & -298 & 7.5 & 0.177 & 47 & R & $\cdots$                                \\
PN G027.6+16.9 & 0.593 & 0.044 & 1692 & 1568 & 1804 & 492 & 55 & 0.451 & $\cdots$ & R & sdO                               \\
PN G029.2-05.9 & 0.33 & 0.047 & 3077 & 2716 & 3509 & -319 & 11.8 & 0.176 & -43 & E & [WO 4]                        \\
PN G029.3-01.2 & 0.781 & 0.061 & 1303 & 1203 & 1399 & -28 & 29.3 & 0.185 & $\cdots$ & B & $\cdots$                                  \\
PN G030.5-18.4 & 7.763 & 0.064 & 129 & 128 & 130 & -41 & 1350 & 0.844 & $\cdots$ & $\cdots$  & DA                                 \\
PN G030.8+03.4a & 1.921 & 0.268 & 546 & 493 & 628 & 33 & $\cdots$ & $\cdots$ & $\cdots$ & S & $\cdots$                                              \\
PN G031.0-10.8 & 0.254 & 0.076 & 4216 & 3425 & 5268 & -797 & 3.5 & 0.071 & 48 & E & $\cdots$                                \\
PN G032.1-06.2 & 8.06 & 0.038 & 124 & 124 & 125 & -14 & 600 & 0.361 & $\cdots$ & $\cdots$  & $\cdots$                                    \\
PN G033.1-06.3 & 1.13 & 0.193 & 1001 & 835 & 1249 & -111 & 37.9 & 0.184 & 0 & B & $\cdots$                                 \\
PN G034.1-10.5 & 0.553 & 0.072 & 1821 & 1618 & 2045 & -332 & 39.5 & 0.349 & $\cdots$ & R & hgO(H)                           \\
PN G034.6+11.8 & 0.572 & 0.071 & 1788 & 1609 & 1991 & 367 & 7 & 0.061 & -9 & B & Of-WR(H)                         \\
PN G034.7-06.2 & 0.209 & 0.027 & 4765 & 4335 & 5238 & -520 & 7.8 & 0.179 & $\cdots$ & R & $\cdots$                                  \\
PN G035.7+19.2 & 0.313 & 0.057 & 3299 & 2807 & 4026 & 1085 & 26.3 & 0.42 & $\cdots$ & E & $\cdots$                                 \\
%PN G036.0+17.6 & 0.478 & 0.034 & 2091 & 1984 & 2206 & 633 & 40 & 0.405 & -42 & R & O7fk                           \\
%PN G036.1-57.1 & 5.036 & 0.046 & 199 & 197 & 200 & -167 & 426.3 & 0.41 & -15 & B & DAO.5                           \\
%PN G037.7-34.5 & 0.852 & 0.069 & 1182 & 1116 & 1262 & -671 & 12.5 & 0.072 & -47 & B & O(H)                          \\
%PN G038.2+12.0 & 0.173 & 0.042 & 5594 & 4592 & 7246 & 1171 & 2.6 & 0.07 & -11 & E & O(H)7 Ib(f)                    \\
%PN G039.0-04.0 & 0.285 & 0.02 & 3529 & 3325 & 3798 & -252 & 16 & 0.274 &  & E & O(H)III-V(e)                     \\
%PN G039.5-02.7 & 0.233 & 0.062 & 4960 & 3997 & 6425 & -242 & 3 & 0.071 & 23 & E & O3-6                            \\
%PN G040.8-09.7 & 0.563 & 0.074 & 1833 & 1622 & 2099 & -311 & 80 & 0.711 &  &   &                                  \\
%PN G041.8-02.9 & 2.032 & 0.071 & 494 & 475 & 514 & -26 & 62.5 & 0.15 & 4 & B & DAO                                 \\

$\cdots$ & $\cdots$ & $\cdots$ &$\cdots$ &$\cdots$ &$\cdots$ &$\cdots$ &$\cdots$ &$\cdots$ &$\cdots$ &$\cdots$ &$\cdots$  \\
\hline \\

\end{tabular}

\tablefoot{Full table is available at CDS.}

\tablebib{Angular radii $<R>_{ang}$ and morphological types taken from HASH database.
Radial velocities $Vel_{rad}$ taken from Simbad database.
Spectral types taken from \citet{2020A&A...640A..10W}.}

\end{table}

\clearpage

% TABLA Velocidades Expansion

\twocolumn

\begin{table}  
\caption{Expansion Velocities and kinematic Ages \newline for a subsample of 65 PNe within GAPN-EDR3 sample.}   
\label{table:vel_exp}    
%\tablehead{hola & hola \\}
%\centering  
\large
%\normalsize
%\scriptsize
%\small
%\footnotesize

\begin{tabular}{l c c c}  

\hline\hline                   
PN Name & Radius & Vel$_{exp}$  & Age$_{kin}$ \\   
 & (pc) & (km/h)  & (kyr)  \\ 
 
\hline   

Abell 24 & 0.687 & 14 & 32              \\              %1
Abell 28 & 0.302 & 4 & 49.3         \\
Abell 29 & 1.17 & 25 & 30.5               \\   %1
Abell 3 & 0.344 & 30 & 7.5          \\
Abell 31 & 1.202 & 29 & 27                \\   %1
Abell 33 & 0.653 & 32 & 13.3              \\   %1
Abell 34 & 0.826 & 35 & 15.4              \\   %1
Abell 39 & 0.455 & 29 & 10.2              \\   %1
Abell 51 & 0.252 & 42 & 3.9               \\   %1
Abell 61 & 0.789 & 32 & 16.1              \\   %1
Abell 7 & 0.972 & 29 & 21.9               \\   %1
Abell 71 & 0.426 & 20 & 13.9        \\
Abell 74 & 1.307 & 27 & 31.6              \\   %1
Cn 3-1 & 0.07 & 10 & 4.6            \\
H 2-1 & 0.026 & 36 & 0.5                  \\   %1
HaWe 4 & 1.219 & 11 & 72.3                \\   %1
Hen 2-107 & 0.077 & 33 & 1.5              \\   %1
Hen 2-108 & 0.114 & 12 & 6.2              \\   %1
Hen 2-131 & 0.061 & 12 & 3.3        \\
Hen 2-138 & 0.068 & 11 & 4          \\
Hen 2-51 & 0.073 & 10 & 4.7               \\   %1
IC 1295 & 0.395 & 27 & 9.6                \\   %1
IC 2448 & 0.168 & 13.5 & 8.1              \\   %1
IC 289 & 0.167 & 25.5 & 4.3               \\   %1
IC 3568 & 0.092 & 8 & 7.5           \\
IC 418 & 0.04 & 16 & 1.6                  \\   %1
IC 4593 & 0.083 & 12 & 4.5                \\   %1
IC 4642 & 0.224 & 34.5 & 4.2        \\
IC 5148/50 & 0.368 & 53 & 4.5             \\   %1
IsWe 2 & 1.711 & 8 & 139.5                \\   %1
Lo 1 & 0.813 & 30 & 17.7                  \\   %1
M 1-26 & 0.041 & 12 & 2.2                 \\   %1
M 1-46 & 0.063 & 7 & 5.9                  \\   %1
M 1-53 & 0.072 & 13 & 3.6           \\
M 1-65 & 0.054 & 4 & 8.9            \\
\hline

\end{tabular}
\end{table}

\begin{table}  
\large
%\caption{Expansion Velocities and kinematic Ages}
\begin{tabular}{l c c c c}  

\hline\hline                   
PN Name & Radius & Vel$_{exp}$  & Age$_{kin}$\\   
 & (pc) & (km/h) & (kyr)  \\ 
\hline  

M 1-77 & 0.05 & 6.5 & 5                   \\       % 1 -- $^_{(1)}$
M 2-33 & 0.049 & 12 & 2.6                 \\   % 1
M 3-34 & 0.071 & 14 & 3.3           \\         
NGC 2022 & 0.139 & 26 & 3.5               \\   % 1
NGC 2792 & 0.108 & 20 & 3.5               \\   % 1
NGC 3587 & 0.405 & 34 & 7.8               \\   % 1
NGC 3699 & 0.147 & 27.5 & 3.5             \\   % 1
NGC 4361 & 0.285 & 32 & 5.8               \\   % 1
NGC 5882 & 0.069 & 11 & 4.1               \\   % 1
NGC 6058 & 0.211 & 27 & 5.1               \\   % 1
NGC 6072 & 0.178 & 6 & 19.3         \\
NGC 6153 & 0.083 & 17 & 3.2               \\   % 1
NGC 6210 & 0.065 & 21 & 2           \\
NGC 6572 & 0.061 & 18 & 2.2               \\   % 1
NGC 6772 & 0.184 & 11 & 10.9              \\   % 1
NGC 6781 & 0.15 & 12 & 8.1                \\   % 1
NGC 6842 & 0.281 & 35 & 5.2         \\
NGC 6891 & 0.077 & 7 & 7.2                \\   % 1
NGC 6894 & 0.203 & 43 & 3.1               \\   % 1
NGC 7009 & 0.072 & 25 & 1.9               \\   % 1
NGC 7094 & 0.393 & 45 & 5.7               \\   % 1
NGC 7293 & 0.41 & 14 & 19.1               \\   % 1
NGC 7354 & 0.164 & 25 & 4.3               \\   % 1
NGC 7662 & 0.117 & 27 & 2.8               \\   % 1
PuWe 1 & 1.156 & 23 & 32.8                \\   % 1
Sh 2-216 & 1.835 & 4 & 299.3              \\   % 1
Vy 2-3 & 0.055 & 12.5 & 2.9               \\   % 1
We 1-10 & 1.1 & 26 & 27.6           \\
WeDe 1 & 1.338 & 17 & 51.3                \\   % 1
Wray 17-31 & 0.5 & 28 & 11.6              \\   % 1
\hline

\end{tabular}

\tablefoot{kinematic ages are calculated by multiplying \\
the expansion velocities by a correction factor of 1.5, \\
as proposed in \citet{Jacob13}. \\
See Sect. 3.3 for more details.}

\tablebib{
Expansion velocities Vel$_{exp}$ taken from \\
\citet{frew08} and \citet{1989A&AS...78..301W}.
%(1): Vel$_{exp}$ value taken from \citet{frew08}. \\
%(2): Vel$_{exp}$ value taken from \citet{1989A&AS...78..301W}.
}

\end{table}

\clearpage

% TABLA Diagrama HR

\onecolumn 
\small

\LTcapwidth=\textwidth

\begin{longtable}{ l c c c c c c c c c l}
\caption{Evolutionary parameters of 74 CSPNe within GAPN-EDR3 sample located in the HR diagram.}
\label{tab:evo}\\

\hline\hline
Name & Radius & $G$  & $V$ & $A_{V}$ & $(G_{BP}-G_{RP})_{\circ}$ & $\log{(\frac{L}{L_{\odot}})}$ & $\log{(T_{eff})}$ & Mass  &  Age$_{evo}$   & Spec. type \\
 &  (pc) & (mag) & (mag) & (mag) & (mag) &  &  & ($M_{\odot}$) &  (kyr) &  \\
\hline
\endfirsthead

\multicolumn{11}{l}
{\tablename\ \thetable\ -- \textit{Continued from previous page}} \\
\hline\hline
Name & Radius (pc) & $G$  & $V$ & $A_{V}$ & $(G_{BP}-G_{RP})_{\circ}$ & $\log{(\frac{L}{L_{\odot}})}$ & $\log{(T)}$ & $M$ ($M_{\odot}$) &  $T_{evo}$ (kyr) & Spec. type \\
 &  (pc) & (mag) & (mag) & (mag) & (mag) &  &  & ($M_{\odot}$) &  (kyr) &  \\
\hline
\endhead

\hline \multicolumn{11}{r}{\textit{Continued on next page}}
\endfoot

\hline
\endlastfoot 

Abell 15 & 0.421 & 15.86 & 15.73 & 0.12 & -0.56 & 3.76 & 5.04 & 1.988 & 6.59     & O(H)                                  \\
Abell 20 & 0.287 & 16.42 & 16.47 & 0.27 & -0.54 & 2.78 & 5.08 & 1.065 & 64.04  & O(H)                   \\
Abell 24 & 0.687 & 17.37 & 17.36 & 0.19 & -0.7 & 1.77 & 5.14 & >3.00 & 5.92  & $\cdots$                         \\
Abell 28 & 0.302 & 16.5 & 16.57 & 0.38 & -0.65 & 0.81 & 4.85 & 2.355 & 395.41 & DAH:                    \\
Abell 29 & 1.17 & 18.23 & 18.33 & 0.38 & -0.71 & 1.48 & 5.01 & 2.856 & 41.24 & $\cdots$                        \\
Abell 31 & 1.202 & 15.47 & 15.54 & 0.12 & -0.54 & 1.69 & 4.96 & 1.621 & 70.16   & DAO                   \\
Abell 33 & 0.653 & 15.93 & 16.03 & 0.16 & -0.23 & 2.17 & 5 & 1.054 & 76.99   & DAO                      \\
Abell 34 & 0.826 & 16.39 & 16.4 & 0.13 & -0.63 & 2.13 & 4.99 & <1.00 & 114.75       & hgO(H)             \\
Abell 36 & 0.378 & 11.49 & 11.55 & 0.09 & -0.53 & 3.28 & 5.05 & 1.054 & 58.43  & sdO7                   \\
Abell 39 & 0.455 & 15.57 & 15.62 & 0.06 & -0.5 & 2.51 & 5.03 & <1.00 & 102.70          & DAO.7           \\
Abell 43 & 0.405 & 14.66 & 14.74 & 0.53 & -0.51 & 3.55 & 5.03 & 1.495 & 8.62   & O7fk                   \\
%Abell 46 & 0.557 & 14.96 & 14.87 & 0.49 & -0.5 & 3.08 & 4.82 & 1.088 & 48.38   & O9k                    \\
Abell 61 & 0.789 & 17.25 & 17.41 & 0.15 & -0.62 & 1.98 & 4.98 & 1.019 & 120.62           & DAO.57       \\
Abell 66 & 0.8 & 18.08 & 18.17 & 0.54 & -0.59 & 1.53 & 4.97 & 2.219 & 33.22 & $\cdots$                         \\
Abell 7 & 0.972 & 15.43 & 15.5 & 0.08 & -0.52 & 1.73 & 4.99 & 1.642 & 62.68  & DAO.6                    \\
Abell 74 & 1.307 & 17.02 & 17.05 & 0.25 & -0.6 & 1.54 & 5.03 & 2.928 & 29.48   & DAO                    \\
BMPJ0642-0417 & 1.097 & 18.76 & 18.5 & 0.86 & -0.13 & 0.58 & 4.78 & 1.823 & 838.06       & Blue         \\
BlDz 1 & 0.329 & 18.27 & 18.4 & 0.42 & -0.52 & 1.94 & 5.11 & >3.00 & 3.21 & O(H)                         \\
%DS 1 & 0.64 & 12.14 & 12.11 & 0.45 & -0.53 & 3.52 & 4.95 & 1.127 & 49.43   & sdO                        \\
DS 2 & 0.359 & 12.33 & 12.37 & 0.68 & -0.63 & 3.45 & 4.93 & 1.061 & 52.54  & O(H)                       \\
DeHt 2 & 0.451 & 14.98 & 15.04 & 0.52 & -0.57 & 3.35 & 5.07 & 1.113 & 53.94 & sdO                       \\
H 2-1 & 0.026 & 13.02 & 12.82 & 2.31 & -0.31 & 3.73 & 4.48 & 1.345 & 4.86   & O(H)5-9                   \\
HaWe 13 & 0.349 & 16.68 & 16.9 & 1.52 & -0.66 & 2.43 & 4.83 & <1.00 & 108.43            & hgO(H)         \\
HaWe 4 & 1.219 & 17.12 & 17.19 & 0.61 & -0.49 & 1.9 & 5.03 & 1.98 & 10.20   & DAO.6                     \\
HaWe 6 & 0.12 & 16.48 & 16.6 & 0.38 & -0.58 & 0.07 & 4.67 & 1.983 & 2136.29 & DA1.0                     \\
%Hb 7 & 0.036 & 13.86 & 13.97 & 0.64 & -0.7 & 3.67 & 4.72 & 1.505 & 6.13     & O3                        \\
HbDs 1 & 0.225 & 12.44 & 12.5 & 0.39 & -0.51 & 3.52 & 5.05 & 1.501 & 8.76 & O(H)3 Vz                    \\
Hen 2-107 & 0.077 & 14.64 & 14.69 & 4.14 & -0.85 & 4.17 & 4.57 & >3.00 & 1.04         & O(H)4Ifc        \\
Hen 2-108 & 0.114 & 12.67 & 12.42 & 1.32 & -0.46 & 4.37 & 4.7 & >3.00 & 1.06  & O(H)                     \\
Hen 2-138 & 0.068 & 10.92 & 10.71 & 0.37 & -0.08 & 4.2 & 4.46 & >3.00 & 1.03         & O(H)7-9 f         \\
Hen 2-187 & 0.109 & 12.51 & 12.49 & 3.08 & -1.21 & 4.01 & 4.28 & 2.601 & 1.26     & O(H)7-9 f       \\
Hen 2-51 & 0.073 & 15.07 & 15.69 & 1.83 & 0.86 & 3.56 & 4.83 & 1.108 & 47.19  & $\cdots$                       \\
IC 1295 & 0.395 & 16.82 & 16.9 & 1.11 & -0.81 & 2.61 & 4.99 & <1.00 & 102.12            & hgO(H)         \\
IC 2149 & 0.042 & 11.26 & 11.34 & 0.77 & -0.44 & 3.63 & 4.59 & 1.146 & 40.50 & O(H)4f                   \\
IC 2448 & 0.168 & 14.21 & 14.26 & 0.22 & -0.59 & 3.84 & 4.98 & 1.977 & 5.84         & O(H)3III-V       \\
IC 3568 & 0.092 & 12.87 & 12.83 & 0.42 & -0.52 & 3.38 & 4.7 & <1.00 & 78.04  & O(H)3                     \\
IC 418 & 0.04 & 10.12 & 10.23 & 0.62 & -0.27 & 3.77 & 4.58 & 1.49 & 5.41 & O7fp                         \\
IC 4593 & 0.083 & 11.22 & 11.33 & 0.21 & -0.41 & 3.73 & 4.61 & 1.395 & 5.61  & O(H)5f                   \\
IC 4642 & 0.224 & 15.9 & 15.66 & 0.79 & -0.66 & 3.92 & 5.05 & 2.449 & 3.03    & abs. lines        \\
IC 5148/50 & 0.368 & 16.07 & 16.16 & 0.02 & -0.6 & 2.31 & 5.04 & 1.122 & 71.59          & hgO(H)        \\
K 1-27 & 0.493 & 16.03 & 16.11 & 0.15 & -0.56 & 3.64 & 5.13 & 1.995 & 7.53  & sdO(He)                   \\
Lo 1 & 0.813 & 15.15 & 15.21 & 0.01 & -0.56 & 2.36 & 5.04 & 1.095 & 71.17    & hgO(H)                   \\
Lo 8 & 0.428 & 12.9 & 12.97 & 0.1 & -0.53 & 3.57 & 4.95 & 1.533 & 7.81   & O(H)3 Vz                     \\
M 1-26 & 0.041 & 12.61 & 12.61 & 3.24 & -0.3 & 4.17 & 4.52 & >3.00 & 1.04   & O(H) f                     \\
%M 1-37 & 0.007 & 14.77 & 14.57 & 2.53 & -0.23 & 3.53 & 4.51 & 1.059 & 39.05    & O(H)                   \\
M 1-46 & 0.063 & 12.84 & 12.76 & 2.44 & -0.52 & 4.26 & 4.7 & >3.00 & 1.06           & O(H)7I(fc)         \\
M 1-53 & 0.072 & 15.43 & 15.52 & 2.44 & -0.95 & 3.9 & 4.72 & 2.182 & 2.41 & O(H)3 If                    \\
M 2-12 & 0.054 & 14.25 & 14.19 & 2.6 & -0.41 & 3.49 & 4.34 & 1.03 & 33.92    & O7-8                     \\
NGC 2022 & 0.139 & 15.7 & 15.75 & 0.99 & -0.72 & 3.27 & 5 & <1.00 & 91.75  & O(H)                        \\
NGC 2792 & 0.108 & 16.74 & 16.89 & 1.24 & -0.43 & 3.36 & 5.1 & 1.163 & 50.86 & $\cdots$                        \\
NGC 3587 & 0.405 & 15.72 & 15.74 & 0.04 & -0.59 & 1.98 & 4.97 & <1.00 & 154.48  & DAO                    \\
NGC 4361 & 0.285 & 13.09 & 13.26 & 0.13 & -0.61 & 3.54 & 5.1 & 1.312 & 9.90  & O(H)6                    \\
NGC 5882 & 0.069 & 13.33 & 13.42 & 0.84 & -0.55 & 3.63 & 4.83 & 1.53 & 6.86 & O(H) f                    \\
NGC 5979 & 0.186 & 16.28 & 16.37 & 1.02 & -0.57 & 3.74 & 5.06 & 1.993 & 6.95            & O(H)3-4       \\
NGC 6058 & 0.211 & 13.73 & 13.94 & 0.57 & -0.84 & 3.73 & 4.89 & 1.412 & 7.48  & O(H)3                   \\
NGC 6072 & 0.178 & 18.55 & 18.47 & 1.83 & -0.53 & 2.31 & 5.15 & 2.409 & 3.94  & $\cdots$                       \\
NGC 6153 & 0.083 & 15.31 & 15.55 & 2.77 & -0.72 & 3.76 & 5.04 & 1.991 & 6.74   & wels                   \\
NGC 6210 & 0.065 & 12.46 & 12.43 & 0.09 & -0.52 & 3.81 & 4.88 & 1.99 & 6.02  & O(H)3                    \\
NGC 6543 & 0.078 & 11.24 & 11.29 & 0.26 & -0.46 & 3.47 & 4.68 & 1.015 & 45.40          & Of-WR(H)       \\
NGC 6563 & 0.122 & 17.28 & 17.49 & 0.22 & -0.81 & 1.84 & 5.09 & 2.932 & 6.35  & $\cdots$                       \\
NGC 6572 & 0.061 & 12.65 & 13 & 0.67 & -0.61 & 3.65 & 4.84 & 1.509 & 6.96           & Of-WR(H)          \\
NGC 6720 & 0.147 & 15.65 & 15.78 & 0.44 & -0.79 & 2.3 & 5.05 & 1.537 & 23.86            & hgO(H)        \\
NGC 6772 & 0.184 & 18.4 & 18.61 & 1.96 & -0.55 & 2.21 & 5.13 & 2.384 & 4.03  & $\cdots$                        \\
NGC 6781 & 0.15 & 16.74 & 16.88 & 1.64 & -0.51 & 1.94 & 5.05 & 2.205 & 5.19    & DAO                    \\
NGC 6853 & 0.382 & 14.04 & 14.09 & 0.14 & -0.62 & 2.27 & 5.06 & 1.491 & 25.29 & DAO.6                   \\
NGC 6891 & 0.077 & 12.29 & 12.43 & 0.59 & -0.54 & 3.73 & 4.7 & 1.361 & 6.15       & O(H)3 Ib(f)        \\
NGC 7009 & 0.072 & 12.77 & 12.87 & 0.25 & -0.55 & 3.45 & 4.94 & 1.066 & 52.68  & O(H)                   \\
NGC 7094 & 0.393 & 13.52 & 13.59 & 0.25 & -0.53 & 3.71 & 5.04 & 1.791 & 7.77 & hybrid                   \\
NGC 7293 & 0.41 & 13.46 & 13.52 & 0.02 & -0.6 & 1.79 & 5.03 & 2.355 & 8.94  & DAO.5                     \\
NGC 7662 & 0.117 & 13.93 & 14 & 0.35 & -0.62 & 3.61 & 5.05 & 1.362 & 9.08   & O(H)                      \\
PB 4 & 0.147 & 16.28 & 15.96 & 1.67 & -0.78 & 3.97 & 4.89 & 2.408 & 2.79  & wels?                       \\
PuWe 1 & 1.156 & 15.51 & 15.55 & 0.35 & -0.59 & 1.73 & 5.04 & 2.627 & 11.94   & DAO.5                   \\
RWT 152 & 0.202 & 12.97 & 13.02 & 0.37 & -0.49 & 3.56 & 4.65 & 1.089 & 43.00    & sdO                   \\
Sh 2-216 & 1.835 & 12.61 & 12.67 & 0.12 & -0.57 & 1.59 & 4.96 & 1.578 & 95.47 & DAO.6                   \\
TK1 & 2.708 & 15.66 & 15.74 & 0.11 & -0.6 & 1.57 & 4.93 & 1.109 & 205.85    & DAO.7                     \\
Vy 2-3 & 0.055 & 14.6 & 14.55 & 3.3 & -1.67 & 4.12 & 4.53 & >3.00 & 1.04    & O(H)3-4 I          \\
WeDe 1 & 1.338 & 17.17 & 17.24 & 0.28 & -0.58 & 1.56 & 5.1 & >3.00 & 17.12  & DA.3              \\
Wray 17-31 & 0.5 & 17.9 & 17.94 & 0.75 & -0.7 & 2.15 & 5.08 & 1.981 & 8.04  & DAO               \\

\end{longtable}

\normalsize

%\onecolumn

\tablefoot{
%The value $G$ is the integrated magnitude in Gaia photometric instrument band and $G_{BP}-G_{RP}$ is the colour in the two Gaia photometric bands. 
Masses and evolutionary ages are estimated from \citet{2016A&A...588A..25M} evolutionary tracks. See Sect. 4 for more details.
}

\tablebib{ $V$ magnitudes are taken from: APASS database, \citet{frew08}, \citet{frew16}, \citet{1991A&AS...89...77T} and \citet{2020A&A...640A..10W}.
Temperatures are taken from \citet{frew08}, \citet{frew16}, \citet{1989A&A...222..237G} and \citet{2013A&A...553A.126G}.
Spectral types are taken from \citet{2020A&A...640A..10W}.
}

%\end{landscape}

%\newpage
%
%
%\section{Images of Binary CSPNe}
%
%%%%%%%%%%%%%%%%%%%%%% Imagenes
%\begin{figure}[h!]
%        \includegraphics[width=9.2cm,height=7.7cm]{images/Abell_24.png}
%        \includegraphics[width=9.2cm,height=7.7cm]{images/Abell_33.png}
%        \includegraphics[width=9.2cm,height=7.7cm]{images/Abell_34.png}
%        \includegraphics[width=9.2cm,height=7.7cm]{images/Fr_2-42.png}
%        \includegraphics[width=9.2cm,height=7.7cm]{images/IPHASJ183531.png}
%        \includegraphics[width=9.2cm,height=7.7cm]{images/NGC_3699.png}
%        \label{fig:images_1}
%\end{figure}
%
%\begin{figure}[h!]
%        \includegraphics[width=9.2cm,height=7.7cm]{images/NGC_6720.png}
%        \includegraphics[width=9.2cm,height=7.7cm]{images/NGC_6781.png}
%        \includegraphics[width=9.2cm,height=7.7cm]{images/NGC_6853.png}
%        \includegraphics[width=9.2cm,height=7.7cm]{images/Sh_2-123.png}
%
%        \caption{Identification images of \object{Abell 24}, \object{Abell 33}, \object{Abell 34}, %\object{Fr 2-42}, \object{IPHASJ183531}, \object{NGC 3699}, \object{NGC 6720}, \object{NGC %6781},  \object{NGC 6853} and \object{Sh 2-123} binary systems (from left to right and from %up to down), showing the location of both the CSPN (with a cross) and the comoving %companion. Images are from \textit{Aladin} sky atlas (CDS).}
%        \label{fig:images_2}
%\end{figure}

%\end{center}

\end{appendix}

\end{document}